\documentclass[a4paper]{article} 
\usepackage{jheppub} 
\usepackage[displaymath]{lineno} 
\usepackage{lineno} 
\usepackage{lscape}

\graphicspath{{figures/}}
\usepackage{subfigure}
\usepackage{mathrsfs}
\usepackage{multirow}
\usepackage{pbox}
\usepackage{rotating}
\usepackage{url}
\usepackage{epstopdf}
\usepackage{graphicx}
\usepackage{atlasphysics}
\usepackage{preprintcover}  
\PreprintCoverPaperTitle{Search for squarks and gluinos with the ATLAS detector in final states with jets and missing transverse momentum using $\sqrt{s}=8$ TeV proton--proton collision data}
\PreprintIdNumber{CERN-PH-EP-2014-093}  
\PreprintCoverAbstract{ A search for squarks and gluinos in final states containing high-$p_{\rm T}$~jets, missing transverse momentum and no electrons or muons is presented. 
The data were recorded in 2012 by the ATLAS experiment in $\sqrt{s}=8$ TeV proton--proton collisions at the Large Hadron Collider, with a total integrated luminosity of $20.3~\mathrm{fb}^{-1}$. No significant excess above the Standard Model expectation is observed.
Results are interpreted in a variety of simplified and specific supersymmetry-breaking
models assuming that $R$-parity is conserved and that the lightest neutralino is the lightest
supersymmetric particle. An exclusion limit at the 95\% confidence level on the mass of the
gluino is set at 1330 GeV for a simplified model incorporating only a gluino and the lightest neutralino. For a simplified model involving the strong production
of first- and second-generation squarks, squark masses below 850~GeV (440~GeV) are excluded for a massless lightest neutralino, assuming mass degenerate (single light-flavour) squarks.
In mSUGRA/CMSSM models with $\tan\beta=30$, $A_0=-2m_0$ and $\mu> 0$, squarks and gluinos of equal mass are excluded for masses below 1700~GeV. Additional limits are set for non-universal Higgs mass models with gaugino mediation and for simplified models involving the pair production of gluinos, each decaying to a top squark and a top quark, with the top squark decaying to a charm quark and a neutralino. These limits extend the region of supersymmetric parameter space excluded by previous searches with the ATLAS detector.}
\PreprintJournalName{JHEP}  

\makeatletter
\g@addto@macro\bfseries{\boldmath}
\makeatother

\newcommand{\meff}{\ensuremath{m_{\mathrm{eff}}}}

\def\Ptmiss{\ensuremath{{\bf E}_\mathrm{T}^{\mathrm{miss}}}}

\def\lsim{\mathrel{\rlap{\lower4pt\hbox{\hskip1pt$\sim$}}
    \raise1pt\hbox{$<$}}}                
\def\gsim{\mathrel{\rlap{\lower4pt\hbox{\hskip1pt$\sim$}}
    \raise1pt\hbox{$>$}}}                
\makeatletter
\renewcommand{\p@subfigure}{\arabic{figure}}
\renewcommand{\thesubfigure}{\alph{subfigure}}
\renewcommand{\@thesubfigure}{(\thesubfigure)}
\makeatother

\title{\boldmath Search for squarks and gluinos with the ATLAS detector in final states with
                  jets and missing transverse momentum using $\sqrt{s}=8$ TeV proton--proton collision data}
\author{ATLAS Collaboration}

\abstract{A search for squarks and gluinos in final states containing high-$p_{\rm T}$~jets, missing
transverse momentum and no electrons or muons is presented. 
The data were recorded in 2012 by the ATLAS experiment in $\sqrt{s}=8$ TeV proton--proton collisions at the Large Hadron Collider, with a total integrated luminosity of $20.3~\mathrm{fb}^{-1}$.
Results are interpreted in a variety of simplified and specific supersymmetry-breaking
models assuming that $R$-parity is conserved and that the lightest neutralino is the lightest
supersymmetric particle. An exclusion limit at the 95\% confidence level on the mass of the
gluino is set at 1330 GeV for a simplified model incorporating only a gluino and the lightest neutralino. For a simplified model involving the strong production
of first- and second-generation squarks, squark masses below 850 GeV (440 GeV) are excluded for a massless lightest neutralino, assuming mass degenerate (single light-flavour) squarks.
In mSUGRA/CMSSM models with $\tan\beta=30$, $A_0=-2m_0$ and $\mu> 0$, squarks and gluinos of equal mass are excluded for masses below 1700~GeV. Additional limits are set for non-universal Higgs mass models with gaugino mediation and for simplified models involving the pair production of gluinos, each decaying to a top squark and a top quark, with the top squark decaying to a charm quark and a neutralino. These limits extend the region of supersymmetric parameter space excluded by previous searches with the ATLAS detector.
}

\newcommand{\mytabref}[1]{Table~\ref{#1}}

\newcommand{\ourvecptmiss}{\Ptmiss}
\newcommand{\ourmagptmiss}{\MET}

\newcommand{\ourdeltaphishort}{\Delta\phi}
\newcommand{\ourdeltaphifull}{\ourdeltaphishort(\textrm{jet},\ourvecptmiss)_\mathrm{min}}

\newcommand{\alpgen}{\textsc{ALPGEN}}
\newcommand{\powheg}{\textsc{POWHEG-BOX}}

\newcommand{\herwig}{\textsc{HERWIG}}
\newcommand{\pythia}{\textsc{PYTHIA}}
\newcommand{\mcatnlo}{\textsc{MC@NLO}}

\newcommand{\madgraph}{\textsc{MADGRAPH}}

\newcommand{\sherpa}{\textsc{SHERPA}}

\newcommand{\Perugia}{{\textsc{Perugia2011C}}}
\newcommand{\AcerMC}{{\textsc{AcerMC}}}
\newcommand{\Madgraph}{{\textsc{MADGRAPH}}}
\newcommand{\ourintlumi}{{$20.3~\ifb$}}
\newcommand{\ourpt}{p_\mathrm{T}}
\newcommand{\mSUGRA}{mSUGRA/CMSSM}

\newcommand{\citemsugraandcmssm}{\cite{Chamseddine:1982jx,Barbieri:1982eh,Ibanez:1982ee,Hall:1983iz,Ohta:1982wn,Kane:1993td}}

\begin{document}

\maketitle 

\section{Introduction}

Many extensions of the Standard Model (SM) include heavy coloured
particles, some of which could be accessible at the Large Hadron Collider (LHC) \cite{LHC:2008}.  The squarks ($\squark$) and
gluinos ($\gluino$) of supersymmetric (SUSY) theories~\cite{Miyazawa:1966,Ramond:1971gb,Golfand:1971iw,Neveu:1971rx,Neveu:1971iv,Gervais:1971ji,Volkov:1973ix,Wess:1973kz,Wess:1974tw} form one class of such
particles.  In these theories the squarks $\squark_{\rm L}$ and $\squark_{\rm R}$ are the partners of the left- and right-handed SM quarks respectively, while the gluinos ($\gluino$) are the partners of the SM gluons. The partners of the neutral and charged SM gauge and Higgs bosons are respectively the neutralinos ($\nino$) and charginos ($\chinopm$). This paper presents a search for these particles in final states containing only jets and large missing transverse momentum.
Interest in this final state is motivated by the large number of $R$-parity-conserving~\cite{Fayet:1976et,Fayet:1977yc,Farrar:1978xj,Fayet:1979sa,Dimopoulos:1981zb} models in which squarks (including anti-squarks) and gluinos can be produced in pairs ($\gluino\gluino$, 
$\squark \squark$, $\squark
\gluino$) and can decay through
$\squark \to q
\ninoone$
and $\gluino
\to q \bar{q} \ninoone$ to weakly interacting lightest neutralinos, $\ninoone$. The $\ninoone$ is the lightest SUSY particle (LSP) in these models and escapes the detector
unseen. Additional decay modes can include the production of charginos via $\squark\to q\chinopm$ (where $\squark$ and $q$ are of different flavour) and $\gluino\to q\bar{q}\chinopm$. Subsequent decay of these charginos to $W^{\pm}\ninoone$ can lead to final states with still larger multiplicities of jets.
The analysis presented here updates previous ATLAS results obtained using similar selections \cite{Aad:2012fqa,Aad:2011ib,daCosta:2011qk}. Further results of relevance to these models were published by the CMS collaboration \cite{Chatrchyan:2013lya,Chatrchyan:2012uea,Chatrchyan:2012jx,Chatrchyan:2012lia}.

In this analysis, events with reconstructed electrons or muons are
vetoed to avoid overlap with a related ATLAS search \cite{Aad:2012ms}. The search strategy is optimised in the $(m_{\gluino},m_{\squark})$-plane (where $m_{\gluino},m_{\squark}$ are the gluino and squark masses respectively) for a range
of models, including simplified models in which all other supersymmetric particles, except for
the lightest neutralino, are assigned masses beyond the reach of the LHC.
Although interpreted in terms of SUSY models, the main results of this analysis (the data and expected background event counts after selection requirements) are relevant for constraining any model of new physics that predicts production of jets in association with missing transverse momentum.  

\section{The ATLAS detector}

The ATLAS detector~\cite{Aad:2008zzm} is a multipurpose particle
physics detector with a forward-backward symmetric cylindrical
geometry and nearly 4$\pi$ coverage in solid angle.\footnote{
ATLAS uses a right-handed coordinate system with its origin at the nominal
interaction point in the centre of the detector and the $z$-axis along the
beam pipe. Cylindrical coordinates $(r,\phi)$ are used in the transverse
plane, $\phi$ being the azimuthal angle around the beam pipe. The pseudorapidity $\eta$ is
defined in terms of the polar angle $\theta$ by $\eta=-\ln\tan(\theta/2)$.
} The detector features four superconducting magnet systems, which comprise a
thin solenoid surrounding inner tracking detectors (covering $|\eta|<2.5$) and, outside a calorimeter system, three large
toroids supporting a muon spectrometer (covering $|\eta|<2.7$, with trigger coverage in the region $|\eta|<2.4$).
The calorimeters are of particular importance to this analysis.  In the pseudorapidity
region $\left|\eta\right| < 3.2$, high-granularity liquid-argon (LAr)
electromagnetic (EM) sampling calorimeters are used.  An iron/scintillator-tile
calorimeter provides hadronic coverage over
$\left|\eta\right| < 1.7$.  The end-cap and forward regions, spanning
$1.5 < \left|\eta\right| < 4.9$, are instrumented with LAr calorimeters
for both EM and hadronic energy measurements.

\section{Dataset and trigger}

The dataset used in this analysis was collected in 2012 with the LHC operating at a centre-of-mass energy of 8~\TeV.
Application of beam, detector and data-quality requirements resulted in
a total integrated luminosity of \ourintlumi. The uncertainty on the integrated luminosity is $\pm 2.8$\%, derived by following the same methodology as that detailed in ref.~\cite{Aad:2013ucp}. During the data-taking period, the peak instantaneous luminosity per LHC fill was typically $7\times 10^{33}$ cm$^{-2}$ s$^{-1}$, while the mean number of proton--proton interactions per LHC bunch crossing was 21.
The trigger required events to contain a jet with an uncorrected transverse momentum ($\ourpt$) above 80 GeV and an uncorrected missing transverse momentum above 100 GeV. The trigger reached its full efficiency for events with a reconstructed jet with $\ourpt$ exceeding $130 \GeV$ and more than 160~\GeV{} of missing transverse momentum, which are requirements of the event selections considered in this analysis. Auxiliary data samples used to estimate the yields of background events in the analysis were selected using triggers requiring a single isolated electron ($\pt>24$ GeV), muon ($\pt>24$ GeV) or photon ($\pt>120$ GeV).

\section{Monte Carlo data samples}
\label{sec:mcsamples}

Monte Carlo (MC) data samples are used to develop the analysis, optimise the selections, estimate backgrounds and assess sensitivity to specific SUSY signal models.  The SM background processes considered are those which can lead to events with jets and missing transverse momentum. The processes considered together with the MC generators, cross-section calculations and parton distribution functions (PDFs) used are listed in table~\ref{tab:MCsamples}. The $\gamma$+jets MC data samples are used to estimate the $Z$+jets background through a data-driven normalisation procedure described in section~\ref{sec:analysisprocedure}. When considering the dominant $W/Z/\gamma^*$+jets and $t\bar{t}$ background processes, two generators are used in each case, with results from the second being used to evaluate systematic uncertainties in background estimates obtained with the first. When using the baseline \powheg+\pythia{} top quark pair production sample, events are reweighted in bins of $\pt(t\bar{t})$ to match the top quark pair differential cross-section observed in ATLAS data 
\cite{Aad:2012hg,Aad:2014zka}. No corrections are applied to the alternative \mcatnlo{} sample used for systematic uncertainty evaluation, which reproduces more accurately the $\pt(t\bar{t})$ distribution measured in data. \mcatnlo{} is nevertheless not used as the default generator for this process as it is observed to reproduce less accurately high jet-multiplicity events. 

\begin{table}[h]
  \centering
  \caption{The Standard Model background Monte Carlo simulation samples used in this article. The generators, the order in $\alpha_{\rm s}$ of cross-section calculations used for yield normalisation (leading order/LO, next-to-leading order/NLO, next-to-next-to-leading order/NNLO, next-to-next-to-leading logarithm/NNLL), tunes used for the underlying event and PDF sets are shown. Samples denoted with (\textbullet) are used for evaluation of systematic uncertainties. For the $\gamma$+jets process the LO cross-section is taken directly from the MC generator.\vspace{2mm}
\label{tab:MCsamples} }

\scriptsize
\renewcommand\arraystretch{1.1}
\begin{tabular}{ ccccc }
\hline
\multirow{2}{*}{Process} &  Generator &Cross-section  & \multirow{2}{*}{Tune}   & \multirow{2}{*}{PDF set}  \\
 & + frag./had. &order in $\alpha_s$ & & \\
\hline

$W$+jets  & \sherpa-1.4.0 ~\cite{Gleisberg:2008ta} &   NNLO~\cite{Catani:2009sm} &\sherpa{} default  & CT10 ~\cite{CT10pdf} \\
\multirow{2}{*}{$W$+jets (\textbullet)}  & \alpgen-2.14 ~\cite{Mangano:2002ea} &  \multirow{2}{*}{NNLO~\cite{Catani:2009sm}} & \multirow{2}{*}{AUET2B~\cite{mc10a}} & \multirow{2}{*}{CTEQ6L1~\cite{Pumplin:2002vw}} \\
&+ \herwig-6.520~\cite{Corcella:2000bw,herwig65long} & & &\\
\hline

 $Z/\gamma^{*}$+jets & \sherpa-1.4.0  &   NNLO~\cite{Catani:2009sm} &\sherpa{} default  & CT10 \\
\multirow{2}{*}{$Z/\gamma^{*}$+jets (\textbullet)}&  \alpgen-2.14  &  \multirow{2}{*}{NNLO~\cite{Catani:2009sm}} & \multirow{2}{*}{AUET2B} & \multirow{2}{*}{CTEQ6L1} \\
&+ \herwig-6.520&&&\\
\hline

$\gamma$+jets & \sherpa-1.4.0  &   LO &\sherpa{} default  & CT10 \\
\multirow{2}{*}{$\gamma$+jets (\textbullet)}&  \alpgen-2.14  &  \multirow{2}{*}{LO} & \multirow{2}{*}{AUET2B} & \multirow{2}{*}{CTEQ6L1} \\
&+ \herwig-6.520&&&\\
\hline

\multirow{2}{*}\ttbar &\powheg-1.0~\cite{Nason:2004rx,Frixione:2007vw,Alioli:2010xd} & \multirow{2}{*}{NNLO+NNLL~\cite{Czakon:2013goa,Czakon:2011xx}}& \Perugia & \multirow{2}{*}{CT10}  \\ 
& + \pythia-6.426 \cite{Sjostrand:2006za} &&\cite{Cooper:2011gk,Skands:2010ak}&\\
\multirow{2}{*}{\ttbar{} (\textbullet)}& \mcatnlo-4.03~\cite{Frixione:2002ik,Frixione:2003ei} &\multirow{2}{*}{NNLO+NNLL~\cite{Czakon:2013goa,Czakon:2011xx}} & \multirow{2}{*}{AUET2B} & \multirow{2}{*}{CT10}  \\ 
& + \herwig-6.520 & & &\\
\hline

Single top & & & & \\
        \multirow{2}{*}{$t$-channel} & \AcerMC-38~\cite{Kersevan:2004yg}  & \multirow{2}{*}{NNLO+NNLL~\cite{Kidonakis:2011wy}}  & \multirow{2}{*}{AUET2B} & \multirow{2}{*}{CTEQ6L1} \\
        &  + \pythia-6.426 &&&\\
        \multirow{2}{*}{$s$-channel, $Wt$} &  \mcatnlo-4.03 & \multirow{2}{*}{NNLO+NNLL~\cite{Kidonakis:2010tc, Kidonakis:2010ux}} & \multirow{2}{*}{AUET2B} & \multirow{2}{*}{CT10} \\
        &  + \herwig-6.520&&&\\
\hline

 \multirow{2}{*}{ \ttbar+EW boson} &\Madgraph-5.0~\cite{Alwall:2011uj}  &   \multirow{2}{*}{NLO~\cite{Lazopoulos:2008de,Garzelli:2012bn,Campbell:2012dh}} &  \multirow{2}{*}{AUET2B} &  \multirow{2}{*}{CTEQ6L1} \\ 
 &+ \pythia-6.426&&&\\
\hline

Dibosons & & & & \\
 $WW$, $WZ$, $ZZ$,  &  \multirow{2}{*}{\sherpa-1.4.0} & \multirow{2}{*}{NLO~\cite{Campbell:1999ah,Campbell:2011bn}} &\multirow{2}{*}{\sherpa{} default}  & \multirow{2}{*}{CT10} \\
   $W\gamma$ and $Z\gamma$&&&&\\
\hline


\hline
\end{tabular} 
     
\end{table}

SUSY signal samples are generated with \herwig++-2.5.2~\cite{Bahr:2008pv} or \Madgraph-5.0 matched to \pythia-6.426, using PDF set CTEQ6L1. The specific generators used for each model are discussed in section~\ref{sec:interpret}. The \Madgraph{} samples are produced using the AUET2B tune (also used for some background samples -- see table~\ref{tab:MCsamples}). The MLM matching scheme \cite{Mangano:2006rw} is used with up to one additional jet in the \Madgraph{} matrix element, and a \Madgraph{} $k_{\rm t}$ measure cut-off and a \pythia{} jet measure cut-off both set to 0.25 times the mass scale of the SUSY particles produced in the hard process, with a maximum value of 500 GeV. Signal cross-sections are calculated to next-to-leading order in the strong coupling constant, including the resummation of soft gluon emission at next-to-leading-logarithmic accuracy (NLO+NLL) \cite{Beenakker:1996ch,Kulesza:2008jb,Kulesza:2009kq,Beenakker:2009ha,Beenakker:2011fu}. In each case the nominal cross-section and its uncertainty are taken from an ensemble of cross-section predictions using different PDF sets and factorisation and renormalisation scales, as described in ref.~\cite{Kramer:2012bx}. For the \mSUGRA{} \citemsugraandcmssm{} and non-universal Higgs mass model with gaugino mediation (NUHMG)~\cite{Covi:2007xj} samples the SUSY particle mass spectra and decay tables are calculated with \textsc{SUSY-HIT} \cite{Djouadi:2006bz} interfaced to the \textsc{SOFTSUSY} spectrum generator \cite{Allanach:2001kg} and \textsc{SDECAY} \cite{Muhlleitner:2003vg}.

The MC samples are generated using the same parameter set as in refs.~\cite{ATL-PHYS-PUB-2011-009,ATL-PHYS-PUB-2011-014,ATLAS:1303025}. SM background samples are passed through either the full ATLAS detector simulation \cite{:2010wqa} based on \textsc{GEANT4} \cite{Agostinelli:2002hh}, or, when larger samples are required, through a fast simulation using a parameterisation of the performance of the ATLAS EM and hadronic calorimeters \cite{ATLAS:2010bfa} and \textsc{GEANT4} elsewhere ($W/Z/\gamma$+jets samples with boson $\pt<280$ GeV and \powheg{}+\pythia{} $t\bar{t}$ samples only). All SUSY signal samples with the exception of \mSUGRA{} model samples (which are produced with the \textsc{GEANT4} simulation) are passed through the fast simulation. The fast simulation of SUSY signal events was validated against full \textsc{GEANT4} simulation for several signal model points. Differing pile-up (multiple proton--proton interactions in the same or neighbouring bunch-crossings) conditions as a function of the instantaneous luminosity are taken into account by overlaying simulated minimum-bias events generated with \pythia-8 onto the hard-scattering process and reweighting them according to the distribution of the mean number of interactions observed in data.

\section{Event reconstruction}\label{sec:evreco}

Jet candidates are reconstructed using the
anti-$k_{\rm T}$ jet clustering algorithm~\cite{Cacciari:2008gp,Cacciari:2005hq} with a
radius parameter of $0.4$. The inputs to this algorithm are the energies of clusters~\cite{Aad:2011he,Lampl:2008}
of calorimeter cells seeded by those with energy significantly above
the measured noise. Jet momenta are constructed by performing a
four-vector sum over these cell clusters, treating each as an
$(E,\vec{p})$ four-vector with zero mass.  
The jets are corrected for energy
from pile-up using a method, suggested in ref.~\cite{Cacciari:2007fd}, which
estimates the pile-up activity in any given event as well as the sensitivity of any given jet
to pile-up. The method subtracts a contribution from the jet energy equal to the product
of the jet area and the average energy density of the event \cite{ATLAS-CONF-2013-083}. The local cluster weighting (LCW) jet calibration
method \cite{Issever:2004qh,Aad:2011he} is used to classify topological cell clusters within the jets as being of either electromagnetic or
hadronic origin, and based on this classification applies specific energy corrections derived
from a combination of MC simulation and data. Further corrections, referred to as `jet energy scale' or `JES' corrections below, are derived from MC simulation and data and used to calibrate  the energies of jets to the scale of their constituent particles \cite{Aad:2011he,ATLAS-CONF-2013-004}.
Only jet candidates with $\ourpt > 20$ \GeV{} and $|\eta|<4.5$ after all corrections are retained. Jets are identified as originating from heavy-flavour ($b$ and $c$ quark) decays using the `MV1' neural-network-based $b$-tagging algorithm, with an operating point with an efficiency of 70\% and a light quark rejection factor of 140 determined with simulated $t\bar{t}$ events \cite{ATLAS-CONF-2012-097}. Candidate $b$-tagged jets must possess $\pt>40$ GeV and $|\eta|<2.5$.

Two different classes of reconstructed leptons (electrons or muons) are used in this analysis. When selecting samples of potential SUSY signal events, events containing any `baseline' electrons or muons are rejected, as described in section~\ref{subsec:srdef}. The selections applied to baseline leptons are designed to maximise the efficiency with which $W$+jet and top quark background events are rejected. When selecting `control region' samples for the purpose of estimating residual $W$+jets and top quark backgrounds, as described in section~\ref{subsec:crdef}, additional requirements are applied to improve the purity of the samples. These leptons will be referred to as `high-purity' leptons and form a subset of the baseline leptons.

Baseline electron candidates are required to have $\ourpt > 10~\GeV$ and $|\eta| <
2.47$, and to satisfy `medium' electron shower shape and track selection criteria based upon those described in
Ref.~\cite{Aad:2014fxa}, but modified to reduce the impact of pile-up and to match tightened trigger requirements in 2012 data.  High-purity electron candidates additionally must have $\pt>25$ GeV, must satisfy tighter selection criteria, must have transverse and longitudinal impact parameters within 1.0 mm and 2.0 mm, respectively, of the primary vertex, which is defined to be the reconstructed vertex with the highest $\sum \pt^2$ of tracks, and must be isolated.\footnote{The scalar sum of the transverse momenta of tracks, other than that from the electron itself, within a cone of $\Delta R\equiv\sqrt{(\Delta\eta)^2+(\Delta\phi)^2}=0.2$ around the electron must be less than 10\% of the $\pt$ of the electron.}
Baseline muon candidates are formed by combining information from the muon spectrometer and inner tracking detectors as described in ref.~\cite{Aad:2014zya} and are required to have $\ourpt > 10$ \GeV{} and $|\eta| <2.4$.  High-purity muon candidates must additionally have $\pt > 25$ GeV, $|\eta|<2.4$, transverse and longitudinal impact parameters within 0.2 mm and 1.0 mm, respectively, of the primary vertex and must be isolated.\footnote{The scalar sum of the transverse momenta of tracks, other than that from the muon itself, within a cone of $\Delta R=0.2$ around the muon must be less than 1.8 GeV.} 

After the selections described above, ambiguities between candidate jets with $|\eta|<2.8$ and leptons are resolved as follows.
First, any such jet candidate lying within a distance $\Delta
R\equiv\sqrt{(\Delta\eta)^2+(\Delta\phi)^2}=0.2$ of a baseline electron is discarded;
then any lepton candidate (baseline or high-purity) remaining within a distance
$\Delta R =0.4$ of any surviving jet candidate is discarded.

The measurement of the missing transverse momentum two-dimensional vector
$\ourvecptmiss$ (and its magnitude $\ourmagptmiss$) is based on
the calibrated transverse momenta of all jet and baseline lepton candidates and all
calorimeter energy clusters not associated with such objects~\cite{Aad:2012re,ATLAS-CONF-2013-082}.  Following the calculation of the value of $\ourvecptmiss$, all jet candidates with $|\eta|>2.8$ are discarded. 
Thereafter, the remaining baseline lepton and jet candidates are considered
``reconstructed'', and the term ``candidate'' is dropped. In the MC simulation, reconstructed baseline or high-purity lepton and $b$-tagged jet identification efficiencies and misidentification probabilities are corrected using factors derived from data control regions.

Reconstructed photons are used to constrain $Z$+jet backgrounds (see section~\ref{subsec:crdef}), although they are not used in the main signal event selection. Photon candidates are required to possess $\ourpt > 130~\GeV$ and $|\eta| <
1.37$ or $1.52<|\eta| <
2.47$, to satisfy photon shower shape and electron rejection criteria \cite{ATLAS-CONF-2012-123}, and to be isolated.\footnote{The transverse energy in the calorimeter, other than from that from the photon itself and corrected for noise and pile-up, within a cone of $\Delta R=0.4$ around the photon must be less than 4 GeV.} Ambiguities between candidate jets and photons (when used in the event selection) are resolved by discarding any jet candidates lying within $\Delta R=0.2$ of a photon candidate. The transverse momenta of the resulting reconstructed photons are taken into account when calculating $\ourvecptmiss$.

Reconstructed $\tau$-leptons are not used in this analysis when selecting potential signal events or control region data samples; however, they are used to validate some of the estimates of $W$+jets and top quark backgrounds, as described in section~\ref{subsec:vrdef}. The $\tau$-leptons are reconstructed using a $\pt$-correlated track counting algorithm described in ref.~\cite{tautracks}. 
The purity of the validation event samples selecting background events containing hadronically decaying $\tau$-leptons ranges from 65\% to 90\%.

\section{Event selection}\label{sec:eventselection}

Events selected by the trigger are discarded if they contain any candidate jets
failing to satisfy quality selection criteria designed to suppress detector noise and
non-collision backgrounds, or if they lack a reconstructed primary
vertex associated with five or more tracks \cite{Aad:2013zwa,ATLAS-CONF-2012-020}. The criteria applied to candidate jets include requirements on the fraction of the transverse momentum of the jet carried by reconstructed charged particle tracks, and on the fraction of the jet energy contained in the EM layers of the calorimeter. A consequence of these requirements is that events containing hard isolated photons have a high probability of failing to satisfy the signal event selection criteria, under which ambiguities between candidate jets and photons are not resolved (see section~\ref{sec:evreco}). 

This analysis aims to search for the production of heavy SUSY particles decaying into jets and stable lightest neutralinos, with the latter creating missing transverse momentum. Because of the high mass scale expected for the SUSY signal, the `effective mass', $\meff$, is a powerful discriminant between the signal and most SM backgrounds. When selecting events with at least $N_{\rm j}$ jets, $\meff(N_{\rm j})$ is defined to be the scalar sum of the transverse momenta of the leading $N_{\rm j}$ jets and \met{}. The final signal selection uses requirements on $\meff({\rm incl.})$, which sums over all jets with $\ourpt>40$ GeV and \met{}. Requirements placed on $\meff$ and \met{}, which suppress the multi-jet background in which jet energy mismeasurement generates missing transverse momentum, formed the basis of the previous ATLAS jets + \met{} + 0-lepton SUSY searches~\cite{Aad:2012fqa,Aad:2011ib,daCosta:2011qk}. The same strategy is adopted in this analysis, and is described below.

\subsection{Signal regions}\label{subsec:srdef}
In order to achieve maximal reach over the
$(m_{\gluino},m_{\squark})$-plane, a variety of signal regions (SRs) are defined. Squarks typically generate at least one jet in their decays, for instance through $\squark \to q
\ninoone$, while gluinos typically generate at least two jets, for instance through $\gluino\to q \bar{q} \ninoone$. Processes contributing to $\squark\squark$, $\squark\gluino$ and $\gluino\gluino$ final states therefore lead to events containing at least two, three or four jets, respectively. Decays of heavy SUSY and SM particles produced in longer $\squark$ and $\gluino$ cascade decays (e.g. $\chinoonepm\to qq'\ninoone$) tend to further increase the jet multiplicity in the final-state. 

Fifteen inclusive SRs characterised by increasing minimum jet-multiplicity from two to six, are defined in table~\ref{tab:srdefs}. In all cases, events are discarded if they contain baseline electrons or muons with $\pt$ $>$ 10 GeV. Several SRs may be defined for the same jet-multiplicity, distinguished by increasing background rejection, ranging from `very loose' (labelled `l-') to `very tight' (labelled `t+'). The lower jet-multiplicity SRs focus on models characterised by squark pair production with short decay chains, while those requiring high jet-multiplicity are optimised for gluino pair production and/or long cascade decay chains. 

Requirements are placed upon $\ourdeltaphifull$, which is defined to be the smallest of the
azimuthal separations between $\ourvecptmiss$ and the reconstructed jets. For the 2-jet and 3-jet SRs the selection requires $\ourdeltaphifull>0.4$ using up to three leading jets with $\pt>40$ GeV if present in the event. For the other SRs an additional requirement $\ourdeltaphifull>0.2$ is placed on all jets with $\ourpt>40$ GeV. 
Requirements on $\ourdeltaphifull$ and $\ourmagptmiss/\meff(N_{\rm j})$ are designed to reduce the background from multi-jet processes.  
\begin{table}[P]
\caption{\label{tab:srdefs} Selection criteria used to define each of the signal regions in the analysis. Each SR is labelled with the inclusive jet-multiplicity considered (`2j', `3j' etc.) together with the degree of background rejection. The latter is denoted by labels `l-' (`very loose'), `l' (`loose'), `m' (`medium'), `t' (`tight') and `t+' (`very tight'). The $\met/\meff(N_{\rm j})$ cut in any $N_{\rm j}$-jet channel uses a value of $\meff$ constructed from only the leading $N_{\rm j}$ jets ($\meff(N_{\rm j})$).  However, the final $\meff$(incl.) selection, which is used to define the signal regions, includes all jets with $\pt>40~\GeV$. In SR 2jW and SR 4jW a requirement 60 GeV $<m(W_{\rm cand})<$ 100 GeV is placed on the masses of candidate resolved or unresolved hadronically decaying $W$ bosons, as described in the text.}
  \footnotesize
  \begin{center}\renewcommand\arraystretch{1.4}
    \begin{tabular}{|l|c |c|c| c|c| c|}
      \hline
      \multirow{2}{*}{Requirement}      &\multicolumn{6}{|c|}{Signal Region} \\
  \cline{2-7}
  & {\bf 2jl} & {\bf 2jm} & {\bf 2jt} & {\bf 2jW} & {\bf 3j} & {\bf 4jW}\\
 \hline
\met [GeV] $>$&\multicolumn{6}{|c|}{ 160 }\\ \hline
$\pt(j_1)$ [GeV] $>$&\multicolumn{6}{|c|}{ 130 }\\ \hline
$\pt(j_2)$ [GeV] $>$&\multicolumn{6}{|c|}{ 60 }\\ \hline
$\pt(j_3)$ [GeV] $>$&\multicolumn{4}{|c|}{--} &60 &40 \\ \hline 
$\pt(j_4)$ [GeV] $>$&\multicolumn{5}{|c|}{--} &40 \\ \hline 
$\ourdeltaphishort(\textrm{jet}_{1,2,(3)},\ourvecptmiss)_\mathrm{min}$ $>$ &\multicolumn{6}{|c|}{0.4} \\ \hline
$\ourdeltaphishort(\textrm{jet}_{i>3},\ourvecptmiss)_\mathrm{min}$ $>$ &\multicolumn{5}{|c|}{--} &0.2 \\ \hline
$W$ candidates &\multicolumn{3}{|c|}{--} &2$(W\to j)$ &--  &$(W\to j)\,+\,(W\to jj)$ \\ \hline
$\met/\sqrt{H_{\rm T}}$ [GeV$^{1/2}$] $>$ &8 &\multicolumn{2}{|c|}{15} &\multicolumn{3}{|c|}{--}  \\ \hline
$\met/\meff(N_{\rm j})$ $>$ &\multicolumn{3}{|c|}{--} & 0.25 & 0.3 & 0.35 \\ \hline
$ \meff({\rm incl.})$ [GeV] $>$ & 800 &1200 &1600 &1800 &2200 &1100 \\ \hline
\end{tabular}
    \begin{tabular}{|l|c |c|c|c|c|c|c|c|c|}
      \hline
      \multirow{2}{*}{Requirement}      &\multicolumn{9}{|c|}{Signal Region} \\
  \cline{2-10}
   & {\bf 4jl-} & {\bf 4jl} & {\bf 4jm} & {\bf 4jt} & {\bf 5j} & {\bf 6jl} & {\bf 6jm} & {\bf 6jt} & {\bf 6jt+} \\
 \hline
\met [GeV] $>$&\multicolumn{9}{|c|}{ 160 }\\ \hline
$\pt(j_1)$ [GeV] $>$&\multicolumn{9}{|c|}{ 130 }\\ \hline
$\pt(j_2)$ [GeV] $>$&\multicolumn{9}{|c|}{ 60 }\\ \hline
$\pt(j_3)$ [GeV] $>$ &\multicolumn{9}{|c|}{60}  \\ \hline
$\pt(j_4)$ [GeV] $>$ &\multicolumn{9}{|c|}{60}  \\ \hline
$\pt(j_5)$ [GeV] $>$&\multicolumn{4}{|c|}{--} &\multicolumn{5}{|c|}{60} \\ \hline
$\pt(j_6)$ [GeV] $>$&\multicolumn{5}{|c|}{--} &\multicolumn{4}{|c|}{60}  \\ \hline
$\ourdeltaphishort(\textrm{jet}_{1,2,(3)},\ourvecptmiss)_\mathrm{min}$ $>$ &\multicolumn{9}{|c|}{0.4} \\ \hline
$\ourdeltaphishort(\textrm{jet}_{i>3},\ourvecptmiss)_\mathrm{min}$ $>$ &\multicolumn{9}{|c|}{0.2}\\ \hline
$\met/\sqrt{H_{\rm T}}$ [GeV$^{1/2}$] $>$  &\multicolumn{2}{|c|}{10} &\multicolumn{7}{|c|}{--} \\ \hline
$\met/\meff(N_{\rm j})$ $>$ &\multicolumn{2}{|c|}{--} &0.4 &0.25 &\multicolumn{3}{|c|}{0.2} &0.25 &0.15 \\ \hline
$ \meff({\rm incl.})$ [GeV] $>$ &700 &1000 &1300 & 2200 &1200 &900 &1200 &1500 &1700\\ \hline
\end{tabular}
  \end{center}
\end{table}

\clearpage

In the SRs 2jl, 2jm, 2jt, 4jl and 4jl- the requirement on $\met/\meff(N_{\rm j})$ is replaced by a requirement on $\MET/\sqrt{H_{\rm T}}$ (where $H_{\rm T}$ is defined as the scalar sum of the transverse momenta of all $\ourpt > 40$ GeV jets), which was found to lead to enhanced sensitivity to models characterised by $\squark\squark$ production. Two of the SRs (2jW and 4jW) place additional requirements on the invariant masses $m(W_{\rm cand})$ of candidate $W$ bosons decaying to hadrons, by requiring 60 GeV $<m(W_{\rm cand})<$ 100 GeV. Candidate $W$ bosons are reconstructed from single high-mass jets (unresolved candidates -- `$W\to j$' in table~\ref{tab:srdefs}) or from pairs of jets (resolved candidates -- `$W\to jj$' in table~\ref{tab:srdefs}). Resolved candidates are reconstructed using an iterative procedure which assigns each jet to a unique pair with minimum separation $\Delta R(j,j)$. SR 2jW requires two unresolved candidates, while SR 4jW requires one resolved candidate and one unresolved candidate. These SRs are designed to improve sensitivity to models predicting enhanced branching ratios for cascade $\squark$ or $\gluino$ decay via $\chinoonepm$ to $W$ and $\ninoone$, in cases where the $\chinoonepm$ is nearly degenerate in mass with the $\squark$ or $\gluino$ (see section~\ref{sec:interpret}). 

Standard Model background processes contribute to the event counts in
the signal regions. The dominant sources are: $Z+$jets, $W+$jets, top quark
pairs, single top quarks, and multiple jets.
The production of boson ($W/Z/\gamma$) pairs in which at least one boson decays to charged leptons and/or neutrinos (referred to as `dibosons' below) is a small component (in most SRs $\lesssim$10\%, up to $\sim$30\% in SR 6jt, predominantly $WZ$) of the total background and is estimated with MC simulated data normalised to NLO cross-section predictions.
The majority of the $W$+jets background is composed 
of $W\to \tau\nu$ events in which the $\tau$-lepton decays to hadrons, with additional contributions from $W\to e\nu, \mu\nu$ events in which  
no baseline electron or muon is reconstructed.
The largest part of the $Z$+jets background comes from the irreducible
component in which $Z\to\nu\bar\nu$ decays generate large $\ourmagptmiss$. 
Top quark pair production followed by semileptonic decays, in particular $t \bar t \to b \bar b \tau \nu q q' $ with the $\tau$-lepton decaying to hadrons, as well as single top quark events,
can also generate large $\ourmagptmiss$ and satisfy the jet
and lepton-veto requirements at a non-negligible rate.
The multi-jet background in the signal regions is caused by
mis-reconstruction of jet energies in the calorimeters generating
missing transverse momentum, as well as by neutrino production in semileptonic decays of heavy-flavour quarks.

\subsection{Control regions}\label{subsec:crdef}
To estimate the backgrounds in a consistent and robust fashion, four control regions (CRs) are defined for each of the 15 signal regions, giving 60 CRs in total. The orthogonal CR event selections are designed to provide independent data samples enriched in particular background sources. The CR selections are optimised to maintain adequate statistical weight and negligible SUSY signal contamination, while minimising as far as possible the systematic uncertainties arising from the extrapolation of the CR event yield to the expectation in the SR. This latter requirement is addressed through the use wherever possible of CR $\meff$(incl.) selections which match those used in the SR.

The CR definitions are listed in table~\ref{tab:crdefs}. The CR$\gamma$ control region is used to estimate the contribution of $Z(\to \nu\nu)$+jets background events to each SR by selecting a sample of $\gamma$+jets events with $\pt(\gamma)>130$ GeV and then treating the reconstructed photon as contributing to $\met$. For $\pt(\gamma)$ greater than $m_Z$ the kinematics of such events strongly resemble those of $Z$+jets events \cite{Aad:2012fqa}. CRQ uses reversed selection requirements placed on $\ourdeltaphifull$ and on $\met/\meff(N_{\rm j})$ ($\met/\sqrt{H_{\rm T}}$ where appropriate) to produce data samples enriched in multi-jet background events. CRW and CRT use respectively a $b$-jet veto or $b$-jet requirement together with a requirement on the transverse mass $m_\mathrm{T}$ of a high-purity lepton with $\pt>25$ GeV and $\Ptmiss$ to select samples of  $W(\to \ell \nu)$+jets and semileptonic $t\bar{t}$ background events. These samples are used to estimate respectively the $W$+jets and combined $t\bar{t}$ and single-top background populations, treating the lepton as a jet with the same momentum to model background events in which a hadronically decaying $\tau$-lepton is produced. With the exception of SR 2jl, the CRW and CRT selections do not use the SR selection requirements applied to $\ourdeltaphifull$ or $\met/\meff(N_{\rm j})$ ($\met/\sqrt{H_{\rm T}}$ where appropriate) in order to increase CR data event statistics without significantly increasing theoretical uncertainties associated with the background estimation procedure. For the same reason, the final $ \meff({\rm incl.})$ requirements are loosened to 1300 GeV in CRW and CRT of SR 6jt. The purity of the control regions for the background process targeted in each case ranges from 48\% to 97\%.

\begin{figure}[htb]
\begin{center}
\includegraphics[height=0.45\textwidth]{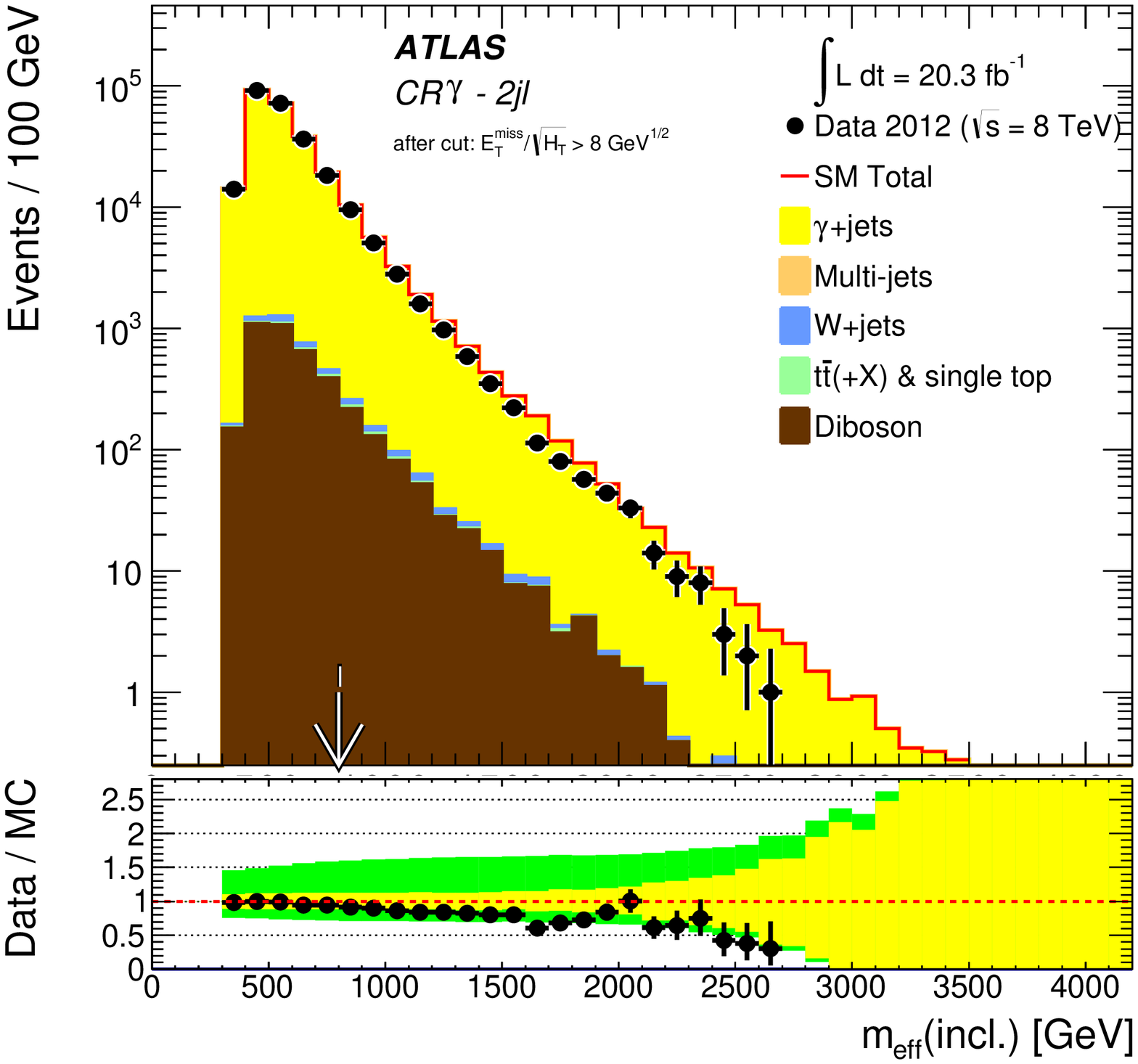}
\includegraphics[height=0.45\textwidth]{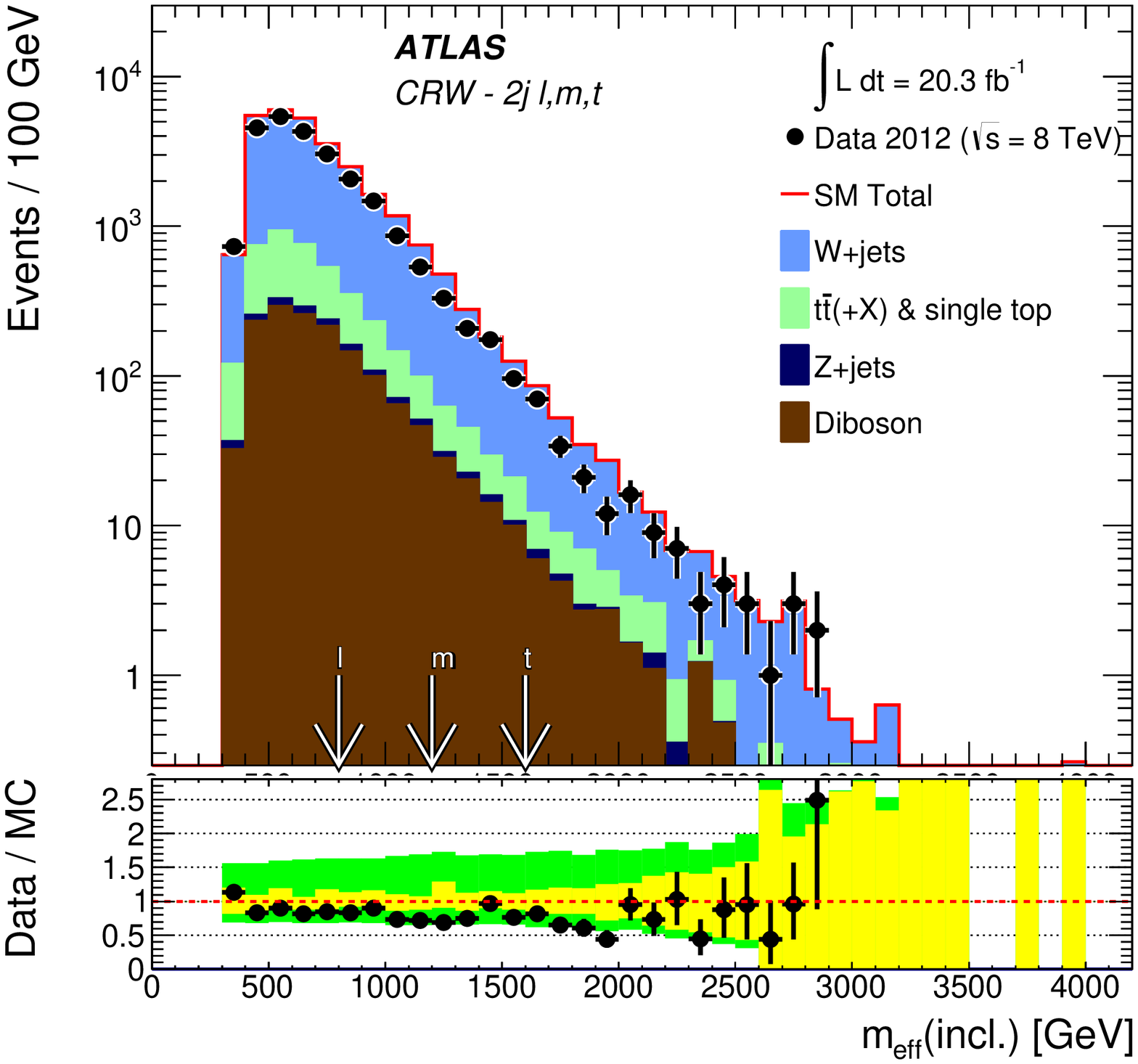}
\includegraphics[height=0.45\textwidth]{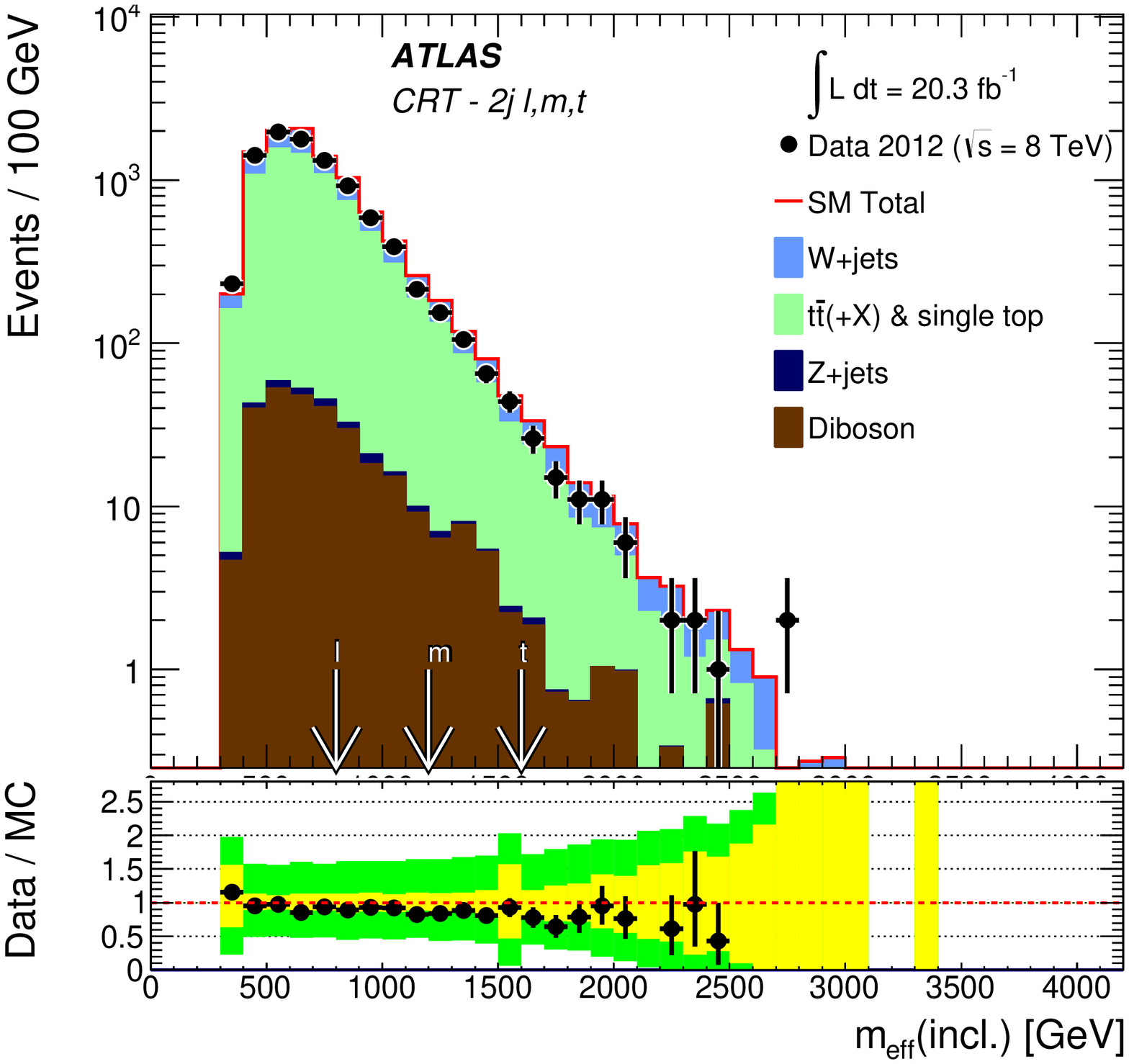}
\includegraphics[height=0.45\textwidth]{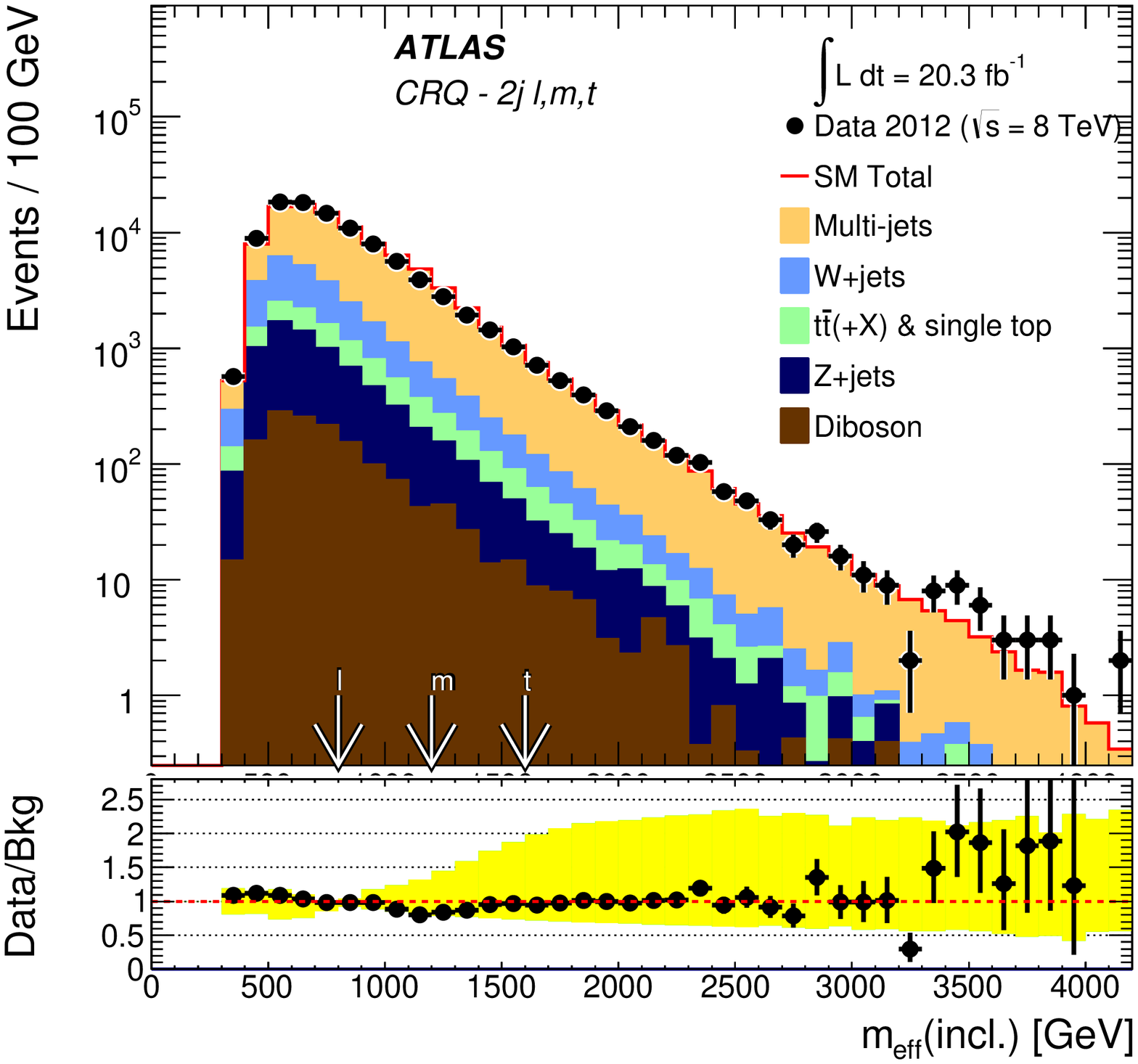}
\end{center}
\caption{\label{fig:sr2jcr_Meff}Observed $\meff({\rm incl.})$ distributions in control regions CR$\gamma$  (top left, for SR 2jl selection criteria only), CRW (top right), CRT (bottom left) and CRQ (bottom right, excluding requirements on $\met/\sqrt{H_{\rm T}}$) corresponding to SRs 2jl, 2jm and 2jt. With the exception of the multi-jet background (which is estimated using the data-driven technique described in the text), the histograms denote the MC background expectations, normalised to cross-section times integrated luminosity. In the lower panels the light (yellow) error bands denote the experimental systematic and MC statistical uncertainties, while the medium dark (green) bands include also the theoretical modelling uncertainty.  The arrows indicate the values at which the requirements on $\meff({\rm incl.})$ are applied. 
}
\end{figure}

\begin{figure}[htb]
\begin{center}
\includegraphics[height=0.45\textwidth]{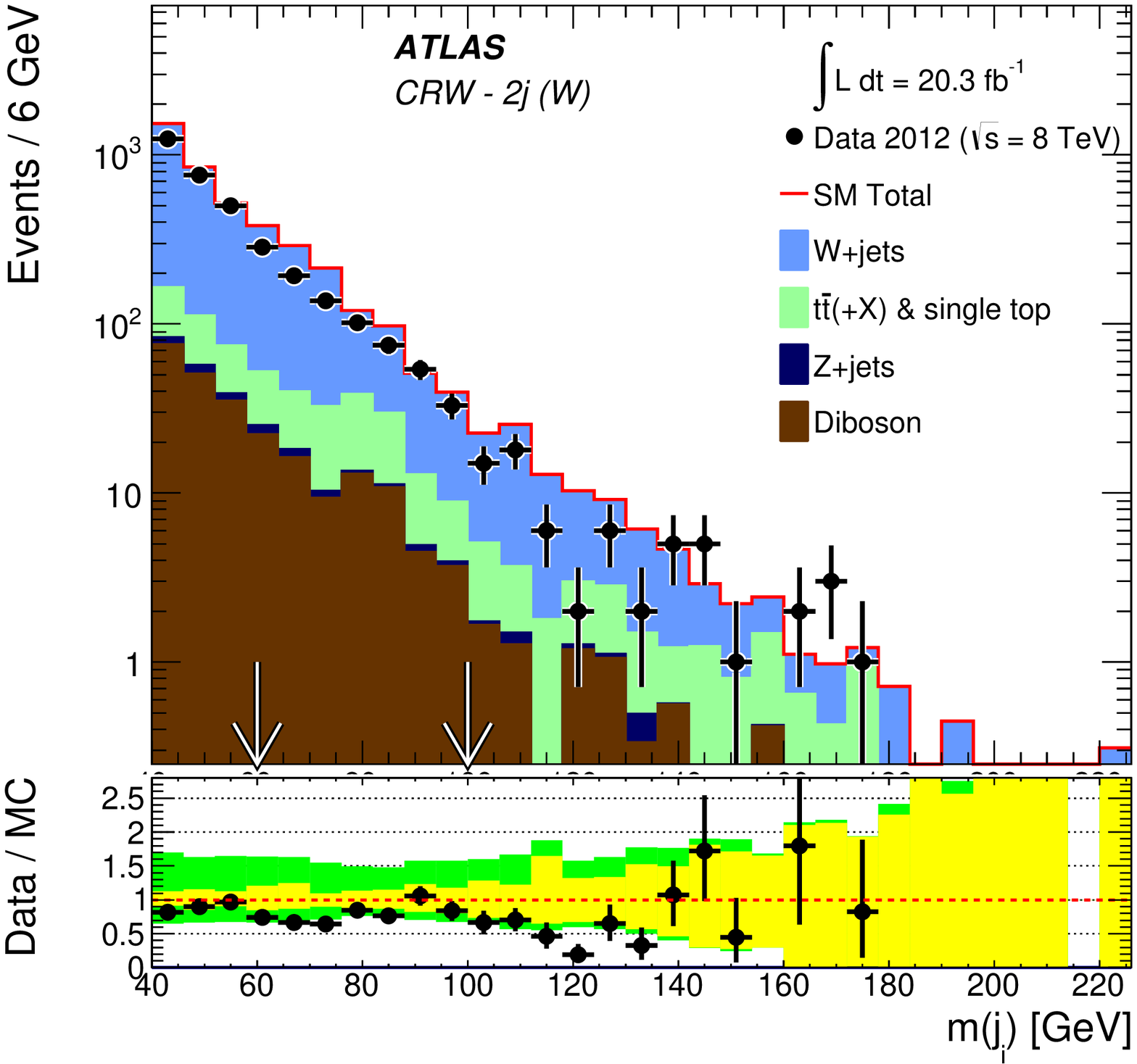}
\includegraphics[height=0.45\textwidth]{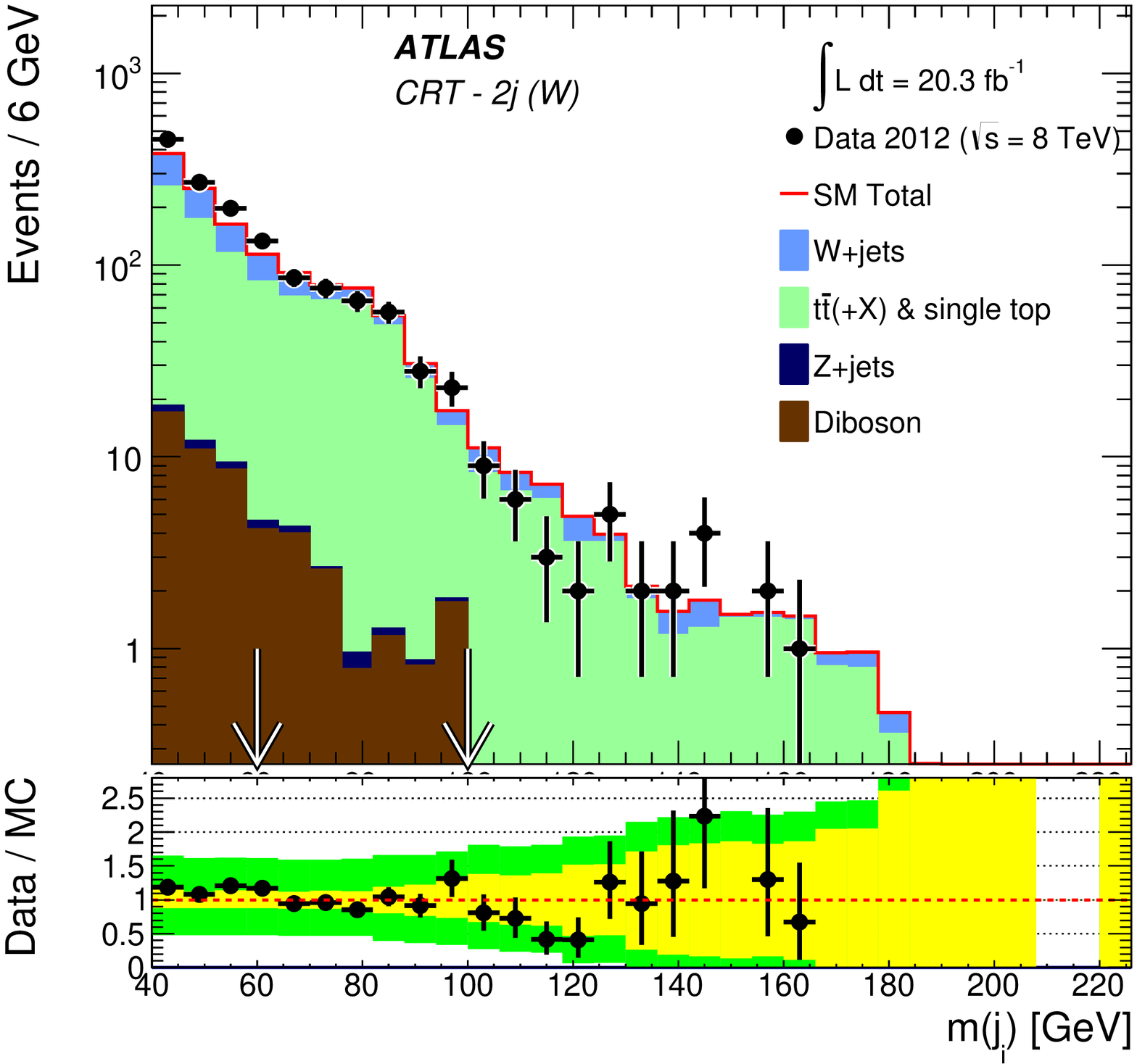}
\includegraphics[height=0.45\textwidth]{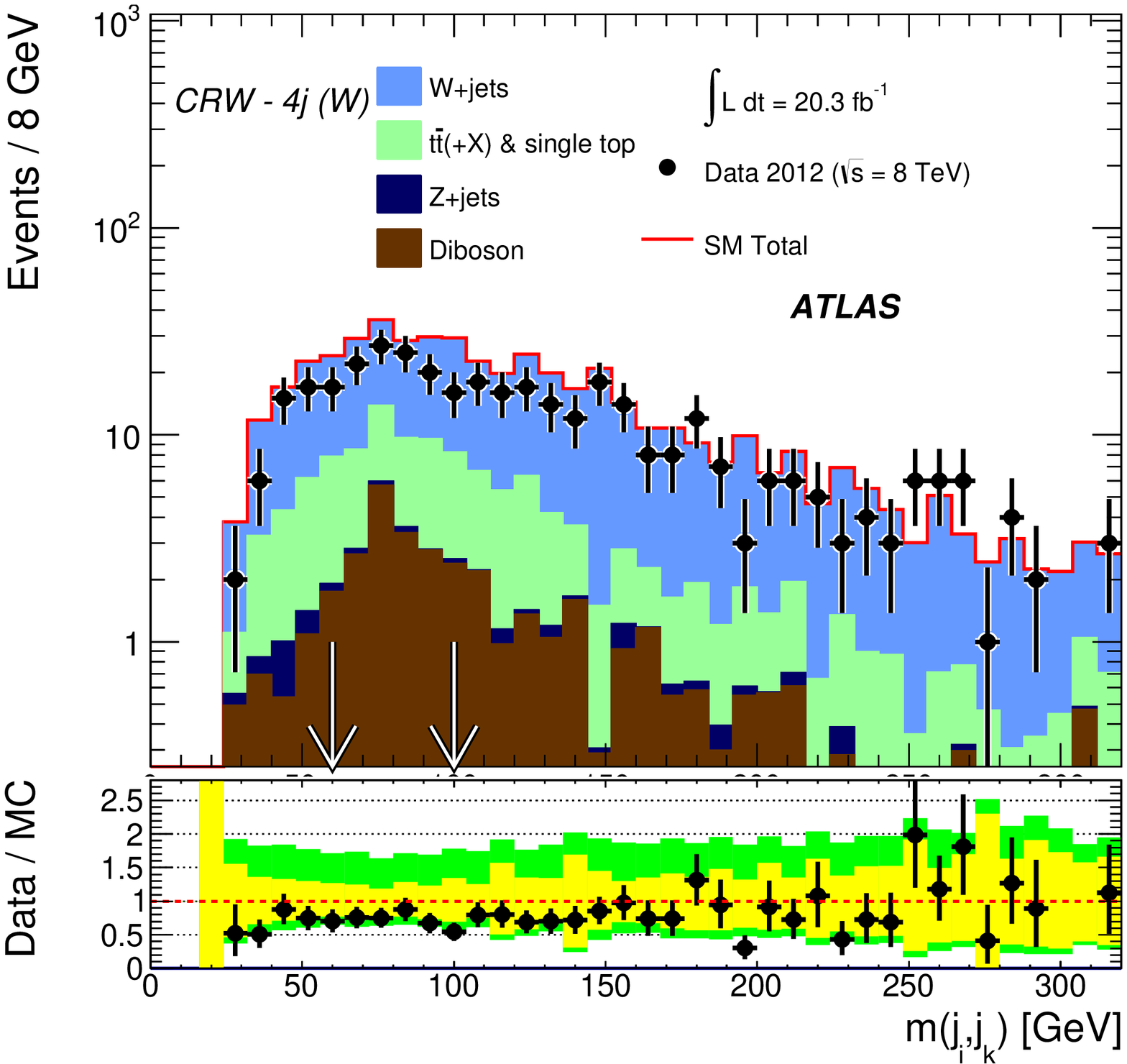}
\includegraphics[height=0.45\textwidth]{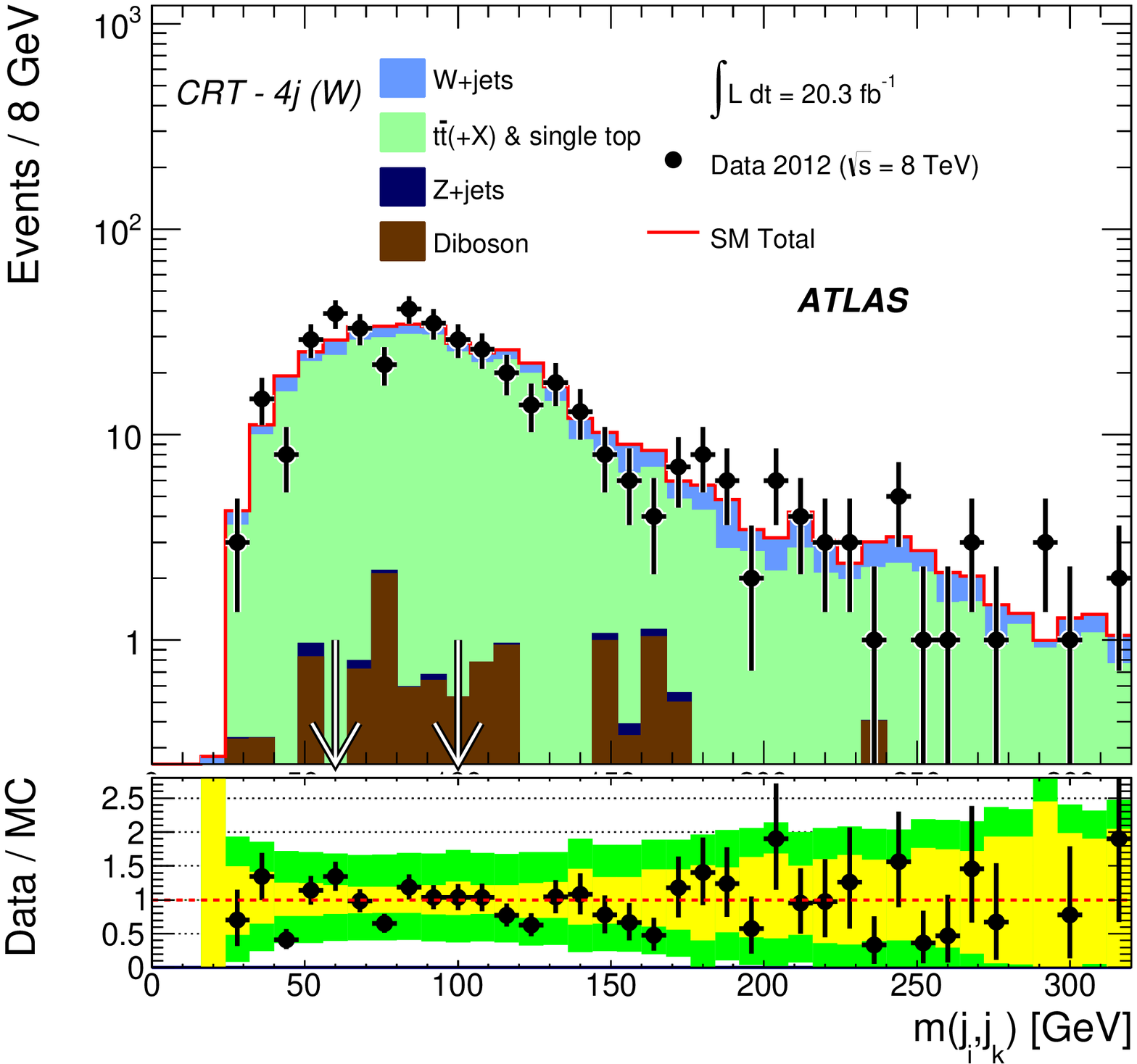}
\end{center}
\caption{\label{fig:crwt_MW} Observed jet (top) or dijet (bottom) mass distributions for the CRW (left) and CRT (right) selections for the 2jW (top) and 4jW (bottom) signal regions.  In the case of the dijet mass distributions for SR 4jW (bottom), events are required to possess at least one unresolved $W$ candidate, with the dijet mass calculated from jets excluding the unresolved $W$ candidate. With the exception of the multi-jet background (which is estimated using the data-driven technique described in the text), the histograms denote the MC background expectations, normalised to cross-section times integrated luminosity. In the lower panels the light (yellow) error bands denote the experimental systematic and MC statistical uncertainties, while the medium dark (green) bands include also the theoretical modelling uncertainty. 
}
\end{figure}

Example CR $ \meff({\rm incl.})$ distributions before the final cut on this quantity for SRs 2jl, 2jm and 2jt are shown in figure~\ref{fig:sr2jcr_Meff}. Jet and dijet mass distributions (respectively for unresolved and resolved $W$ candidates) in CRW and CRT of SRs 2jW and 4jW are shown in figure~\ref{fig:crwt_MW}. The MC $\meff({\rm incl.})$ distributions in figure~\ref{fig:sr2jcr_Meff} are somewhat harder than the data, with better agreement seen at low values of $\meff({\rm incl.})$. This issue is seen also in the SR $\meff({\rm incl.})$ distributions (see section~\ref{sec:results}) and is ameliorated in the SR background estimates using a combined fit to the CR observations (see section~\ref{sec:strategy}). The discrepancy is most pronounced for CR$\gamma$ and CRW and may be related to the overestimation by \sherpa{} (and also \alpgen) of the $Z$ boson differential cross-section at high $\pt$ observed in Ref.~\cite{Aad:2014xaa}.
\begin{table}
\caption{\label{tab:crdefs} Control regions used in the analysis. Also listed are the main targeted background in the SR in each case, the process used to model the background, and the main CR requirement(s) used to select this process. The transverse momenta of high-purity leptons (photons) used to select CR events must exceed 25 (130) GeV.}
  \footnotesize
  \begin{center}\renewcommand\arraystretch{1.2}
    \begin{tabular}{ l  c  c  c }
      \hline
      CR & SR background &  CR process & CR selection \\ \hline
CR$\gamma$ & $Z(\to\nu\nu)$+jets & $\gamma$+jets & Isolated photon \\
CRQ & Multi-jets & Multi-jets & SR with reversed requirements on (i) $\ourdeltaphifull$  \\
& & & and (ii) $\met/\meff(N_{\rm j})$ or $\met/\sqrt{H_{\rm T}}$\\
CRW & $W(\to\ell\nu)$+jets & $W(\to\ell\nu)$+jets & 30 GeV $<m_{\rm T}(\ell,\met) < 100$ GeV, $b$-veto\\
CRT & $t\bar{t}$ and single-$t$ & $t\bar{t}\to b\bar{b}qq'\ell\nu$ & 30 GeV $<m_{\rm T}(\ell,\met) < 100$ GeV, $b$-tag\phantom{o}\\ 
\hline
\end{tabular}
  \end{center}
\end{table}

\clearpage

\clearpage

\subsection{Validation regions}\label{subsec:vrdef}
Cross-checks of the background estimates (see section \ref{sec:valid}) are performed using several `validation region' (VR) samples selected with requirements, distinct from those used in the control regions, which maintain a low probability of signal contamination. CR$\gamma$ estimates of the $Z(\to \nu\bar{\nu})$+jets background are validated with samples of $Z(\to \ell\ell)$+jets events selected by requiring high-purity lepton pairs of opposite sign and identical flavour for which the dilepton invariant mass lies within 25 GeV of the mass of the $Z$ boson (VRZ). In VRZ the leptons are treated as contributing to $\met$. CRW and CRT estimates of the $W$+jets and top quark background are validated with CRW and CRT events with the signal region $\ourdeltaphifull$ and $\met/\meff(N_{\rm j})$ or $\met/\sqrt{H_{\rm T}}$ (as appropriate) requirements reinstated, and with the lepton treated either as a jet (VRW, VRT) or as contributing to $\met$ (VRW$\nu$, VRT$\nu$). Further validation of CRW and CRT estimates is provided by validation regions in which at least one hadronically decaying $\tau$-lepton is reconstructed, without (VRW$\tau$) or with (VRT$\tau$) a requirement of a $b$-tagged jet. CRQ estimates of the multi-jet background are validated with validation regions for which the CRQ selection is applied with the signal region $\met/\meff(N_{\rm j})$ ($\met/\sqrt{H_{\rm T}}$) requirement reinstated (VRQa), or with a requirement of an intermediate value of $\ourdeltaphifull$ applied (VRQb).

\section{Background estimation}
\label{sec:analysisprocedure}
\subsection{Overview}
\label{sec:strategy}
The observed numbers of events in the CRs for each SR are used to generate consistent SM background estimates for the SR via a likelihood fit \cite{Cowan:2010js}. This procedure enables CR correlations due to common systematic uncertainties and contamination by other SM processes and/or SUSY signal events to be taken into account. Poisson likelihood functions are used for event counts in signal and control regions. Systematic uncertainties are treated as Gaussian-distributed nuisance parameters in the likelihood function. Key ingredients in the fit are the ratios of expected event counts from each background process between the SR and each CR, and between CRs. These ratios, referred to as transfer factors or `TFs', enable observations in the CRs to be converted into background estimates in the SR using:
\begin{equation}
\label{eq:tf}
N\mathrm{(SR, scaled)} = N\mathrm{(CR, obs)} \times \left [\frac{N\mathrm{(SR, unscaled)}}{N\mathrm{(CR, unscaled)}}\right ],
\end{equation}
where $N$(SR, scaled) is the estimated background contribution to the SR by a given process, $N$(CR, obs) is the observed number of data events in the CR for the process, and $N$(SR, unscaled) and $N$(CR, unscaled) are a priori estimates of the contributions from the process to the SR and CR, respectively. The TF is the ratio in the square brackets in eq.~(\ref{eq:tf}). Similar equations containing inter-CR TFs enable the background estimates to be normalised coherently across all the CRs associated with a given SR.

Background estimation requires determination of the central expected values of the TFs for each SM process, together with their associated correlated and uncorrelated uncertainties. Some systematic uncertainties, for instance those arising from the jet energy scale (JES), or theoretical uncertainties in MC cross-sections, largely cancel when calculating the event-count ratios constituting the TFs. The use of similar kinematic selections for the CRs and the SR minimises residual uncertainties correlated between these regions. The multi-jet TFs are estimated using a data-driven technique \cite{Aad:2012fqa}, which applies a resolution function to well-measured multi-jet events in order to estimate the impact of jet energy mismeasurement and heavy-flavour semileptonic decays on \met{} and other variables. The other TF estimates use MC samples. Corrections are applied to the CR$\gamma$ TFs which reduce the theoretical uncertainties in the SR $Z/\gamma^*$+jets background expectations arising from the use of LO $\gamma$+jets cross-sections (see table~\ref{tab:MCsamples}) when evaluating the denominator of the TF ratio in Eqn.~\ref{eq:tf}. These corrections are determined by comparing CR$\gamma$ observations with observations in a highly populated auxiliary control region selecting events containing a low \pt{} $Z$ boson (160 GeV $\lesssim \pt(Z) \lesssim 300$ GeV) decaying to electrons or muons, together with at least two jets. 

Three different classes of likelihood fit are employed in this analysis. The first is used to determine the compatibility of the observed event yield in each SR with the corresponding SM background expectation. In this case (the `background-only fit') the fit is performed using only the observed event yields from the CRs associated with the SR, but not the SR itself, as constraints. It is assumed that signal events from physics beyond standard model (BSM) do not contribute to these yields. The significance of an excess of events observed in the SR above the resulting SM background expectation is quantified by the probability (the one-sided p-value, $p_0$) that the SR event yield obtained in a single hypothetical background-only experiment is greater than that observed in this dataset. The background-only fit is also used to estimate the background event yields in the VRs.

If no excess is observed, then a second class of likelihood fit (the `model-independent fit') is used to set `model-independent' upper limits on the number of BSM signal events in each SR. These limits, when normalised by the integrated luminosity of the data sample, may be interpreted as upper limits on the visible cross-section of BSM physics ($\langle\epsilon\sigma\rangle$) defined as the product of acceptance, reconstruction efficiency and production cross-section. The model-independent fit proceeds in the same way as the background-only fit, except that the number of events observed in the SR is added as an input to the fit and the BSM signal strength, constrained to be non-negative, is added as a free parameter. Possible contamination of the CRs by BSM signal events is neglected.

A third class of likelihood fit (the `SUSY-model exclusion fit') is used to set limits on the signal cross-sections for specific SUSY models. The SUSY-model exclusion fit proceeds in the same way as the model-independent fit, except that signal contamination in the CR is taken into account as well as theoretical and experimental uncertainties on the SUSY production cross-section and kinematic distributions. Correlations between signal and background systematic uncertainties are also taken into account where appropriate.

\subsection{Systematic uncertainties}\label{sec:systematics}

Systematic uncertainties in background estimates arise through the use of the transfer factors relating observations in the control regions to background expectations in the signal regions, and from the MC modelling of minor backgrounds. The total background uncertainties for all SRs, broken down into the main contributing sources, are presented in table~\ref{tab:BreakdownSysSRCompressed}. The overall background uncertainties range from 5\% in SR 4jl-, where the loose selection minimises theoretical uncertainties and the impact of statistical fluctuations in the CRs, to 61\% in SR 2jW, where the opposite is true.

\begin{table}

\caption[Breakdown of uncertainty on background estimates]{
	Breakdown of the systematic uncertainties on background estimates obtained from the fits described in the text.
	Note that the individual uncertainties can be correlated, and do not necessarily sum in quadrature to
	the total background uncertainty. Uncertainties relative to the total expected background yield are shown in parenthesis. When a dash is shown, the resulting relative uncertainty is lower than 0.1\%. Rows labelled `CR stats' refer to uncertainties arising from finite data statistics in the main CR for the background process specified. \label{tab:BreakdownSysSRCompressed}}

\scriptsize
\begin{center}
\begin{tabular}{lrrrrr}
\hline
Channel  &  {\bf 2jl}  &  {\bf 2jm}  &  {\bf 2jt}  &  {\bf 2jW}  &  {\bf 3j} \\
Total bkg  &  $13000$  &  $760$  &  $125$  &  $2.3$  &  $5.0$ \\
Total bkg unc.  &  $\pm 1000$  [$8\%$]  &  $\pm 50$  [$7\%$]  &  $\pm 10$  [$8\%$]  &  $\pm 1.4$  [$61\%$]  &  $\pm 1.2$  [$24\%$] \\
\hline
CR stats: $Z/\gamma^*$+jets  &  $\pm 100$ [$0.8\%$]  &  $\pm 15$ [$2.0\%$]  &  $\pm 5$ [$4.0\%$]  &  $\pm 0.4$ [$17.4\%$]  &  $\pm 0.7$ [$14.0\%$] \\
CR stats: $W$+jets  &  $\pm 300$ [$2.3\%$]  &  $\pm 21$ [$2.8\%$]  &  $\pm 5$ [$4.0\%$]  &  $\pm 0.7$ [$30.4\%$]  &  $\pm 0.8$ [$16.0\%$] \\
CR stats: top quark  &  $\pm 200$ [$1.5\%$]  &  $\pm 5$ [$0.7\%$]  &  $\pm 1.6$ [$1.3\%$]  &  $\pm 0.35$ [$15.2\%$]  &  $\pm 0.5$ [$10.0\%$] \\
CR stats: multi-jets  &  -- &  --  &  $\pm 0.1$ [$0.1\%$]  &  --  &  $\pm 0.1$ [$2.0\%$]\\
MC statistics   &  $\pm 130$ [$1.0\%$]  &  $\pm 6$ [$0.8\%$]  &  $\pm 2.1$ [$1.7\%$]  &  $\pm 0.34$ [$14.8\%$]  &  $\pm 0.35$ [$7.0\%$] \\
Jet/MET   &  $\pm 140$ [$1.1\%$]  &  $\pm 8$ [$1.1\%$]  &  $\pm 0.7$ [$0.6\%$]  &  $\pm 0.27$ [$11.7\%$]  &  $\pm 0.23$ [$4.6\%$] \\
Leptons   &  $\pm 80$ [$0.6\%$]  &  $\pm 2.5$ [$0.3\%$]  &  $\pm 0.6$ [$0.5\%$]  &  $\pm 0.04$ [$1.7\%$]  &  $\pm 0.06$ [$1.2\%$] \\
$Z/\gamma$ TF   &  $\pm 500$ [$3.8\%$]  &  $\pm 35$ [$4.6\%$]  &  $\pm 5$ [$4.0\%$]  &  $\pm 0.028$ [$1.2\%$]  &  $\pm 0.14$ [$2.8\%$] \\

Theory: $Z/\gamma^*$+jets   &  $\pm 800$ [$6.2\%$]  &  $\pm 5$ [$0.7\%$]  &  $\pm 4$ [$3.2\%$]  &  $\pm 0.03$ [$1.3\%$]  &  $\pm 0.29$ [$5.8\%$]\\
Theory: $W$+jets    &  $\pm 270$ [$2.1\%$]  &  $\pm 10$ [$1.3\%$]  &  $\pm 1.4$ [$1.1\%$]  &  $\pm 0.1$ [$4.3\%$]  &  $\pm 0.35$ [$7.0\%$] \\
Theory: top quark   &  $\pm 13$ [$0.1\%$]  &  $\pm 1.8$ [$0.2\%$]  &  $\pm 0.11$ [$0.1\%$]  &  $\pm 0.9$ [$39.1\%$]  &  $\pm 0.05$ [$1.0\%$] \\
Theory: diboson   &  $\pm 400$ [$3.1\%$]  &  $\pm 40$ [$5.3\%$]  &  $\pm 6$ [$4.8\%$]  &  $\pm 0.2$ [$8.7\%$]  &  $\pm 0.18$ [$3.6\%$] \\
Theory: scale unc.   &  $\pm 90$ [$0.7\%$]  &  $\pm 4$ [$0.5\%$]  &  $\pm 0.7$ [$0.6\%$]  &  $\pm 0.13$ [$5.7\%$]  &  $\pm 0.12$ [$2.4\%$] \\

Multi-jets method   &  $\pm 140$ [$1.1\%$]  &  $\pm 1.4$ [$0.2\%$]  &  $\pm 0.4$ [$0.3\%$]  &  $\pm 0.04$ [$1.7\%$]  &  $\pm 0.06$ [$1.2\%$] \\
Other   &  $\pm 32$ [$0.2\%$]  &  $\pm 0.6$ [$0.1\%$]  &  $\pm 0.4$ [$0.3\%$]  &  $\pm 0.24$ [$10.4\%$]  &  $\pm 0.02$ [$0.4\%$] \\
\hline
\hline
Channel  &  {\bf 4jl-}  &  {\bf 4jl}  &  {\bf 4jm}  &  {\bf 4jt}  &  {\bf 4jW} \\
Total bkg  &  $2120$  &  $630$  &  $37$  &  $2.5$  &  $14$ \\
Total bkg unc.  &  $\pm 110$  [$5\%$]  &  $\pm 50$  [$8\%$]  &  $\pm 6$  [$16\%$]  &  $\pm 1.0$  [$40\%$]  &  $\pm 4$  [$29\%$] \\
\hline
CR stats: $Z/\gamma^*$+jets  &  $\pm 22$ [$1.0\%$]  &  $\pm 12$ [$1.9\%$]  &  $\pm 2.3$ [$6.2\%$]  &  $\pm 0.5$ [$20.0\%$]  &  $\pm 1.3$ [$9.3\%$] \\
CR stats: $W$+jets  &  $\pm 60$ [$2.8\%$]  &  $\pm 25$ [$4.0\%$]  &  $\pm 1.3$ [$3.5\%$]  &  $\pm 0.4$ [$16.0\%$]  &  $\pm 1.0$ [$7.1\%$] \\
CR stats: top quark  &  $\pm 40$ [$1.9\%$]  &  $\pm 16$ [$2.5\%$]  &  $\pm 0.5$ [$1.4\%$]  &  $\pm 0.4$ [$16.0\%$]  &  $\pm 0.5$ [$3.6\%$] \\
CR stats: multi-jets  &  -- &  --  &  --  &  --  &  -- \\
MC statistics   &  $\pm 18$ [$0.8\%$]  &  $\pm 6$ [$1.0\%$]  &  $\pm 1.3$ [$3.5\%$]  &  $\pm 0.26$ [$10.4\%$]  &  $\pm 0.7$ [$5.0\%$] \\
Jet/MET   &  $\pm 40$ [$1.9\%$]  &  $\pm 7$ [$1.1\%$]  &  $\pm 0.15$ [$0.4\%$]  &  $\pm 0.06$ [$2.4\%$]  &  $\pm 0.6$ [$4.3\%$] \\
Leptons   &  $\pm 20$ [$0.9\%$]  &  $\pm 5$ [$0.8\%$]  &  $\pm 0.27$ [$0.7\%$]  &  $\pm 0.08$ [$3.2\%$]  &  $\pm 0.06$ [$0.4\%$] \\
$Z/\gamma$ TF   &  $\pm 50$ [$2.4\%$]  &  $\pm 19$ [$3.0\%$]  &  $\pm 1.3$ [$3.5\%$]  &  $\pm 0.06$ [$2.4\%$]  &  $\pm 0.5$ [$3.6\%$] \\
Theory: $Z/\gamma^*$+jets   &  --  &  $\pm 18$ [$2.9\%$]  &  $\pm 2.4$ [$6.5\%$]  &  $\pm 0.4$ [$16.0\%$]  &  $\pm 1.3$ [$9.3\%$] \\
Theory: $W$+jets    &  $\pm 33$ [$1.6\%$]  &  $\pm 7$ [$1.1\%$]  &  $\pm 2.3$ [$6.2\%$]  &  $\pm 0.07$ [$2.8\%$]  &  $\pm 0.9$ [$6.4\%$] \\
Theory: top quark   &  $\pm 29$ [$1.4\%$]  &  $\pm 12$ [$1.9\%$]  &  $\pm 1.6$ [$4.3\%$]  &  $\pm 0.4$ [$16.0\%$]  &  $\pm 2.8$ [$20.0\%$] \\
Theory: diboson   &  $\pm 90$ [$4.2\%$]  &  $\pm 35$ [$5.6\%$]  &  $\pm 4$ [$10.8\%$]  &  $\pm 0.17$ [$6.8\%$]  &  $\pm 1.0$ [$7.1\%$] \\
Theory: scale unc.   &  $\pm 23$ [$1.1\%$]  &  $\pm 7$ [$1.1\%$]  &  $\pm 0.4$ [$1.1\%$]  &  $\pm 0.13$ [$5.2\%$]  &  $\pm 0.12$ [$0.9\%$] \\
Multi-jets method   &  $\pm 4$ [$0.2\%$]  &  $\pm 1.6$ [$0.3\%$]  &  --  &  --  &  -- \\

Other   &  $\pm 5$ [$0.2\%$]  &  $\pm 5$ [$0.8\%$]  &  $\pm 0.23$ [$0.6\%$]  &  $\pm 0.06$ [$2.4\%$]  &  $\pm 0.12$ [$0.9\%$] \\
\hline
\hline
Channel  &  {\bf 5j}  &  {\bf 6jl}  &  {\bf 6jm}  &  {\bf 6jt}  &  {\bf 6jt+} \\
Total bkg  &  $126$  &  $111$  &  $33$  &  $5.2$  &  $4.9$ \\
Total bkg unc.  &  $\pm 13$  [$10\%$]  &  $\pm 11$  [$10\%$]  &  $\pm 6$  [$18\%$]  &  $\pm 1.4$  [$27\%$]  &  $\pm 1.6$  [$33\%$] \\
\hline
CR stats: $Z/\gamma^*$+jets  &  $\pm 3.0$ [$2.4\%$]  &  $\pm 1.4$ [$1.3\%$]  &  $\pm 0.7$ [$2.1\%$]  &  $\pm 0.33$ [$6.3\%$]  &  $\pm 0.31$ [$6.3\%$] \\
CR stats: $W$+jets  &  $\pm 6$ [$4.8\%$]  &  $\pm 4$ [$3.6\%$]  &  $\pm 2.4$ [$7.3\%$]  &  $\pm 0.5$ [$9.6\%$]  &  $\pm 0.7$ [$14.3\%$] \\
CR stats: top quark  &  $\pm 7$ [$5.6\%$]  &  $\pm 7$ [$6.3\%$]  &  $\pm 2.3$ [$7.0\%$]  &  $\pm 0.31$ [$6.0\%$]  &  $\pm 1.1$ [$22.4\%$] \\
CR stats: multi-jets  &  $\pm 0.08$ [$0.1\%$]  &  $\pm 0.19$ [$0.2\%$]  &  $\pm 0.08$ [$0.2\%$]  &  --  &  $\pm 0.04$ [$0.8\%$] \\
MC statistics   &  $\pm 2.8$ [$2.2\%$]  &  $\pm 2.8$ [$2.5\%$]  &  $\pm 1.5$ [$4.5\%$]  &  $\pm 0.7$ [$13.5\%$]  &  $\pm 0.4$ [$8.2\%$] \\
Jet/MET   &  $\pm 4$ [$3.2\%$]  &  $\pm 6$ [$5.4\%$]  &  $\pm 1.2$ [$3.6\%$]  &  $\pm 0.5$ [$9.6\%$]  &  $\pm 0.29$ [$5.9\%$] \\
Leptons   &  $\pm 1.8$ [$1.4\%$]  &  $\pm 1.8$ [$1.6\%$]  &  $\pm 0.7$ [$2.1\%$]  &  $\pm 0.05$ [$1.0\%$]  &  $\pm 0.32$ [$6.5\%$] \\
$Z/\gamma$ TF   &  $\pm 2.5$ [$2.0\%$]  &  $\pm 0.8$ [$0.7\%$]  &  $\pm 0.27$ [$0.8\%$]  &  $\pm 0.04$ [$0.8\%$]  &  $\pm 0.04$ [$0.8\%$] \\

Theory: $Z/\gamma^*$+jets   &  $\pm 7$ [$5.6\%$]  &  $\pm 3.0$ [$2.7\%$]  &  $\pm 2.0$ [$6.1\%$]  &  $\pm 0.5$ [$9.6\%$]  &  $\pm 0.7$ [$14.3\%$] \\
Theory: $W$+jets    &  $\pm 2.2$ [$1.7\%$]  &  $\pm 1.7$ [$1.5\%$]  &  $\pm 2.8$ [$8.5\%$]  &  $\pm 0.4$ [$7.7\%$]  &  $\pm 0.08$ [$1.6\%$] \\
Theory: top quark   &  $\pm 5$ [$4.0\%$]  &  $\pm 2.7$ [$2.4\%$]  &  $\pm 3.5$ [$10.6\%$]  &  $\pm 0.08$ [$1.5\%$]  &  $\pm 0.5$ [$10.2\%$] \\
Theory: diboson   &  $\pm 8$ [$6.3\%$]  &  $\pm 4$ [$3.6\%$]  &  $\pm 1.9$ [$5.8\%$]  &  $\pm 0.8$ [$15.4\%$]  &  $\pm 0.1$ [$2.0\%$] \\
Theory: scale unc.   &  $\pm 2.5$ [$2.0\%$]  &  $\pm 1.1$ [$1.0\%$]  &  $\pm 0.8$ [$2.4\%$]  &  $\pm 0.11$ [$2.1\%$]  &  $\pm 0.5$ [$10.2\%$] \\

Multi-jets method   &  $\pm 2.6$ [$2.1\%$]  &  $\pm 2.9$ [$2.6\%$]  &  $\pm 0.8$ [$2.4\%$]  &  $\pm 0.032$ [$0.6\%$]  &  $\pm 0.4$ [$8.2\%$] \\
Other   &  $\pm 0.9$ [$0.7\%$]  &  $\pm 2.5$ [$2.3\%$]  &  $\pm 0.9$ [$2.7\%$]  &  $\pm 0.14$ [$2.7\%$]  &  $\pm 0.03$ [$0.6\%$] \\
\hline
\end{tabular}
\end{center}
\end{table}

\clearpage

For the backgrounds estimated with MC simulation-derived transfer factors the primary common sources of systematic uncertainty are the jet energy scale (JES) calibration, jet energy resolution (JER), theoretical uncertainties, MC and CR data statistics and the reconstruction performance in the presence of pile-up. Correlations between uncertainties (for instance between JES uncertainties in CRs and SRs) are taken into account where appropriate. 

The JES uncertainty was measured using the techniques
described in refs.~\cite{Aad:2011he,Aad:2012vm}, leading to a slight dependence upon $\ourpt$ and $\eta$. The JER uncertainty is estimated using the methods discussed in ref.~\cite{Aad:2012ag}. Contributions are added to both the JES and the JER uncertainties to account for the effect of pile-up at the relatively high luminosity delivered by the LHC in the 2012 run. A further uncertainty on the low-$\pt$ calorimeter activity not associated with jets or baseline leptons but included in the $\met$ calculation is taken into account. The jet mass scale and resolution uncertainties applicable in SR 2jW and SR 4jW are estimated using a sample of tagged $W\to qq'$ decays reconstructed as single jets in selected $t\bar{t}$ events. For the specific selections used in these SRs these uncertainties are estimated to be of order 10\% (scale) and 20\% (resolution). The JES, JER, $\met$ and jet mass scale and resolution (SR 2jW and SR 4jW) uncertainties are taken into account in the combined `Jet/MET' uncertainty quoted in table~\ref{tab:BreakdownSysSRCompressed}. This uncertainty ranges from less than 1\% of the expected background in SR 4jm to 12\% in SR 2jW. 

Uncertainties arising from theoretical models of background processes are evaluated by comparing TFs obtained from samples produced with different MC generators, as described in section~\ref{sec:mcsamples}.  Renormalisation and factorisation scale uncertainties are also taken into account by increasing and decreasing the scales used in the MC generators by a factor of two. The largest uncertainties are associated with the modelling of top quark production ($t\bar{t}$ and single top quark production) in the higher jet-multiplicity SRs (e.g. SR 4jW), and with the modelling of $Z/\gamma^*$+jets in SR 4jt. Uncertainties associated with PDF modelling for background processes were checked with dedicated MC samples and found to be negligible. Uncertainties on diboson production due to scale and PDF errors are found to be $\lesssim$50\% for all SRs, and a conservative uniform 50\% uncertainty is applied. The uncertainty on diboson production arising from the error on the integrated luminosity of the data sample is negligible, while for other processes this uncertainty cancels in the TF ratio between CR and SR event yields.

The statistical uncertainty arising from the use of finite-size MC samples is largest (15\%) in SR 2jW. Uncertainties arising from finite data statistics in the control regions are most important for the tighter signal region selections, reaching 20\% for $Z/\gamma^*$+jets (estimated with CR$\gamma$) in SR 4jt and 22\% for top quark production processes (estimated with CRT) in SR 6jt+. 

The experimental systematic uncertainties associated with CR event reconstruction include photon and lepton reconstruction efficiency, energy scale and resolution (CR$\gamma$, CRW and CRT) and $b$-tag/$b$-veto efficiency (CRW and CRT). The photon reconstruction uncertainties associated with CR$\gamma$, together with uncertainties arising from the data-driven CR$\gamma$ TF correction procedure described in section~\ref{sec:strategy}, are included in table~\ref{tab:BreakdownSysSRCompressed} under `$Z/\gamma$ TF'. The impact of lepton reconstruction uncertainties on the overall background uncertainty is found to be negligible for all SRs. Uncertainties in the $b$-tag/$b$-veto efficiency are included in table~\ref{tab:BreakdownSysSRCompressed} under `other', together with additional small uncertainties such as those associated with the modelling of pile-up in MC events.

Uncertainties related to the multi-jet background estimates are determined by varying the width and tails of the jet resolution function within the appropriate experimental uncertainties and then repeating the background estimation procedure described in section~\ref{sec:strategy}. The maximum resulting contribution to the overall background uncertainty is 8\% in SR 6jt+.

\subsection{Validation}\label{sec:valid}
\begin{figure}[p]
\begin{center}
\includegraphics[height=0.5\textwidth]{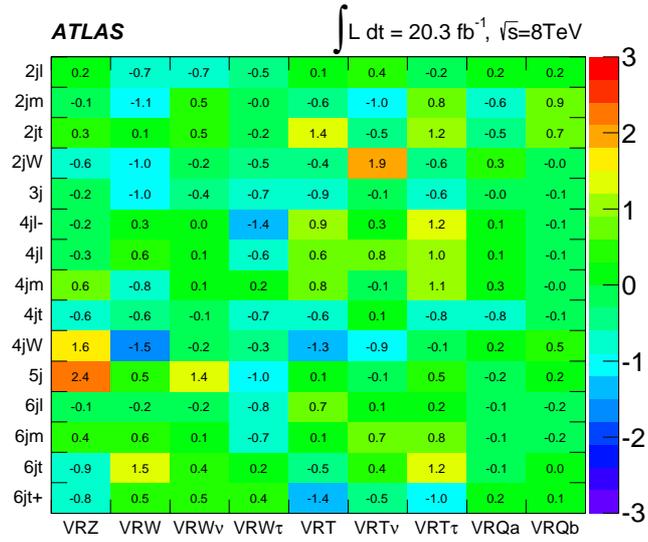}
    \caption{\label{fig:vrpull}Differences between the numbers of observed events in data and SM background expectations for each VR, expressed as fractions of the uncertainties on the latter.}
\end{center}
\end{figure}

The background estimation procedure is validated by comparing the numbers of events observed in the VRs (see section~\ref{subsec:vrdef}) in the data to the corresponding SM background expectations obtained from the background-only fits. The results are shown in figure~\ref{fig:vrpull}. The entries in the table are the differences between the numbers of observed and expected events expressed as fractions of the one-standard deviation ($1\sigma$) uncertainties on the latter. Most VR observations lie within 1$\sigma$ of the background expectations, with the largest discrepancy out of the 135 VRs being 2.4$\sigma$ (13 events observed, $6.1\pm1.3$ expected) for the VRZ region associated with SR 5j.

\section{Results}
\label{sec:results}

\begin{figure}[h!]
\begin{center}
\includegraphics[height=0.45\textwidth]{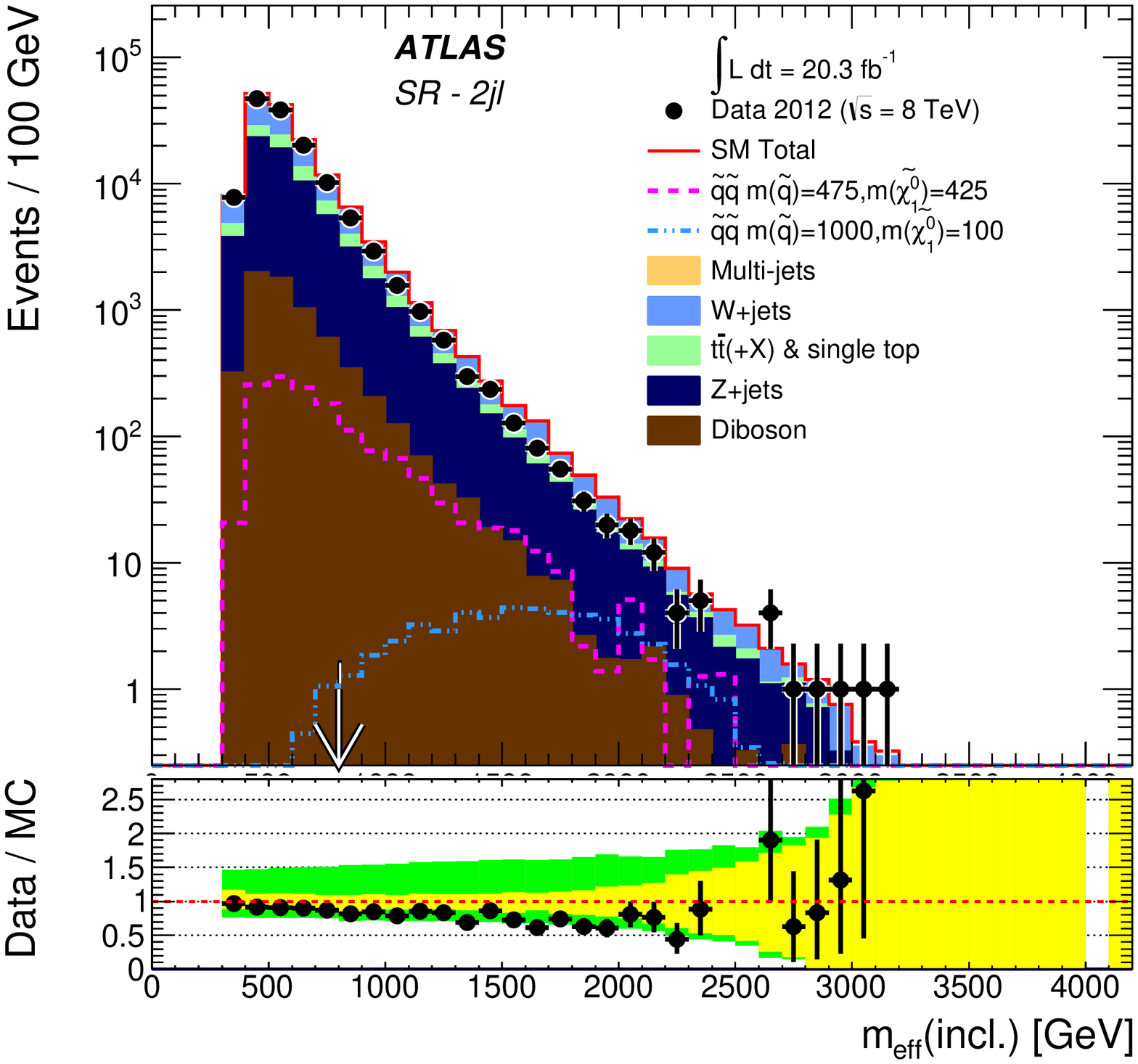}
\includegraphics[height=0.45\textwidth]{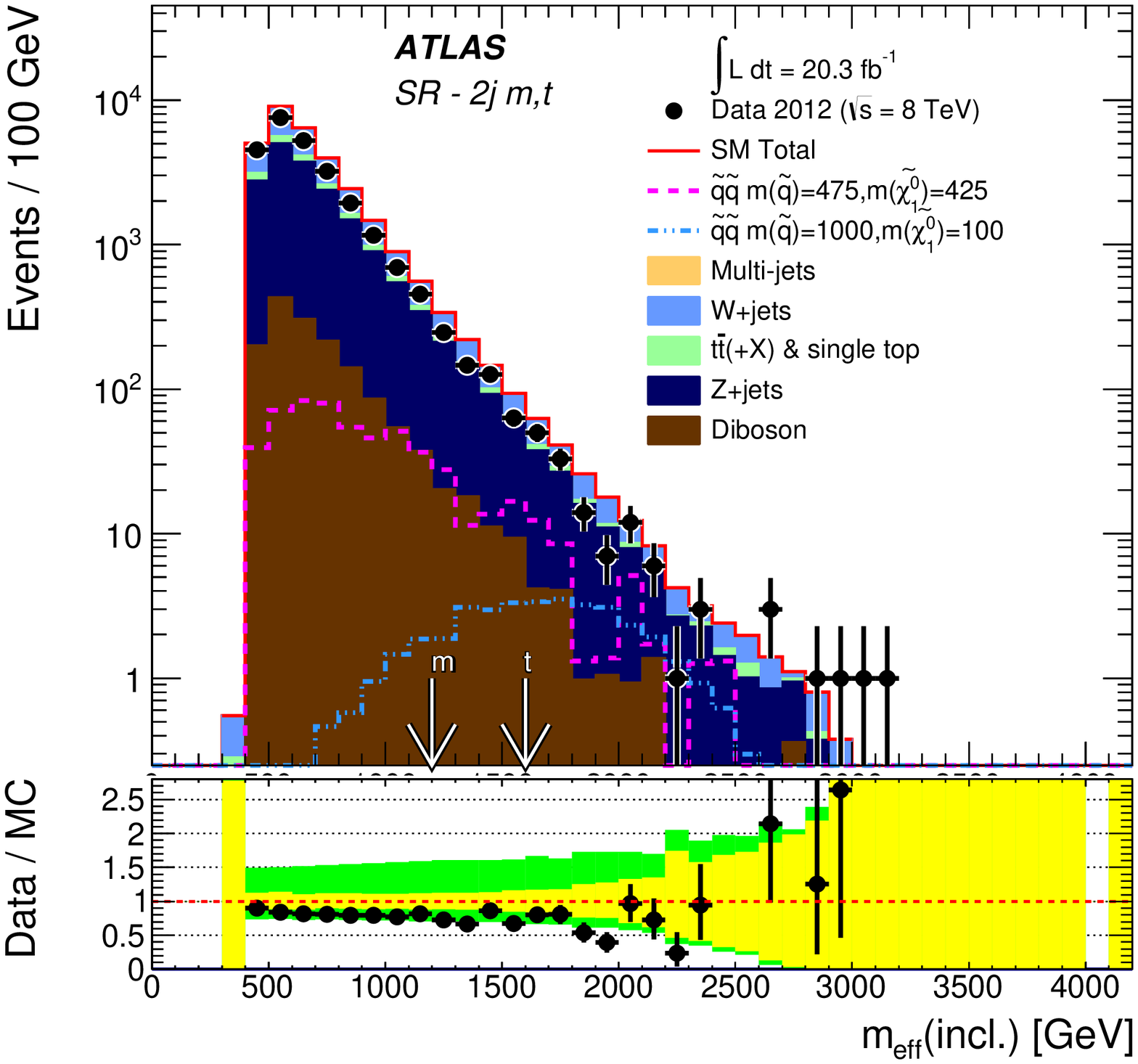}
\includegraphics[height=0.45\textwidth]{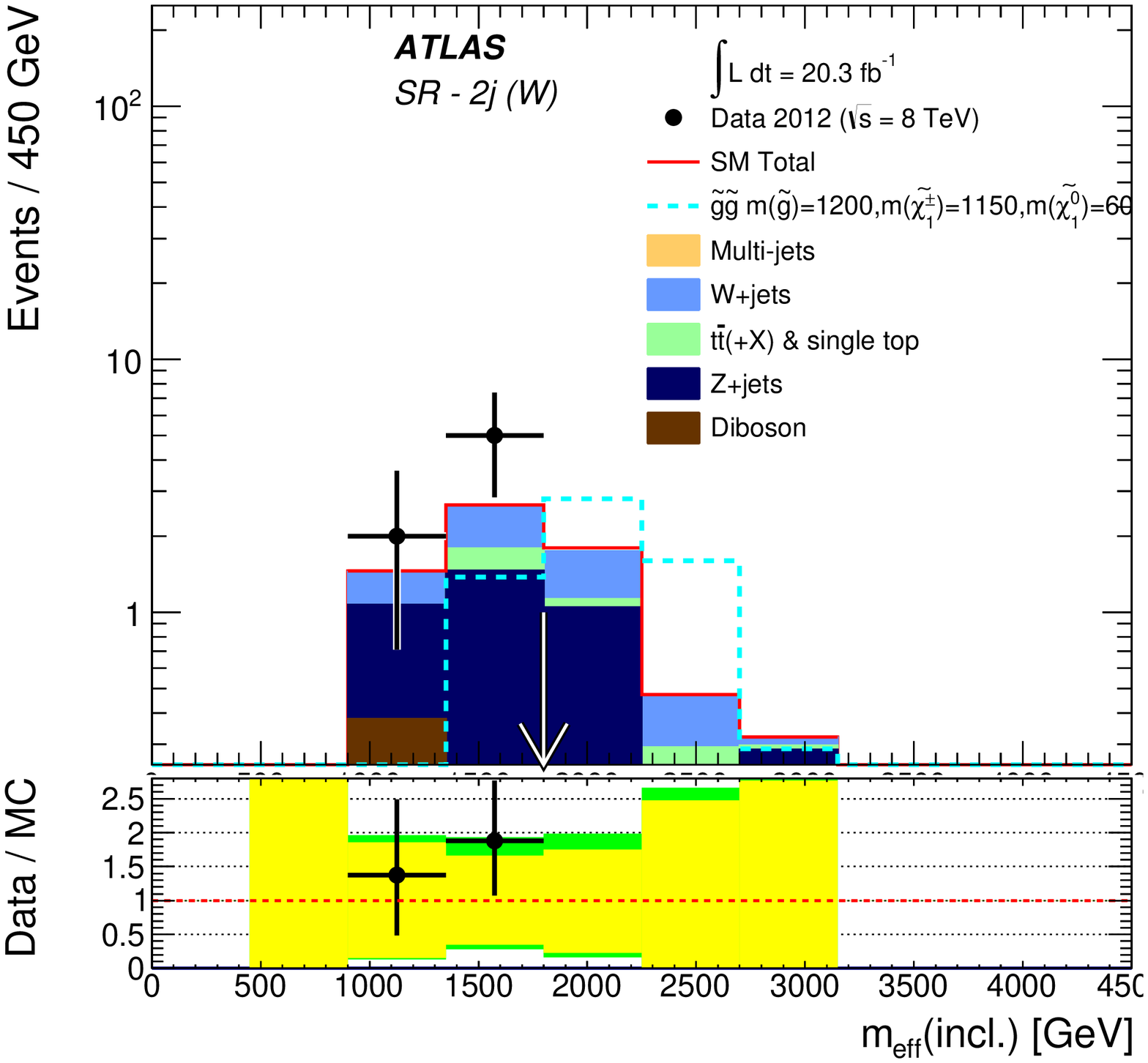}
\includegraphics[height=0.45\textwidth]{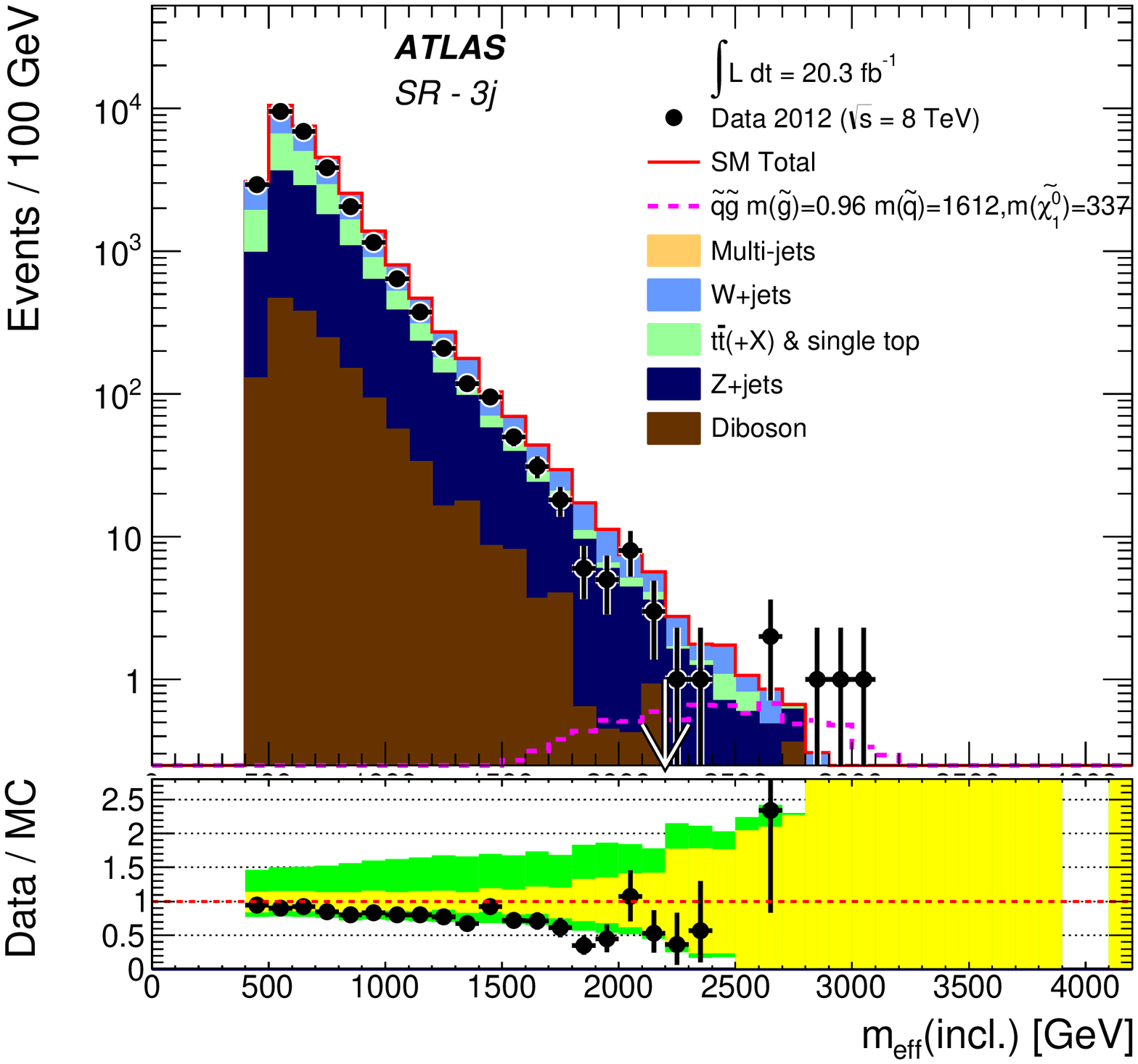}
\includegraphics[height=0.45\textwidth]{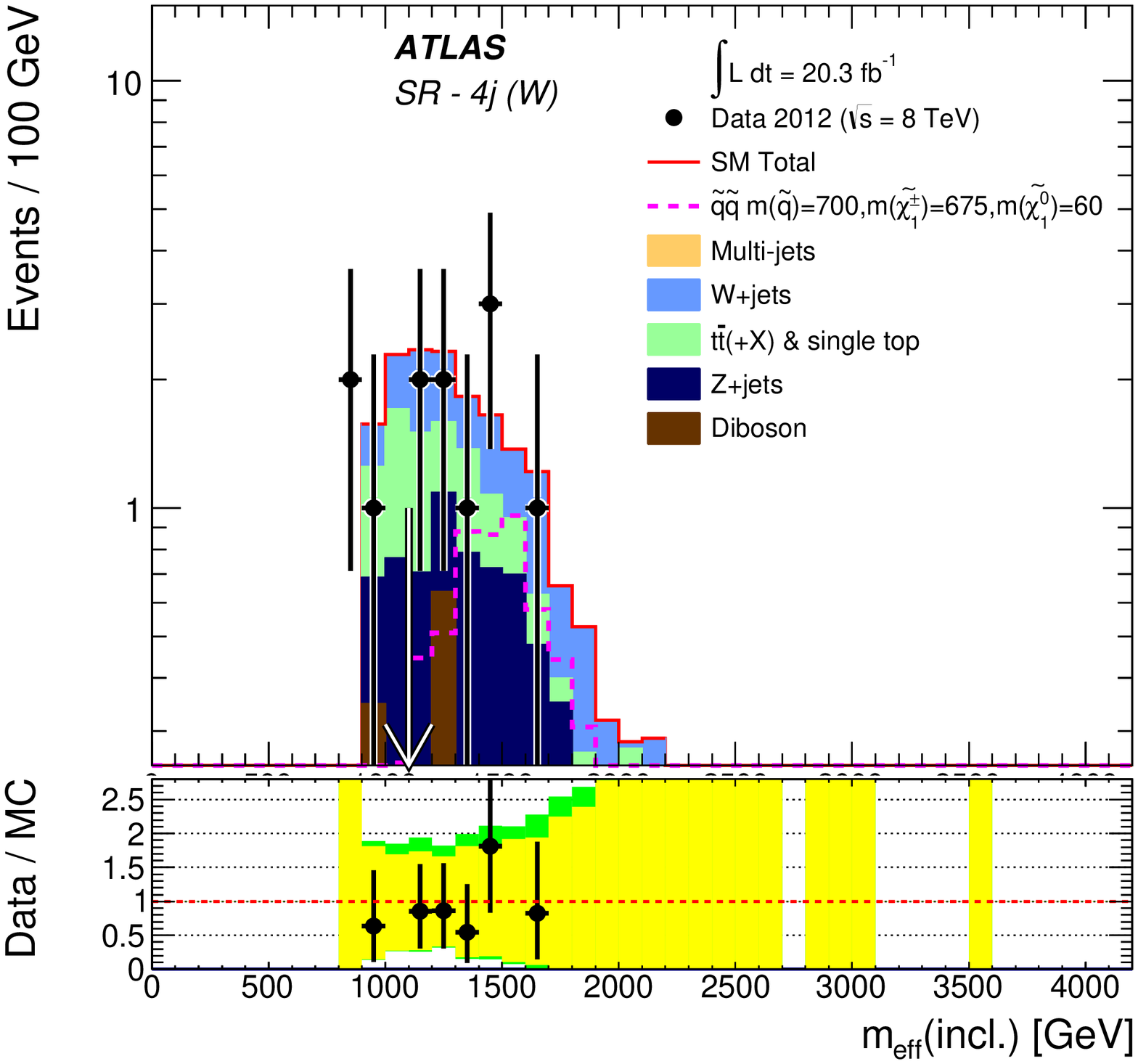}
\includegraphics[height=0.45\textwidth]{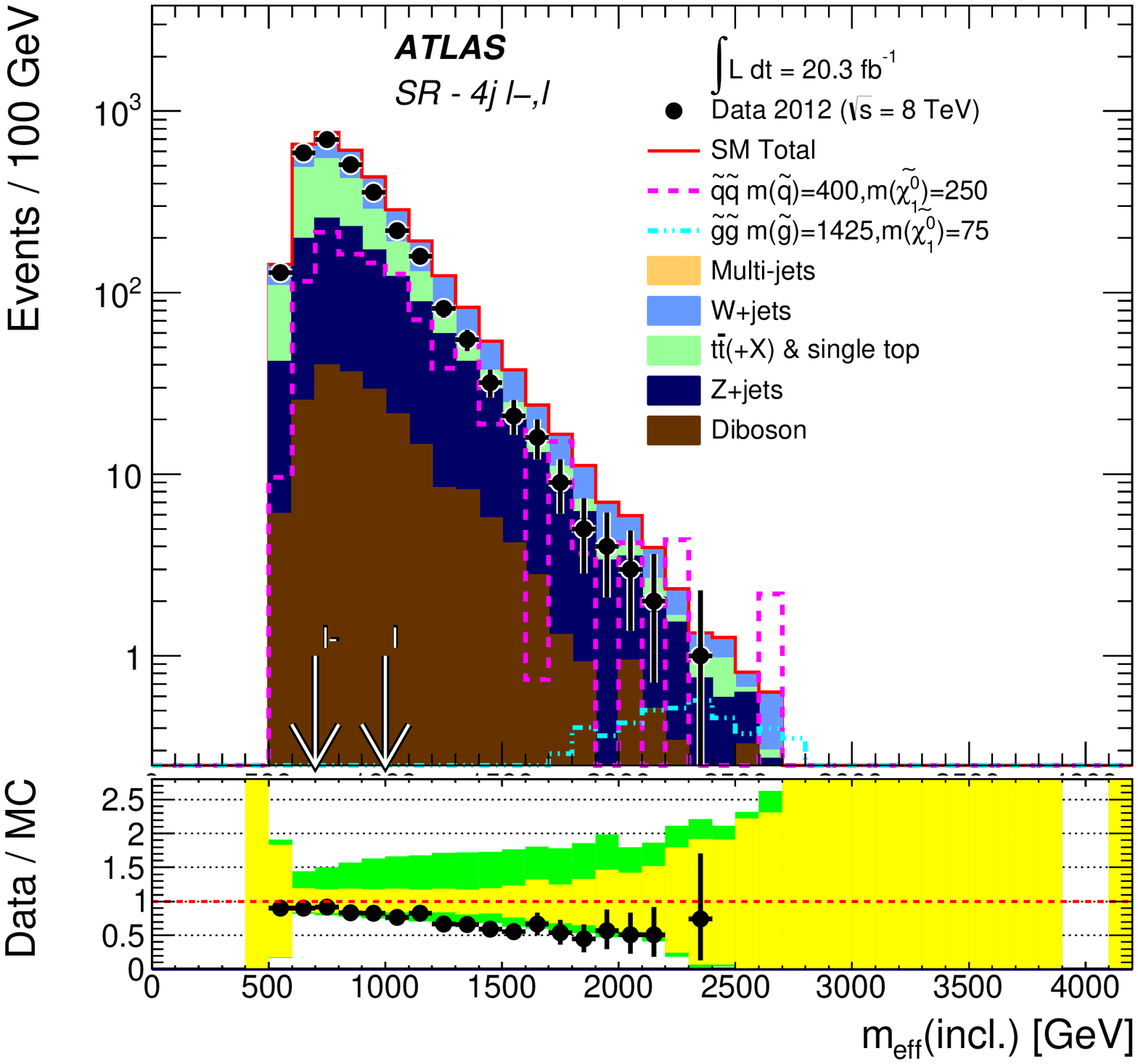}
\end{center}
\vspace*{-0.5cm}
\caption{\label{fig:sra}Observed $\meff({\rm incl.})$ distributions for the 2-jet (top and middle-left), 3-jet (middle-right) and 4-jet (4jW, 4jl- and 4jl) signal regions (bottom). With the exception of the multi-jet background (which is estimated using the data-driven technique described in the text), the histograms denote the MC background expectations prior to the fits described in the text, normalised to cross-section times integrated luminosity. In the lower panels the light (yellow) error bands denote the experimental systematic and MC statistical uncertainties, while the medium dark (green) bands include also the theoretical modelling uncertainty. The arrows indicate the values at which the requirements on $\meff({\rm incl.})$ are applied.
Expected distributions for benchmark model points are also shown for comparison (masses in GeV). See text for discussion of compatibility of data with MC background expectations. 
}
\end{figure}

\begin{figure}[h]
\begin{center}
\includegraphics[height=0.45\textwidth]{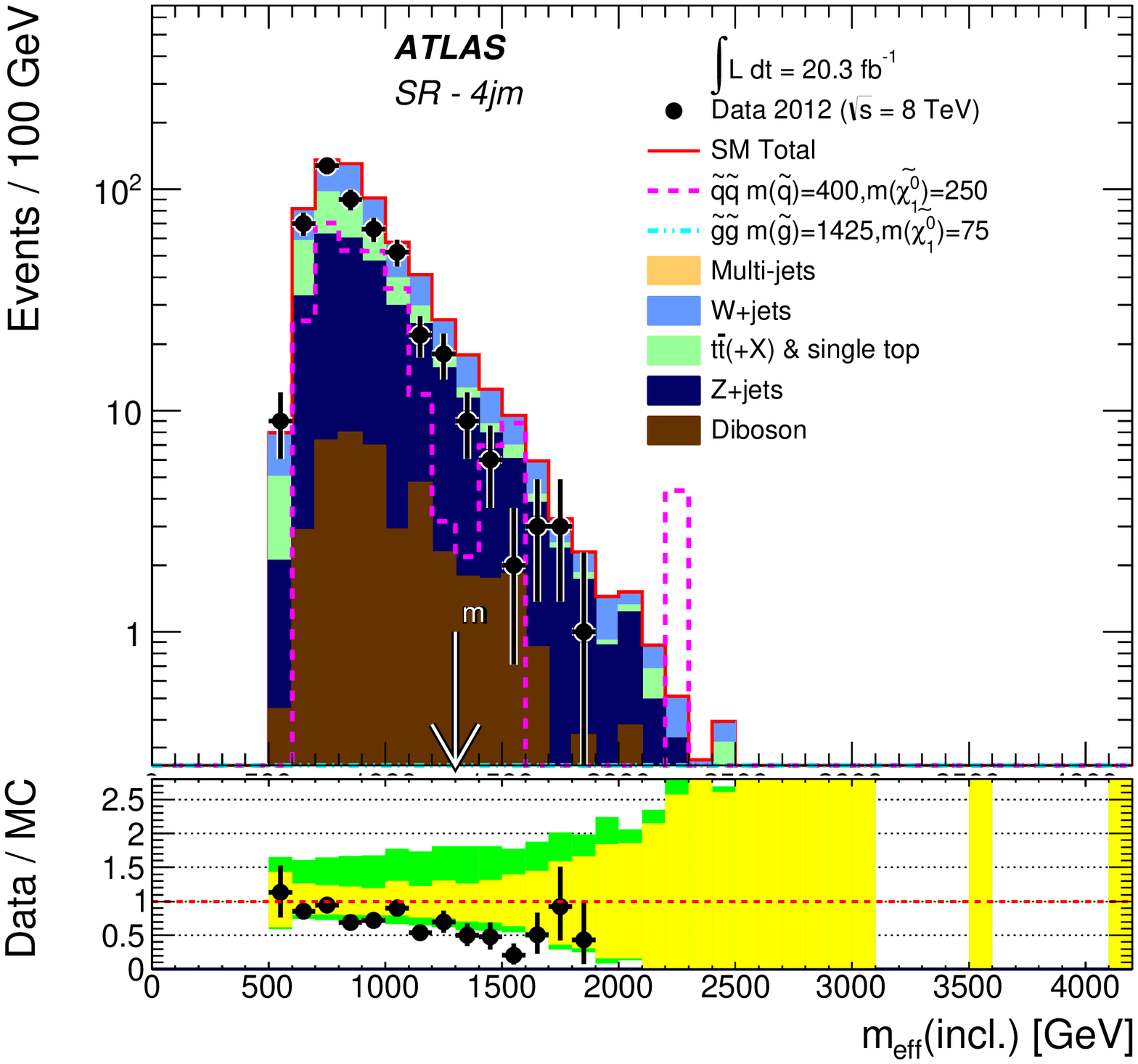}
\includegraphics[height=0.45\textwidth]{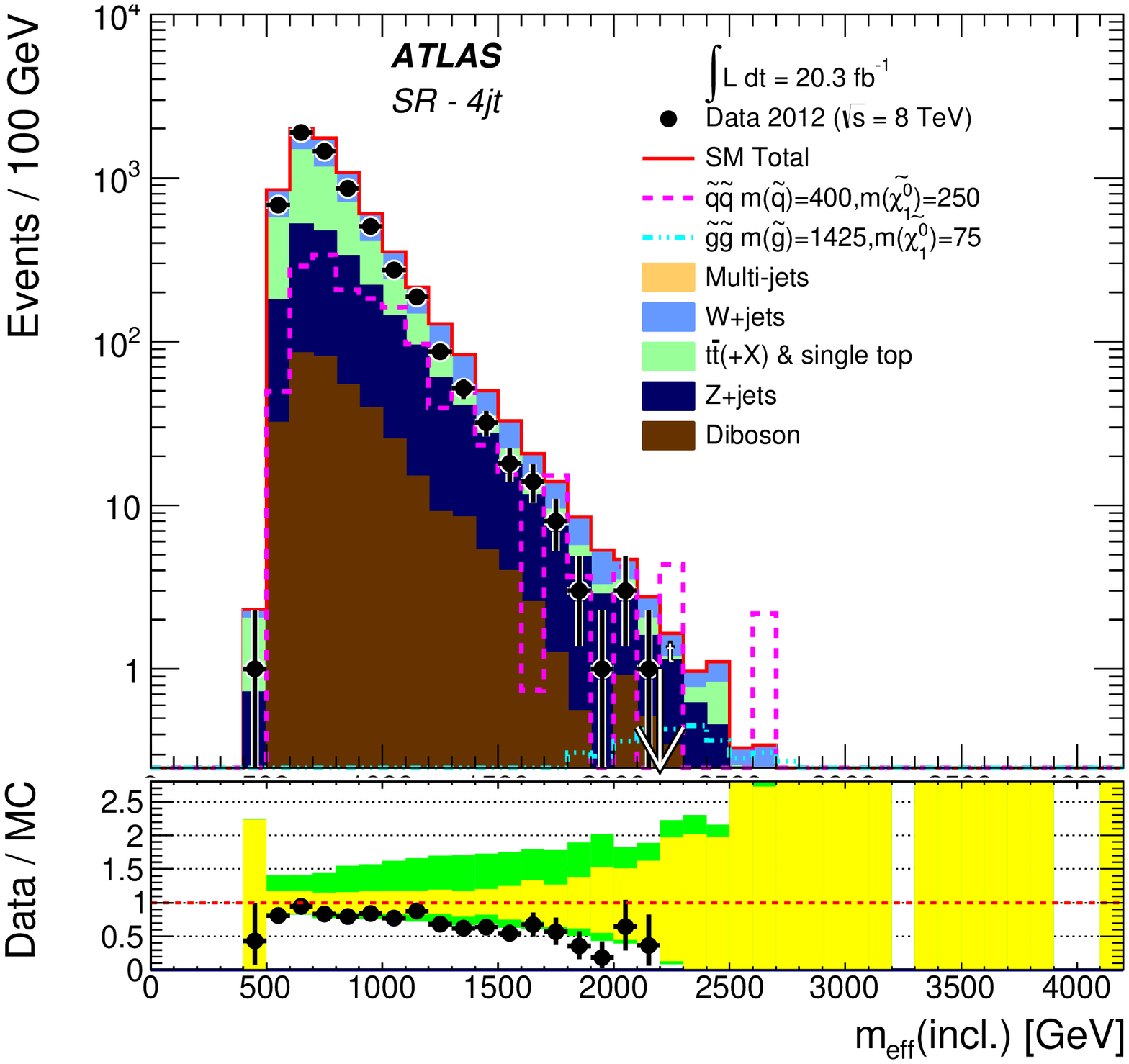}
\includegraphics[height=0.45\textwidth]{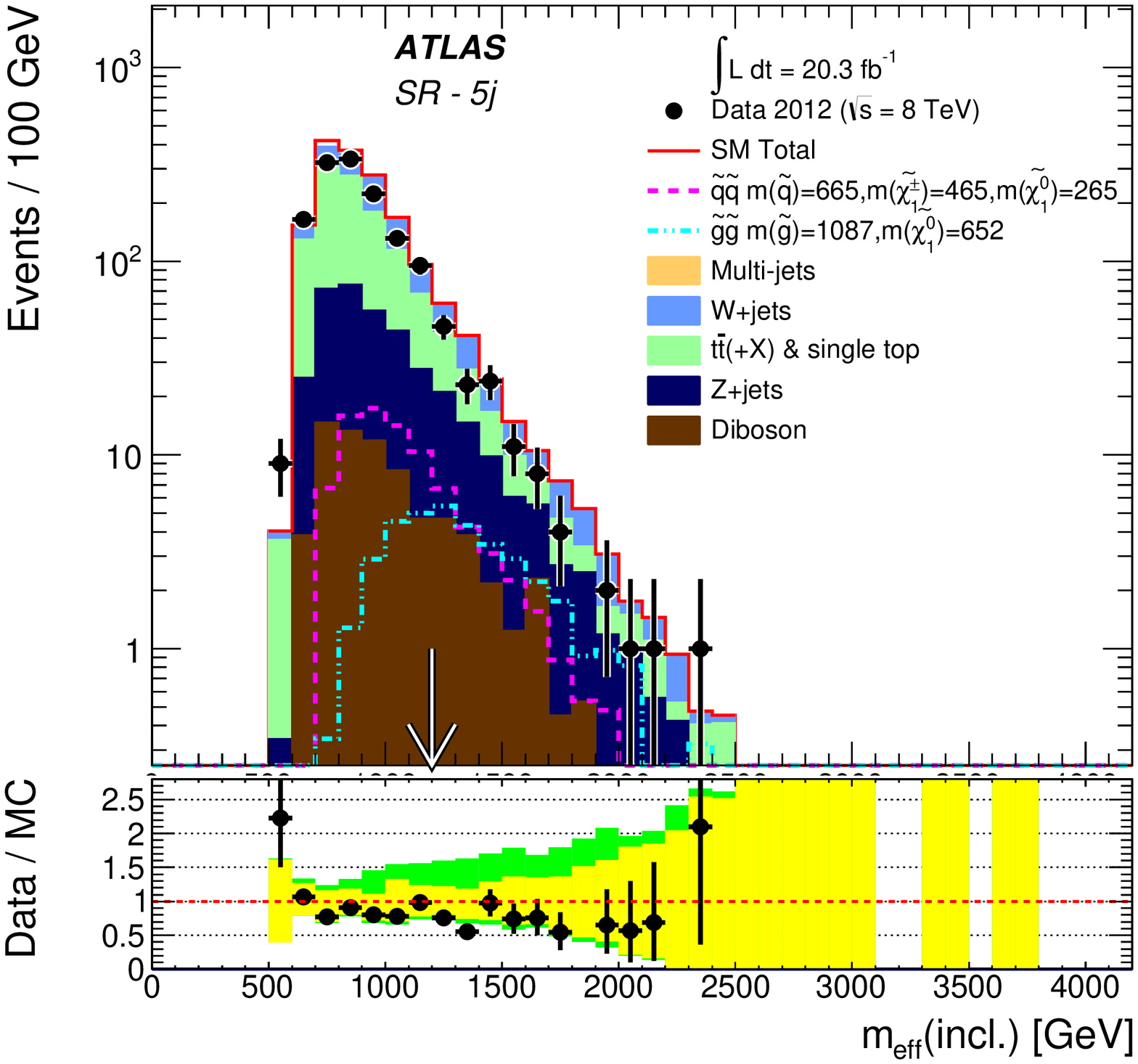}
\includegraphics[height=0.45\textwidth]{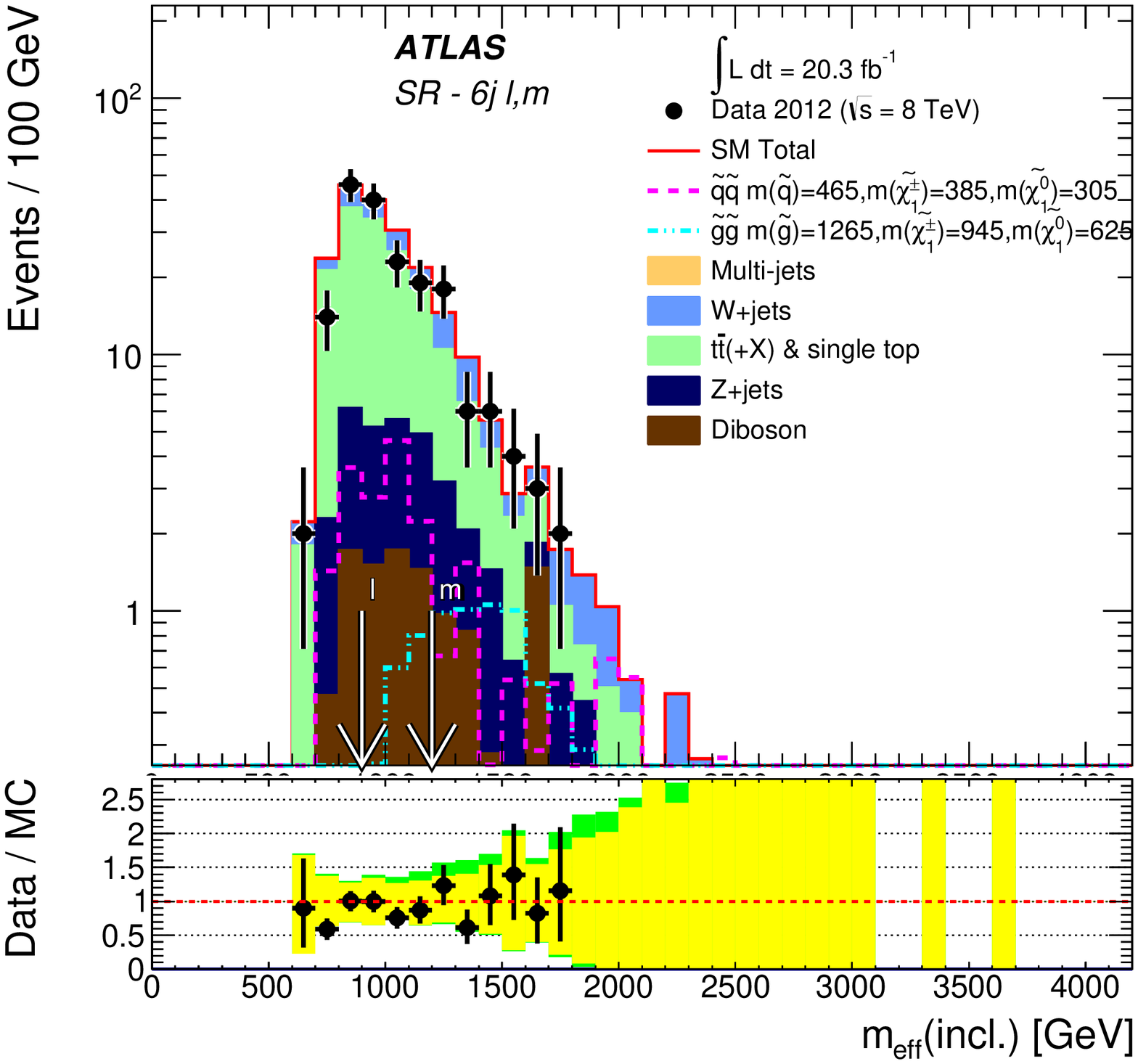}
\includegraphics[height=0.45\textwidth]{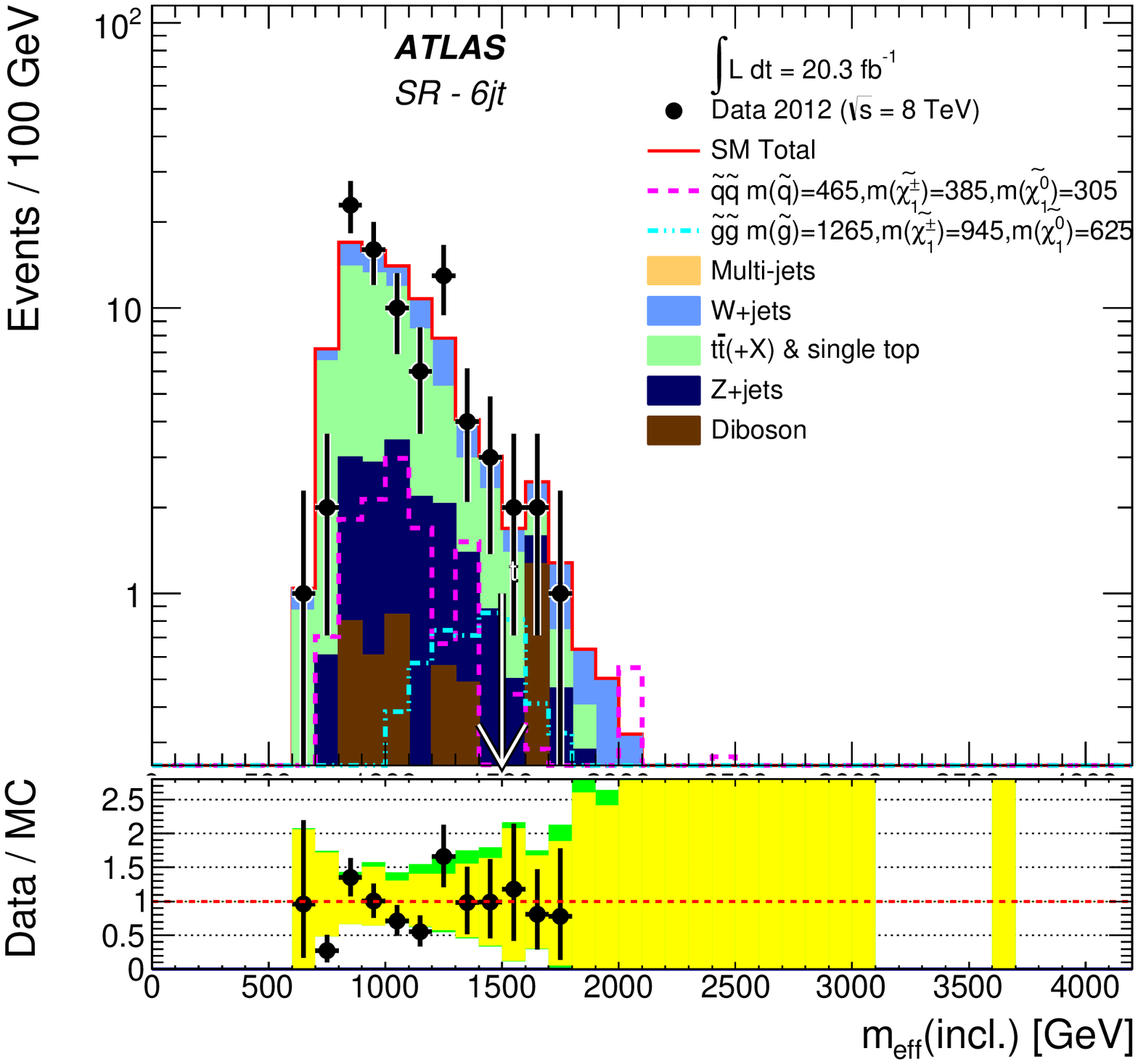}
\includegraphics[height=0.45\textwidth]{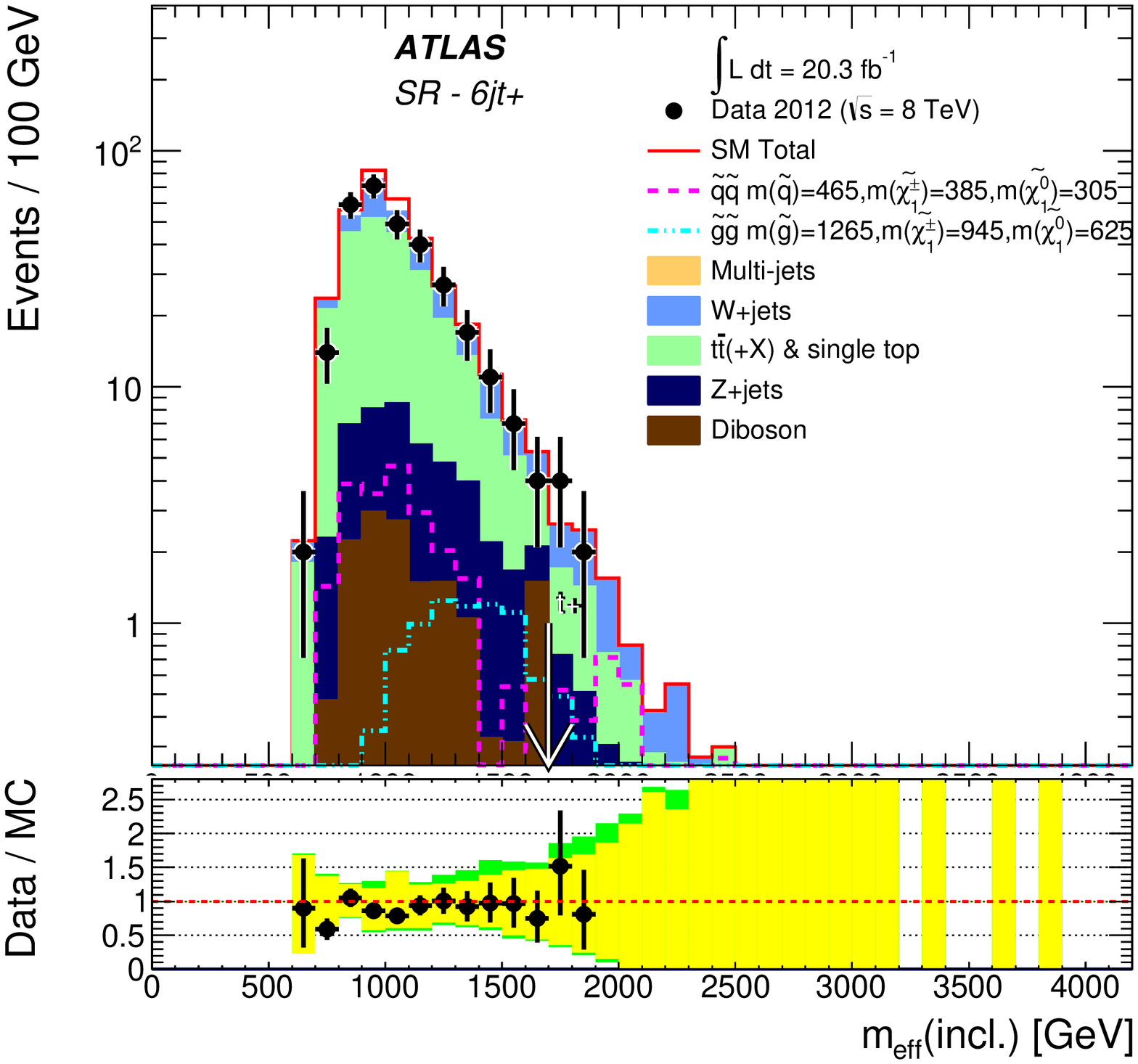}
\end{center}
\vspace*{-0.5cm}
\caption{\label{fig:srb}Observed $\meff({\rm incl.})$ distributions for the medium and tight 4-jet (top), 5-jet (middle-left) and 6-jet (middle-right and bottom) signal regions. With the exception of the multi-jet background (which is estimated using the data-driven technique described in the text), the histograms denote the MC background expectations prior to the fits described in the text, normalised to cross-section times integrated luminosity. In the lower panels the light (yellow) error bands denote the experimental systematic and MC statistical uncertainties, while the medium dark (green) bands include also the theoretical modelling uncertainty. The arrows indicate the values at which the requirements on $\meff({\rm incl.})$ are applied.
Expected distributions for benchmark model points are also shown for comparison (masses in GeV). See text for discussion of compatibility of data with MC background expectations. 
}
\end{figure}

\begin{figure}[h]
\begin{center}
\includegraphics[height=0.45\textwidth]{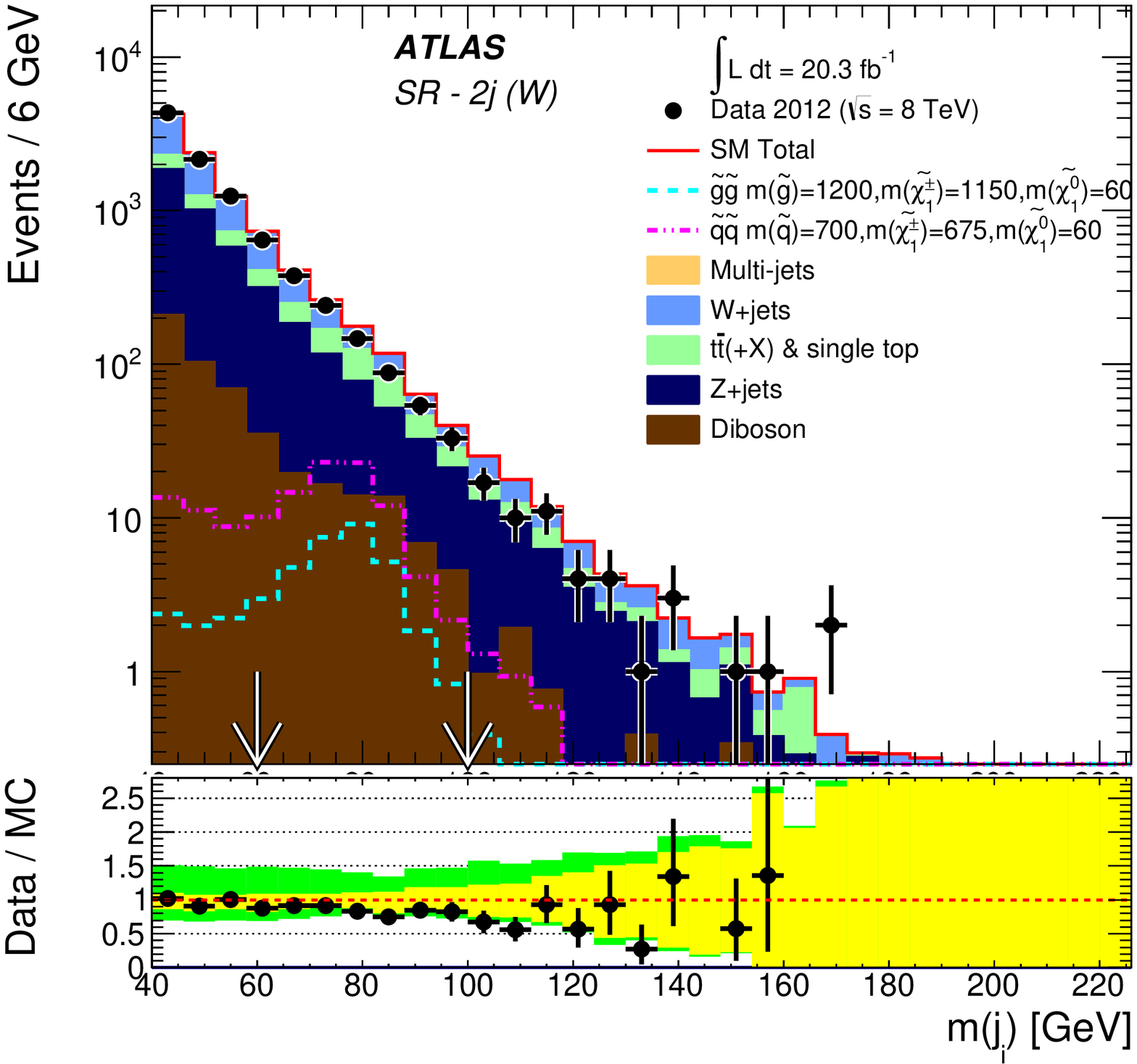}
\includegraphics[height=0.45\textwidth]{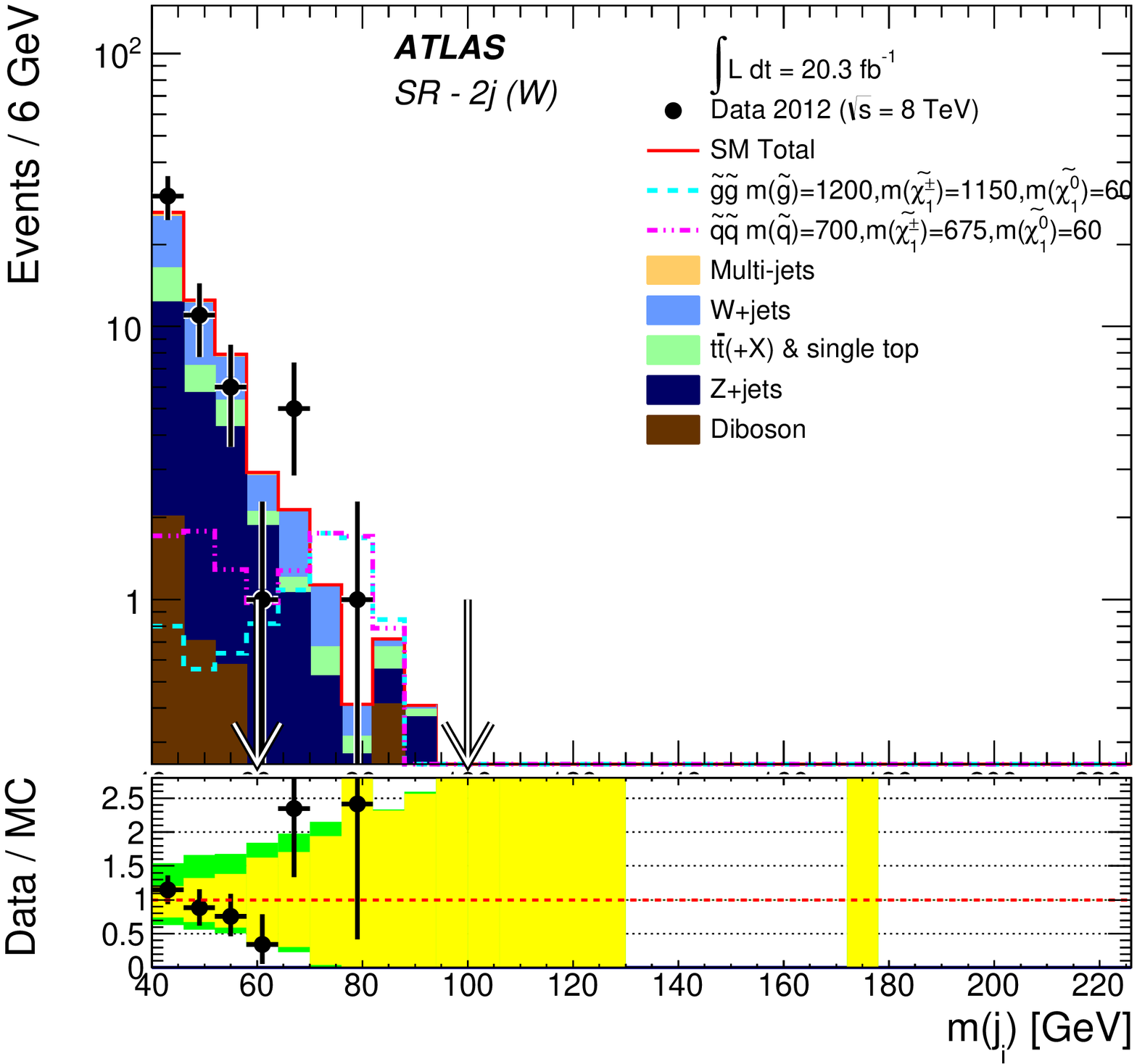}
\includegraphics[height=0.45\textwidth]{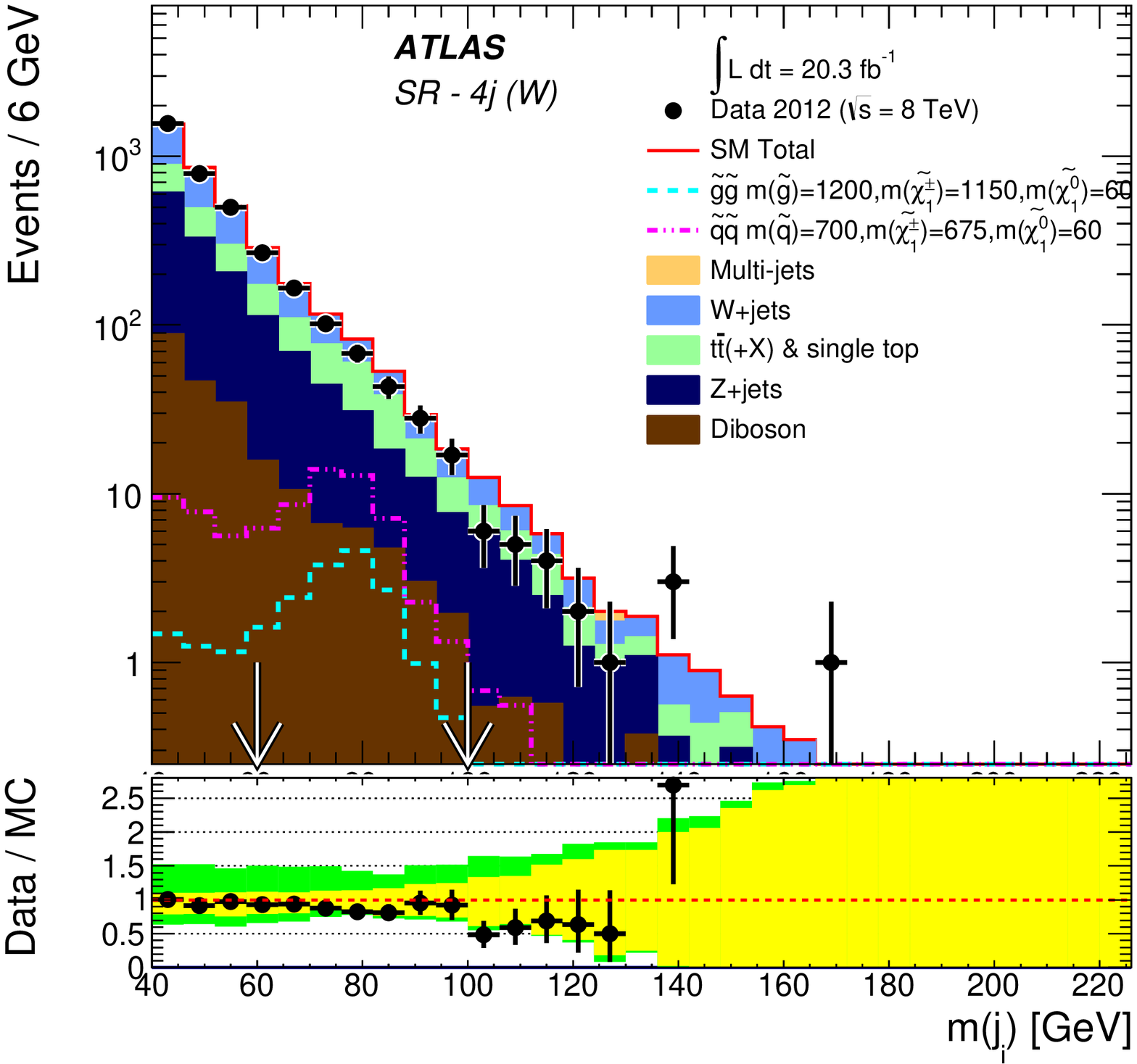}
\includegraphics[height=0.45\textwidth]{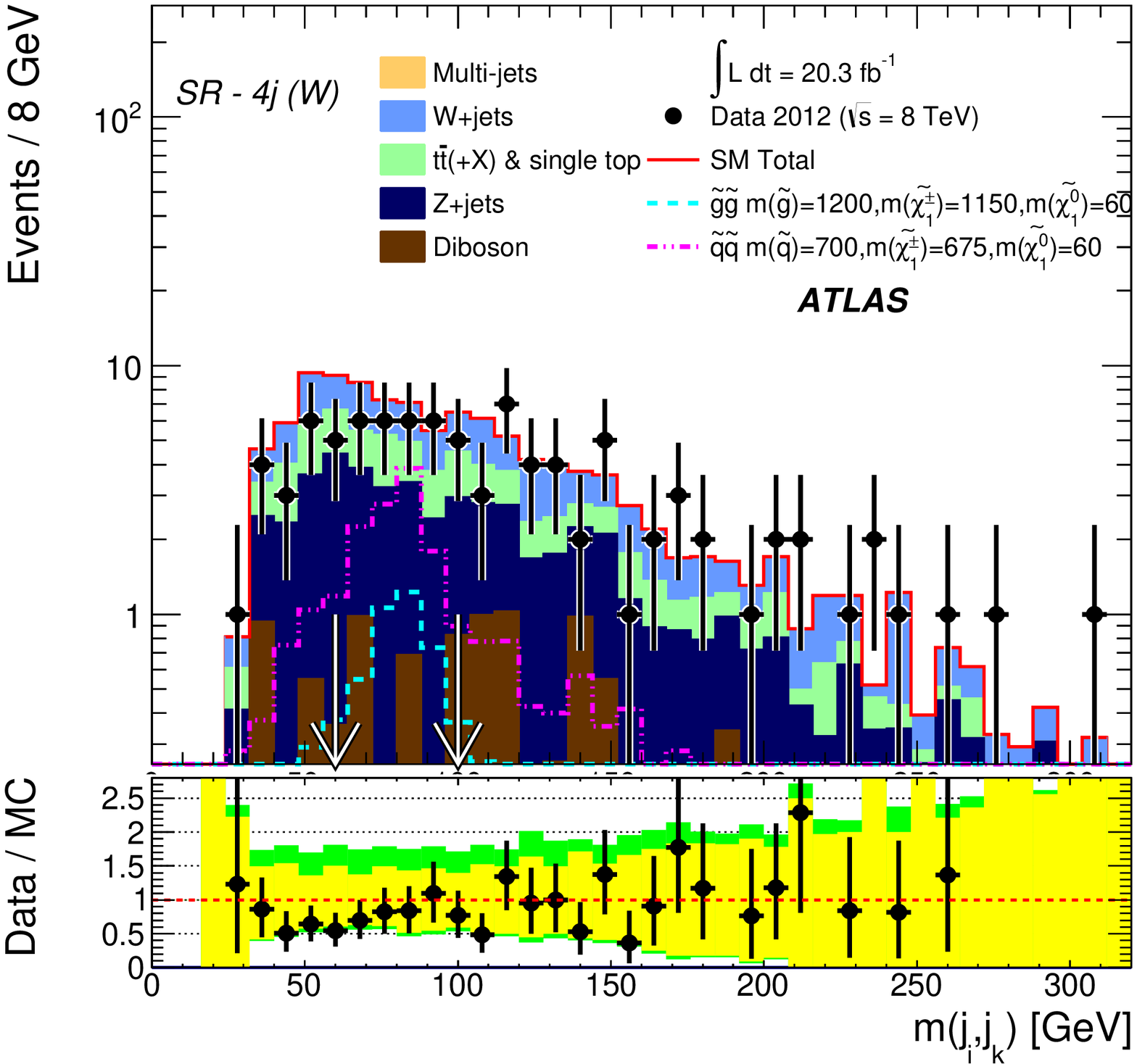}
\end{center}
\vspace*{-0.5cm}
\caption{\label{fig:src}Observed jet and dijet mass distributions for the 2jW (top) and 4jW (bottom) signal regions for all unresolved $W$ candidates (left) and for an additional $W$ candidate after requiring at least one unresolved $W$ candidate (right). The additional $W$ candidate is unresolved (SR 2jW, top-right) or resolved (SR 4jW, bottom-right). With the exception of the multi-jet background (which is estimated using the data-driven technique described in the text), the histograms denote the MC background expectations prior to the fits described in the text, normalised to cross-section times integrated luminosity. In the lower panels the light (yellow) error bands denote the experimental systematic and MC statistical uncertainties, while the medium dark (green) bands include also the theoretical modelling uncertainty. Expected distributions for benchmark model points are also shown for comparison (masses in GeV). Arrows indicate the location of the mass window used in the final selection. See text for discussion of compatibility of data with MC background expectations. 
}
\end{figure}

Distributions of $\meff$(incl.) and jet and dijet masses (the latter for SR 2jW and SR 4jW) obtained before the final selections on these quantities (but after applying all other selections), for data and the different MC samples normalised with the theoretical cross-sections (with the exception of the multi-jet background, which is estimated using the data-driven technique described in section~\ref{sec:strategy}), are shown in figures~\ref{fig:sra}--\ref{fig:src}. Examples of typical expected SUSY signals are shown for illustration. These signals correspond to the processes to which each SR is primarily sensitive -- $\squark\squark$ production for the lower jet-multiplicity SRs, $\squark\gluino$ associated production for intermediate jet-multiplicity SRs, and $\gluino\gluino$ production for the higher jet-multiplicity SRs. In these figures data and background distributions largely agree within uncertainties; however, there is a systematic difference between the data and the background prediction which increases towards larger values of the kinematic variables considered. This difference does not affect the background expectations in the signal regions used in the analysis, however, due to the use of the likelihood fits to the CR event yields discussed in section~\ref{sec:strategy}.

\begin{figure}
\begin{center}
\includegraphics[height=0.5\textwidth]{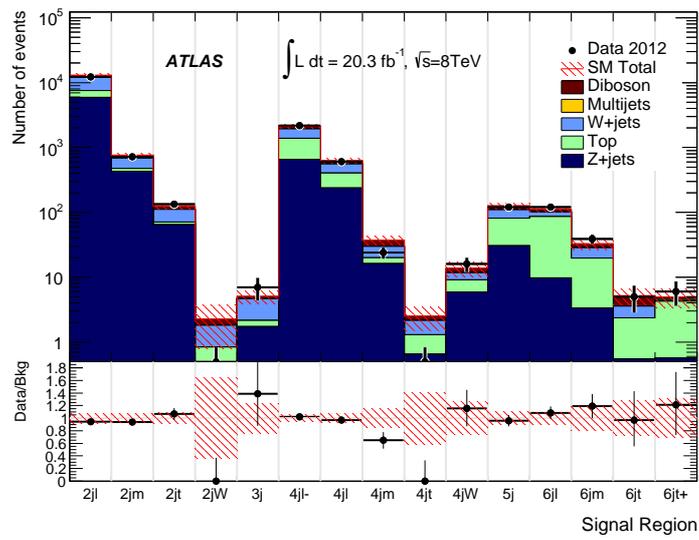}
\end{center}
\caption{\label{fig:PlotSR} Comparison of the observed and expected event yields as a function of signal region. The background expectations are those obtained from the background-only fits presented in table~\ref{tab:p0_UL}. In SRs 2jW and 4jt no events are observed in the data.}
\end{figure}

The number of events observed in the data and the number of SM events expected to enter each of the signal regions, determined using the background-only fit, are shown in \mytabref{tab:p0_UL} and figure~\ref{fig:PlotSR}. The pre-fit background expectations are also shown in \mytabref{tab:p0_UL} to aid comparison. The fit to the CRs for each SR compensates for the disagreement between data and pre-fit background expectations seen in figures~\ref{fig:sra}--\ref{fig:src}, leading to good agreement between data and post-fit expectations. The most significant observed excess across the 15 SRs, with a p-value for the background-only hypothesis of 0.24, occurs in SR 3j. 

\clearpage

\begin{table}
\scriptsize
\begin{center}
\caption[p0 and UL]{Numbers of events observed in the signal regions used in the analysis compared with background expectations obtained from the fits described in the text. When a dash is shown, the entry is less than $0.01$. Combined uncertainties on the predicted background event yields are quoted as symmetric except where the negative uncertainty reaches down to zero predicted events, in which case the negative uncertainty is truncated.
The p-values ($p_0$) for the background-only hypothesis are truncated at 0.5 and are also interpreted in terms of the equivalent Gaussian significance ($Z$).
Also shown are 95\% CL upper limits on the visible cross-section ($\langle\epsilon\sigma\rangle_{\rm obs}^{95}$), 
the observed number of signal events ($S_{\rm obs}^{95}$ ) and the number of signal events ($S_{\rm exp}^{95}$) 
given the expected number (and $\pm 1\sigma$ excursions on the expectation) of background events. Limits are evaluated using MC pseudo-experiments as well as asymptotic formulae.
\label{tab:p0_UL}}
\scriptsize
\begin{tabular}{lrrrrr}
\hline
Signal Region              & {\bf 2jl }            & {\bf 2jm }                         & {\bf 2jt }                  & {\bf 2jW }                  & {\bf 3j } \\
\hline
\multicolumn{6}{c}{MC expected events} \\ \hline
Diboson                    &  $879$                &  $72$                               &  $13$                       &  $0.41$                     &  $0.36$ \\
${Z/\gamma^{*}}$+jets      &  $6709$               &  $552$                              &  $103$                      &  $1.2$                      &  $5.5$  \\
$W$+jets                   &  $5472$               &  $303$                              &  $59$                       &  $0.82$                     &  $3.1$  \\
$\ttbar$(+EW) + single top &  $1807$               &  $54$                               &  $9$                        &  $0.14$                     &  $0.85$ \\
\hline
\multicolumn{6}{c}{Fitted background events} \\ \hline
Diboson                     & $900 \pm 400$           & $70 \pm 40$                      & $13 \pm 6$                   & $0.41 \pm 0.21$             & $0.36 \pm 0.18$ \\
${Z/\gamma^{*}}$+jets       & $5900 \pm 900$          & $430 \pm 40$                     & $65 \pm 8$                   & $0.4 \pm 0.4$         & $1.7 \pm 1.0$ \\
$W$+jets                    & $4500 \pm 600$          & $216 \pm 26$                     & $40 \pm 6$                   & $1.0 \pm 1.0$               & $2.5 \pm 0.9$ \\
$\ttbar$(+EW) + single top  & $1620 \pm 320$          & $47 \pm 8$                       & $6.5 \pm 2.2$                & $0.4_{-0.4}^{+0.8}$         & $0.4_{-0.4}^{+0.5}$ \\
Multi-jets                  & $115_{-120}^{+140}$     & $0.4_{-0.4}^{+1.4}$              & $0.1_{-0.1}^{+0.4}$          & $0.03 \pm 0.03$             & $0.03_{-0.03}^{+0.06}$ \\
\hline
Total bkg                   & $13000 \pm 1000$        & $760 \pm 50$                     & $125 \pm 10$                 & $2.3 \pm 1.4$               & $5.0 \pm 1.2$ \\
\hline
Observed                    & $12315$                 & $715$                            & $133$                        & $0$                         & $7$  \\
\hline
$\langle\epsilon{\rm \sigma}\rangle_{\rm obs}^{95}$ [fb]              &  $60$                    &  $4.3$                &  $1.9$              &  $0.16$                 &  $0.40$  \\
$\langle\epsilon{\rm \sigma}\rangle_{\rm obs}^{95}$ [fb] (asymptotic) &  $62$                    &  $4.0$               &  $1.8$              &  $0.12$                 &  $0.40$  \\
$S_{\rm obs}^{95}$                                                    &  $1200$                  &  $90$                 &   $38$              &  $3.2$                  &  $8.2$   \\
$S_{\rm obs}^{95}$ (asymptotic)                                       &  $1300$                  &  $80$                 &   $37$              &  $2.5$                  &  $8.1$   \\
$S_{\rm exp}^{95}$                                                    &  ${1700}^{+600}_{-500}$  &  ${110}^{+40}_{-30}$  &  ${32}^{+11}_{-10}$ &  ${4.0}^{+1.7}_{-0.7}$  &  ${6.4}^{+2.9}_{-1.3}$  \\
$S_{\rm exp}^{95}$ (asymptotic)                                       &  ${1600}^{+600}_{-400}$  &  ${110}^{+40}_{-30}$  &  ${31}^{+12}_{-8}$  &  ${4.1}^{+2.4}_{-1.4}$  &  ${6.3}^{+3.2}_{-2.0}$  \\
$p_{0}$ ($Z$)                                                         &  $0.50$ ($0.0$)          &  $0.49$  ($0.0$)      &   $0.29$  ($0.5$)   &  $0.50$ ($0.0$)         &  $0.24$  ($0.7$)   \\
\hline \hline
Signal Region               & {\bf 4jl- }              & {\bf 4jl }                       & {\bf 4jm }                   & {\bf 4jt }                  & {\bf 4jW } \\
\hline
\multicolumn{6}{c}{MC expected events} \\ \hline
Diboson                     &  $175$                   &  $70$                            &  $7.2$                       &  $0.34$                     &  $2.1$  \\
${Z/\gamma^{*}}$+jets       &  $885$                   &  $333$                           &  $30$                        &  $2.9$                      &  $11$   \\
$W$+jets                    &  $832$                   &  $284$                           &  $16$                        &  $1.2$                      &  $6.1$   \\
$\ttbar$(+EW) + single top  &  $764$                   &  $167$                           &  $4.0$                       &  $0.6$                      &  $3.1$  \\
\hline
\multicolumn{6}{c}{Fitted background events} \\ \hline
Diboson                     & $180 \pm 90$             & $70 \pm 34$                      & $7 \pm 4$                    & $0.34 \pm 0.17$             & $2.1 \pm 1.0$ \\
${Z/\gamma^{*}}$+jets       & $660 \pm 60$             & $238 \pm 28$                     & $16 \pm 4$                   & $0.7_{-0.7}^{+0.8}$         & $5.9 \pm 2.1$ \\
$W$+jets                    & $560 \pm 80$             & $151 \pm 28$                     & $10 \pm 4$                   & $0.9 \pm 0.4$               & $2.7 \pm 1.6$ \\
$\ttbar$(+EW) + single top  & $730 \pm 50$             & $167 \pm 18$                     & $4 \pm 2$                    & $0.6 \pm 0.6$               & $3.2 \pm 3.1$ \\
Multi-jets                  & $1.7_{-1.7}^{+4.0}$      & $0.7_{-0.7}^{+1.6}$              & --                           & --                          & -- \\
\hline
Total bkg                   & $2120 \pm 110$           & $630 \pm 50$                     & $37 \pm 6$                   & $2.5 \pm 1.0$               & $14 \pm 4$ \\
\hline
Observed                    & $2169$                   & $608$                            & $24$                         & $0$                         & $16$ \\
\hline
$\langle\epsilon{\rm \sigma}\rangle_{\rm obs}^{95}$ [fb]              &  $13$                 &  $4.5$               &  $0.52$            &  $0.15$                 &  $0.68$  \\
$\langle\epsilon{\rm \sigma}\rangle_{\rm obs}^{95}$ [fb] (asymptotic) &  $13$                 &  $4.3$               &  $0.45$            &  $0.12$                 &  $0.63$  \\
$S_{\rm obs}^{95}$                                                    &  $270$                &  $91$                &  $10$              &  $3.1$                  &  $14$   \\
$S_{\rm obs}^{95}$ (asymptotic)                                       &  $270$                &  $87$                &  $9$             &  $2.5$                  &  $13$   \\
$S_{\rm exp}^{95}$                                                    &  ${240}^{+90}_{-70}$  &  ${103}^{+34}_{-29}$ &  ${16}^{+6}_{-4}$  &  ${4.0}^{+1.8}_{-0.9}$  &  ${11}^{+5}_{-3}$  \\
$S_{\rm exp}^{95}$ (asymptotic)                                       &  ${240}^{+90}_{-70}$  &  ${97}^{+35}_{-25}$  &  ${15}^{+6}_{-4}$  &  ${4.0}^{+2.4}_{-1.4}$  &  ${11}^{+5}_{-3}$  \\
$p_{0}$ ($Z$)                                                         &  $0.35$  ($0.4$)      &  $0.50$ ($0.0$)      &  $0.50$ ($0.0$)    &  $0.50$  ($0.0$)        &  $0.34$  ($0.4$)   \\

\hline \hline
Signal Region               & {\bf 5j }                & {\bf 6jl }                        & {\bf 6jm }                  & {\bf 6jt }                  & {\bf 6jt+ } \\
\hline
\multicolumn{6}{c}{MC expected events} \\ \hline
Diboson                     &  $16$                    &  $9$                              &  $4$                        &  $1.6$                      &  $0.21$  \\
${Z/\gamma^{*}}$+jets       &  $51$                    &  $18$                             &  $7$                        &  $1.8$                      &  $2.1$   \\
$W$+jets                    &  $54$                    &  $26$                             &  $12$                       &  $2.1$                      &  $3.4$   \\
$\ttbar$(+EW) + single top  &  $52$                    &  $80$                             &  $19$                       &  $2.2$                      &  $3.4$   \\
\hline 
\multicolumn{6}{c}{Fitted background events} \\ \hline
Diboson                     & $16 \pm 8$               & $9 \pm 4$                         & $4 \pm 2$                   & $1.6 \pm 0.8$               & $0.2 \pm 0.1$ \\
${Z/\gamma^{*}}$+jets       & $31 \pm 8$               & $9 \pm 4$                         & $3 \pm 2$                   & $0.6 \pm 0.6$         & $0.6_{-0.6}^{+0.8}$ \\
$W$+jets                    & $28 \pm 8$               & $15 \pm 7$                        & $9 \pm 5$                   & $1.2 \pm 0.9$               & $0.3_{-0.3}^{+1.2}$ \\
$\ttbar$(+EW) + single top  & $51 \pm 9$               & $76 \pm 7$                        & $16 \pm 4$                  & $1.8 \pm 0.6$               & $3.7 \pm 1.7$ \\
Multi-jets                  & $1.0_{-1.0}^{+2.6}$      & $1.7_{-1.7}^{+3.0}$               & $0.4_{-0.4}^{+0.8}$         & $0.01_{-0.01}^{+0.03}$      & $0.3_{-0.3}^{+0.4}$ \\
\hline
Total bkg                   & $126 \pm 13$             & $111 \pm 11$                      & $33 \pm 6$                  & $5.2 \pm 1.4$               & $4.9 \pm 1.6$ \\
\hline
Observed                    &  $121$                   &  $121$                            &  $39$                       &  $5$                        &  $6$                     \\
\hline
$\langle\epsilon{\rm \sigma}\rangle_{\rm obs}^{95}$ [fb]              &  $1.7$  &  $1.9$     &  $1.2$  &  $0.32$   &  $0.39$  \\
$\langle\epsilon{\rm \sigma}\rangle_{\rm obs}^{95}$ [fb] (asymptotic) &  $1.6$  &  $1.8$     &  $1.1$  &  $0.30$   &  $0.36$  \\
$S_{\rm obs}^{95}$                                                    &  $35$   &  $39$      &  $25$   &  $6.6$    &  $7.9$   \\
$S_{\rm obs}^{95}$ (asymptotic)                                       &  $32$   &  $37$      &  $22$   &  $6.1$    &  $7.3$   \\
$S_{\rm exp}^{95}$                                                    &  ${37}^{+13}_{-10}$  &  ${31}^{+12}_{-6}$  &  ${20}^{+6}_{-4}$  &  ${6.2}^{+2.6}_{-1.3}$  &  ${6.6}^{+2.6}_{-1.6}$  \\
$S_{\rm exp}^{95}$ (asymptotic)                                       &  ${35}^{+13}_{-10}$  &  ${30}^{+12}_{-8}$  &  ${18}^{+7}_{-5}$  &  ${6.3}^{+3.1}_{-2.0}$  &  ${6.4}^{+3.2}_{-2.0}$  \\
$p_{0}$ ($Z$)                                                         &  $0.50$ ($0.0$)      &  $0.27$  ($0.6$)    &  $0.25$  ($0.7$)   &  $0.50$ ($0.0$)         &  $0.36$  ($0.4$)   \\
\hline
\end{tabular}

\end{center}
\end{table}

\clearpage

\clearpage

\section{Interpretation}
\label{sec:interpret}

In the absence of a statistically significant excess, limits are set on contributions to the SRs from BSM physics. Upper limits at 95\% CL on the number of BSM signal events in each SR and the corresponding visible BSM cross-section are derived from the model-independent fits described in section~\ref{sec:strategy} using the CL$_{\rm s}$ prescription~\cite{clsread}. The limits are evaluated using MC pseudo-experiments as well as asymptotic formulae  \cite{Cowan:2010js}. The results are presented in table~\ref{tab:p0_UL}. Asymptotic limits differ appreciably from those evaluated using MC pseudo-experiments only for the tightest signal regions (2jW and 4jt), where the small expected number of events limits the accuracy of the former.

The SUSY-model exclusion fits in all the SRs are then used to set limits on specific classes of SUSY models, using the result from the SR with the best expected sensitivity at each point in each model parameter space. These limits are evaluated using asymptotic formulae only. `Observed limits' are calculated from the observed SR event yields for both the nominal signal cross-section and with its $\pm1 \sigma$ uncertainties. Numbers quoted in the text are evaluated in a conservative fashion from the observed exclusion limit based on the nominal signal cross-section minus its $1\sigma$ theoretical uncertainty. `Expected limits' are calculated by setting the nominal event yield in each SR to the corresponding mean expected background.

Theoretical uncertainties on both the signal cross-section (discussed in section~\ref{sec:mcsamples}) and signal acceptance are taken into account when setting limits on specific SUSY models using the SUSY-model exclusion fits. An important consideration is that initial state radiation (ISR) can significantly affect the signal acceptance for SUSY models with small mass splittings, $\Delta m$, between the strongly interacting states ($\squark$ or $\gluino$) and the $\ninoone$. Systematic uncertainties arising from the treatment of ISR are studied with MC data samples by varying the value of $\alpha_{\rm s}$, renormalisation and factorisation scales, and the \Madgraph/\pythia{} matching parameters. For mass splittings $\Delta m < 100$~GeV the uncertainty ranges from 10\% to 40\% depending on the signal region. For fixed $\Delta m$ the uncertainty is found to be independent of the SUSY particle mass, while for fixed mass it falls approximately exponentially with increasing $\Delta m$, with a characteristic decay constant $\sim$ 200--300 GeV.

\begin{figure*}[htb]
\begin{center}
\subfigure{\includegraphics[width=0.49\textwidth]{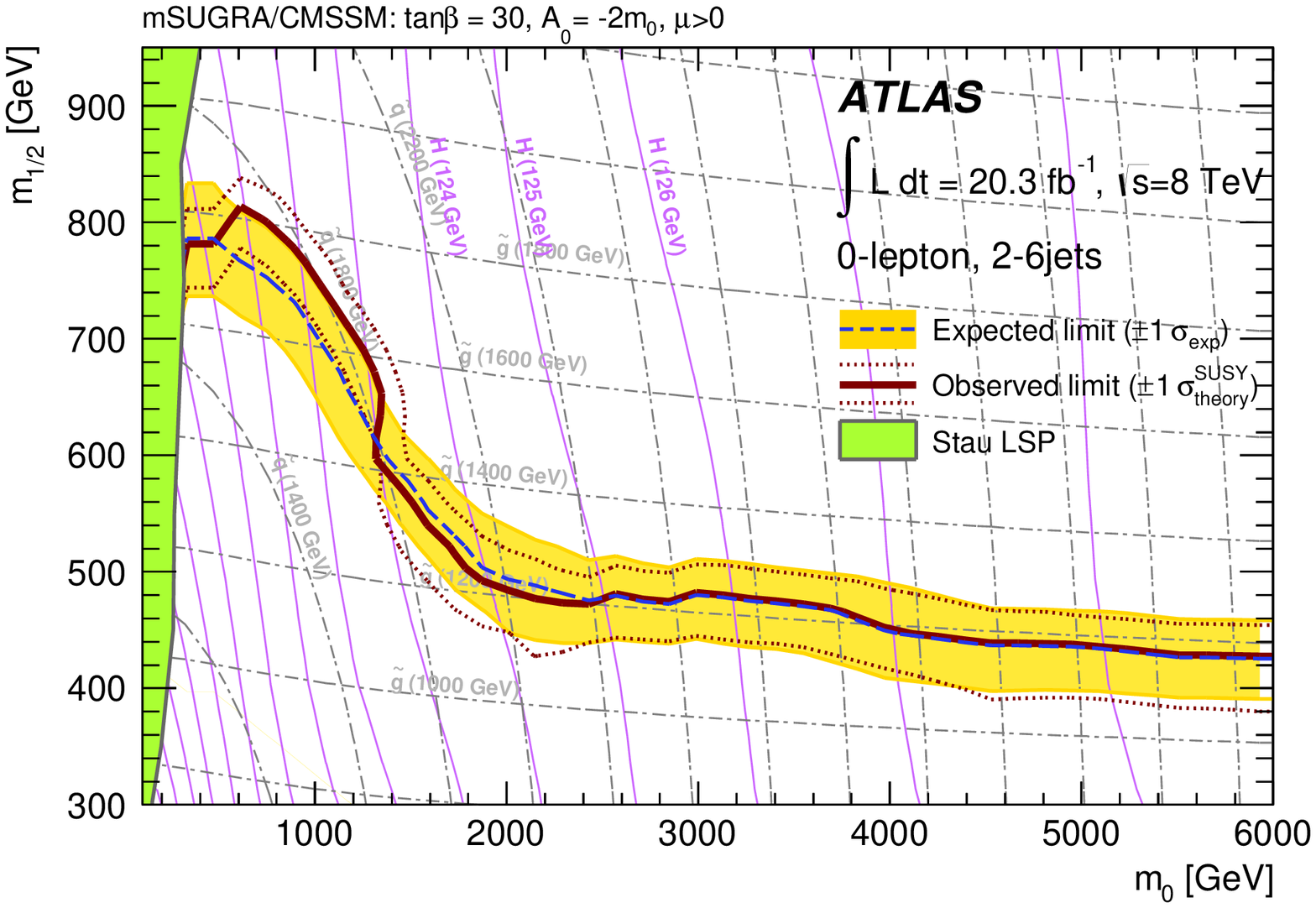}}
\subfigure{\includegraphics[width=0.49\textwidth]{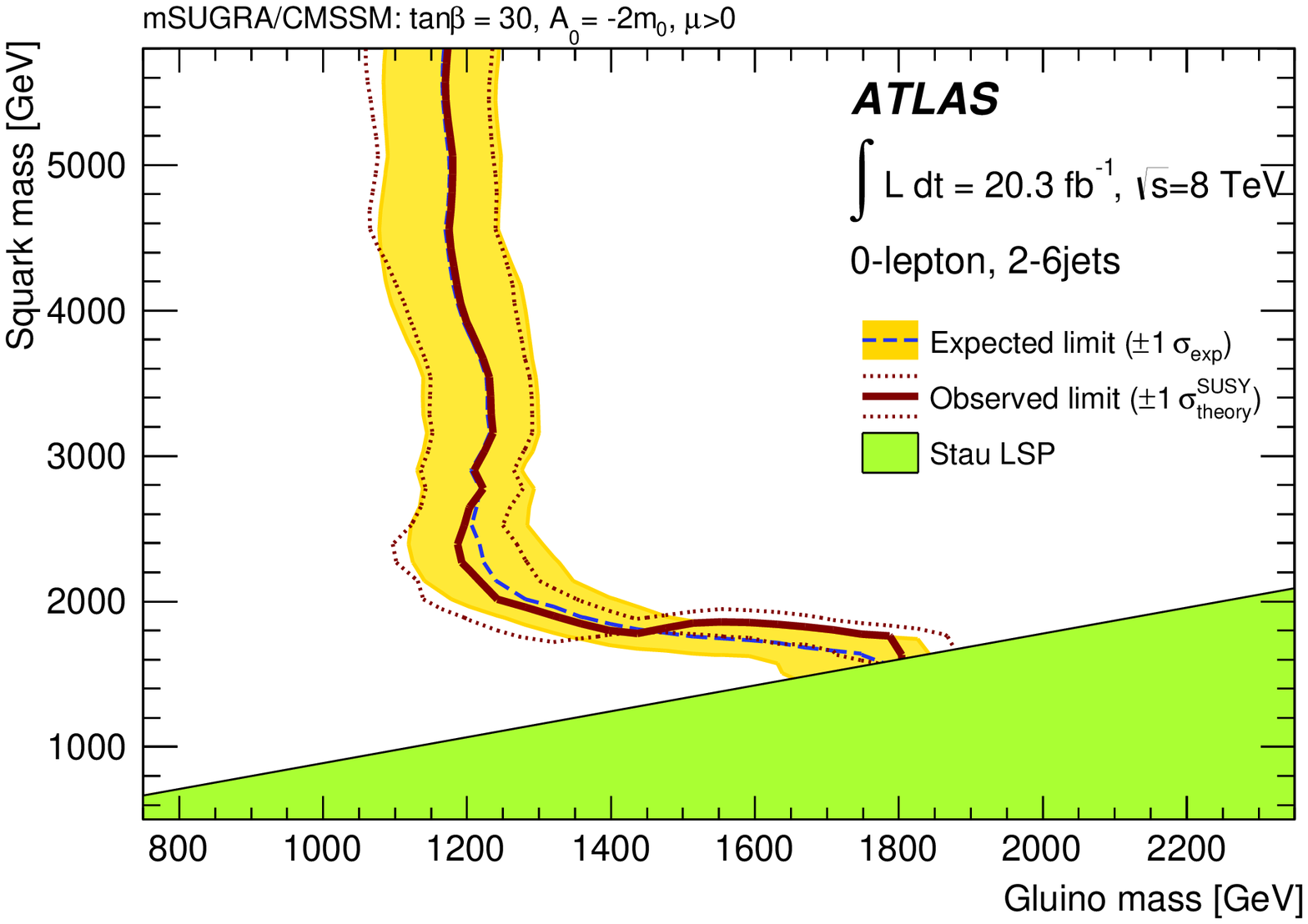}}
 \caption{
Exclusion limits for \mSUGRA{} models with $\tan\beta=30$, $A_0=-2m_0$ and $\mu>0$ presented (left) in the $(m_0,m_{1/2})$-plane and (right) in the $(m_{\gluino},m_{\squark})$-plane. Exclusion limits are obtained by using the signal region with the best expected sensitivity at each point. The blue dashed lines show the expected limits at 95\% CL, with the light (yellow) bands indicating the $1\sigma$ excursions due to experimental and background-only theory uncertainties.
Observed limits are indicated by medium dark (maroon) curves, where the solid contour represents the nominal limit, and the dotted lines are obtained by varying the signal cross-section by the renormalisation and factorisation scale and PDF uncertainties. 
\label{fig:limitcombined}}
\end{center}
\end{figure*}

In figure~\ref{fig:limitcombined} the results are interpreted in the
$\tan\beta=30$, $A_0=-2m_0$, $\mu>0$ slice of mSUGRA /CMSSM \citemsugraandcmssm{} models.\footnote{Five parameters are needed to specify a particular \mSUGRA{}\ model: the universal scalar mass, $m_0$, the universal gaugino mass $m_{1/2}$, the universal trilinear scalar coupling, $A_0$, the ratio of the vacuum expectation values of the two Higgs fields, $\tan\beta$, and the sign of the higgsino mass parameter, $\mu=\pm$. } The mass of the lightest neutral Higgs boson predicted by such models is enhanced relative to that predicted by the \mSUGRA{} models used in previous related ATLAS publications~\cite{Aad:2012fqa,Aad:2011ib,daCosta:2011qk}.
The best-performing signal regions are 6jt for $m_{0}\gtrsim$ 1300 GeV and 4jt for $m_{0}\lesssim$ 1300 GeV, with SR 3j providing additional sensitivity for $m_{0}\lesssim$ 400 GeV. Results  are presented in both the $(m_0,m_{1/2})$-plane and the $(m_{\gluino},m_{\squark})$-plane. For this model SUSY signal events are generated with \herwig++-2.5.2. The lower limit on $m_{1/2}$ is greater than 380~GeV for $m_0<6$ TeV and reaches 770~GeV for
low values of $m_0$. Equal mass light-flavour squarks and gluinos are excluded below 1700~GeV in this scenario. 

\begin{figure*}[htb]
\begin{center}
\subfigure{\includegraphics[width=0.70\textwidth]{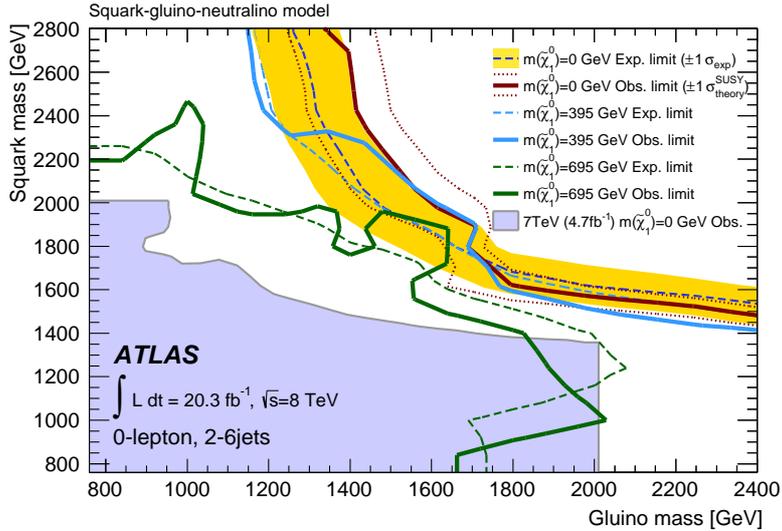}}
\caption{Exclusion limits for a simplified phenomenological MSSM scenario with only strong production of gluinos and first- and second-generation squarks (of common mass), with direct decays to quarks and lightest neutralinos. Three values of the lightest neutralino mass are considered: $m_{\ninoone}=$ 0, 395 GeV and 695 GeV. Exclusion limits are obtained by using the signal region with the best expected sensitivity at each point. The dashed lines show the expected limits at 95\% CL, with the light (yellow) band indicating the $1\sigma$ experimental and background-only theory uncertainties on the $m_{\ninoone}=0$ limit.
Observed limits are indicated by solid curves. The dotted lines represent the $m_{\ninoone}=$ 0 observed limits obtained by varying the signal cross-section by the renormalisation and factorisation scale and PDF uncertainties. Previous results for $m_{\ninoone}= 0$ from ATLAS at 7 TeV~\cite{Aad:2012fqa} are represented by the shaded (light blue) area. Results at 7 TeV are valid for squark or gluino masses below 2000 GeV, the mass range studied for that analysis.
\label{fig:phenoMSSM}}
\end{center}
\end{figure*}

An interpretation of the results is also presented in figure~\ref{fig:phenoMSSM} as a 95\% CL exclusion region in the $(m_{\gluino},m_{\squark})$-plane for a simplified set of phenomenological MSSM (Minimal Supersymmetric extension of the SM) models \cite{Alwall:2008ag,Alves:2011wf} with $m_{\ninoone}$ equal to 0, 395 GeV or 695 GeV.  In these models the gluino mass and the masses of the `light'-flavour squarks (of the first two generations, including both $\squark_{\rm R}$ and $\squark_{\rm L}$, and assuming mass degeneracy) are set to the values shown on the axes of figure~\ref{fig:phenoMSSM}. All other supersymmetric particles, including the squarks of the third generation, have their masses set to very high values (`decoupled'). SUSY signal events are generated with \madgraph-5.0 interfaced to \pythia-6.426. A lower limit of 1650 GeV for equal mass light-flavour squarks and gluinos is found for the scenario with a massless $\ninoone$. 

\clearpage

\begin{figure*}[htb]
\begin{center}
\subfigure{\includegraphics[width=0.49\textwidth]{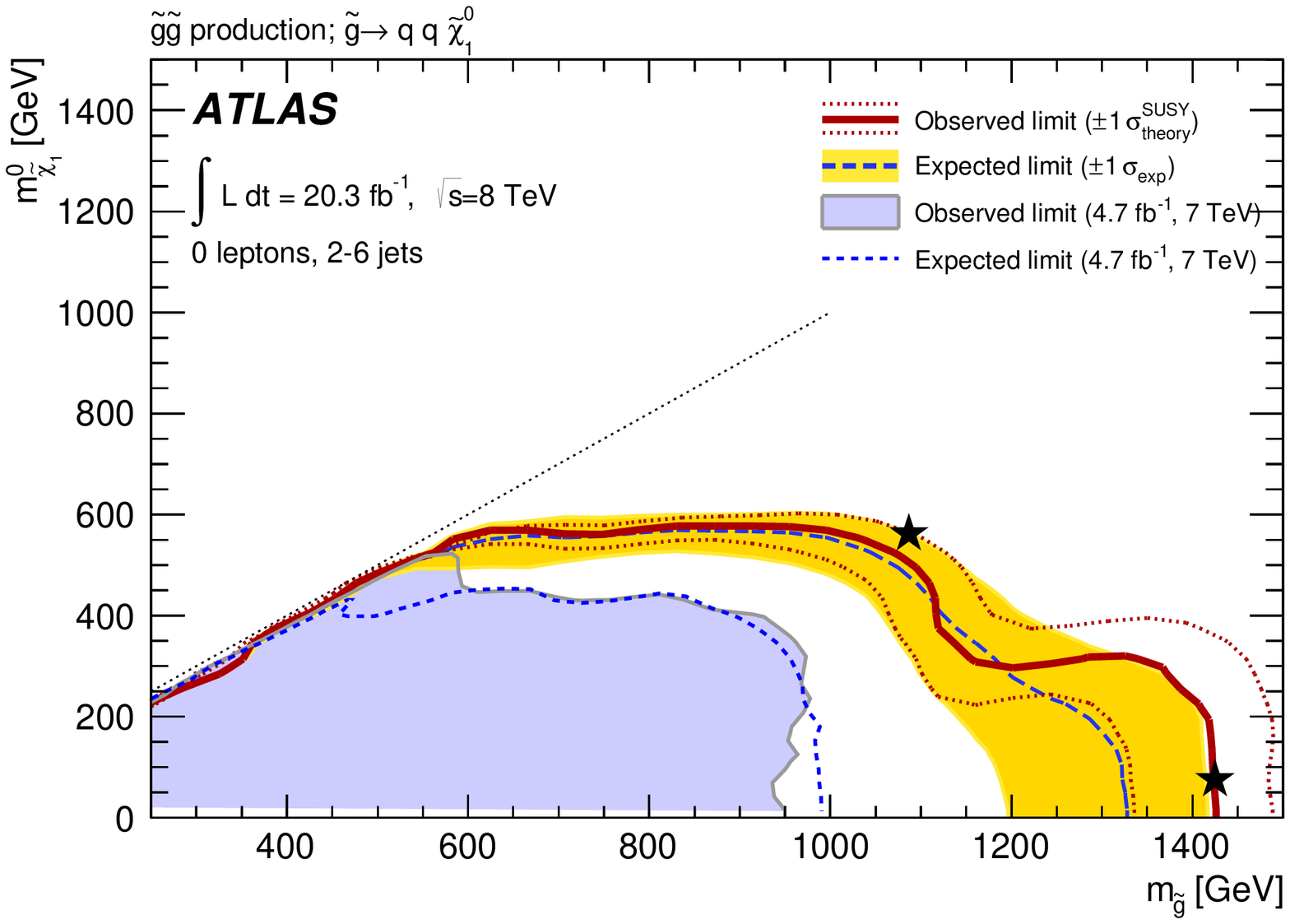}}
\subfigure{\includegraphics[width=0.49\textwidth]{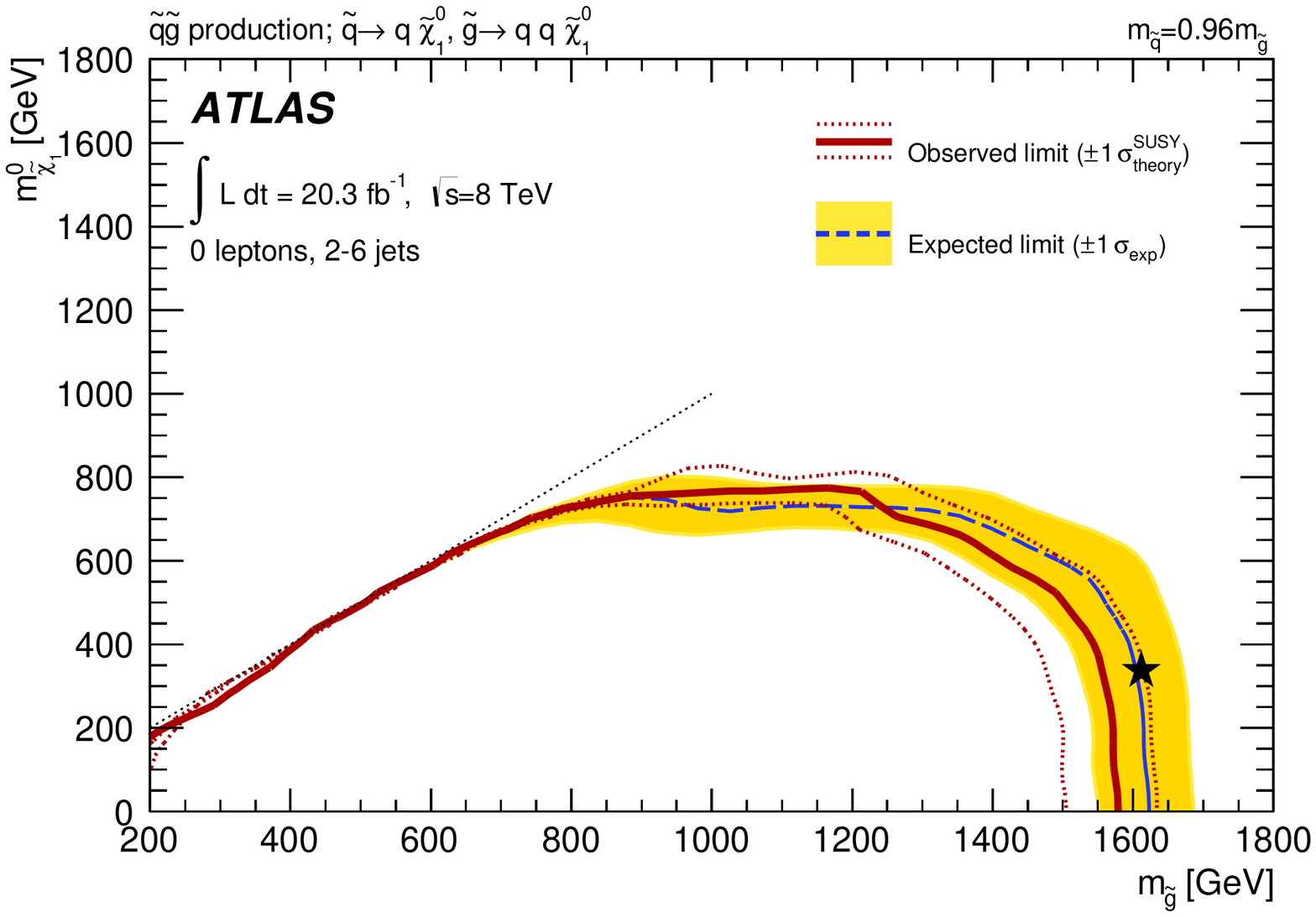}}
\subfigure{\includegraphics[width=0.49\textwidth]{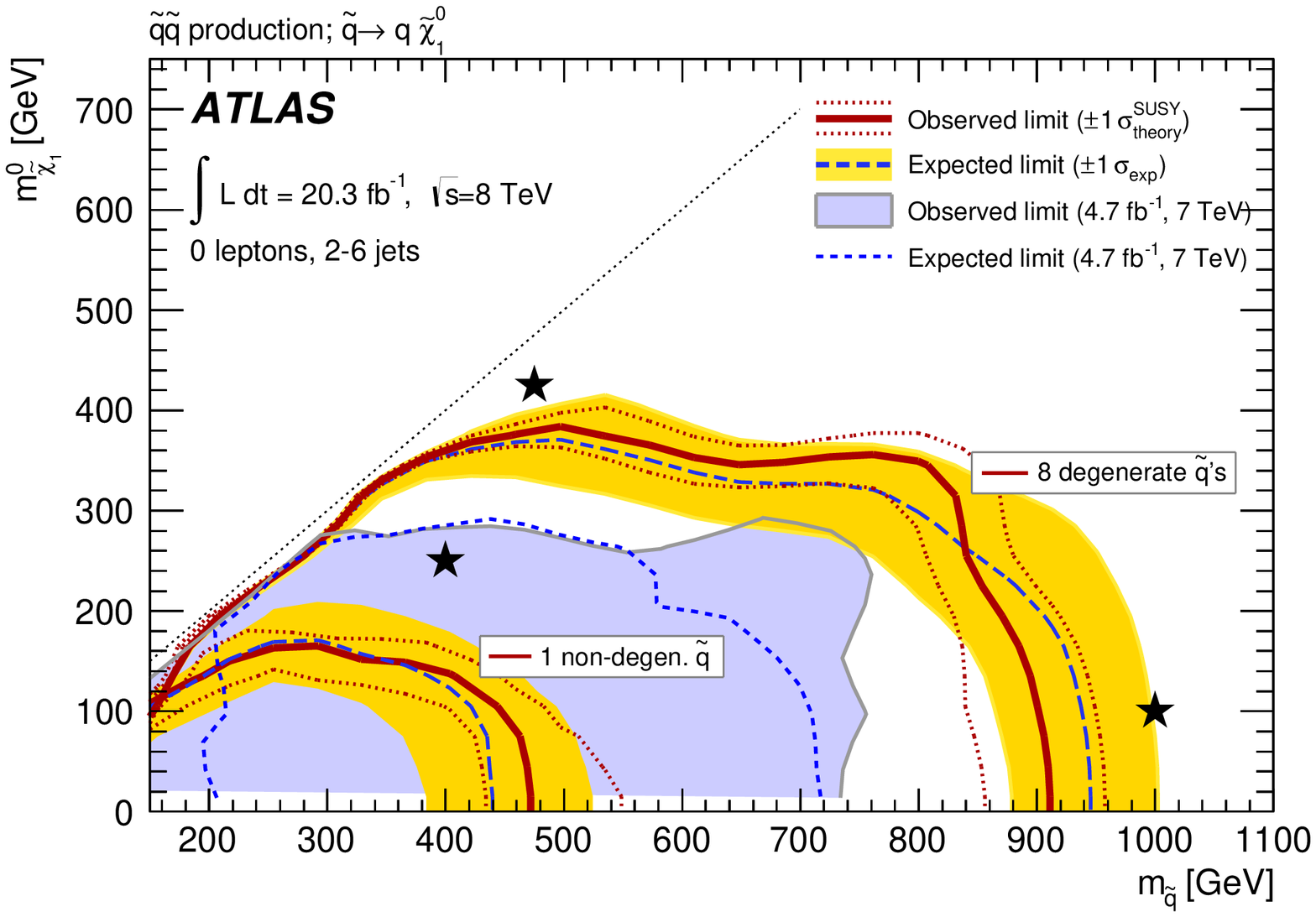}}
 \caption{ Exclusion limits for direct production of (case (a) -- top left) gluino pairs with decoupled squarks, (case (b) -- top right) light-flavour squarks and gluinos and (case (c) -- bottom) light-flavour squark pairs with decoupled gluinos. Gluinos (light-flavour squarks) are required to decay to two quarks (one quark) and a neutralino LSP. In the bottom figure (case (c)) limits are shown for scenarios with eight degenerate light-flavour squarks ($\squark_{\rm L}+\squark_{\rm R}$), or with only one non-degenerate light-flavour squark produced. Exclusion limits are obtained by using the signal region with the best expected sensitivity at each point.  
The blue dashed lines show the expected limits at 95\% CL, with the light (yellow) bands indicating the $1\sigma$ excursions due to experimental and background-only theory uncertainties.
Observed limits are indicated by medium dark (maroon) curves, where the solid contour represents the nominal limit, and the dotted lines are obtained by varying the signal cross-section by the renormalisation and factorisation scale and PDF uncertainties. Previous results from ATLAS~\cite{Aad:2012fqa} are represented by the shaded (light blue) areas and light blue dotted lines. The black stars indicate benchmark models used in figures~\ref{fig:sra}--\ref{fig:src}. 
\label{fig:directLimit}}
\end{center}
\end{figure*}

\begin{figure}[htb]
\begin{center}
\subfigure{\includegraphics[width=0.49\textwidth]{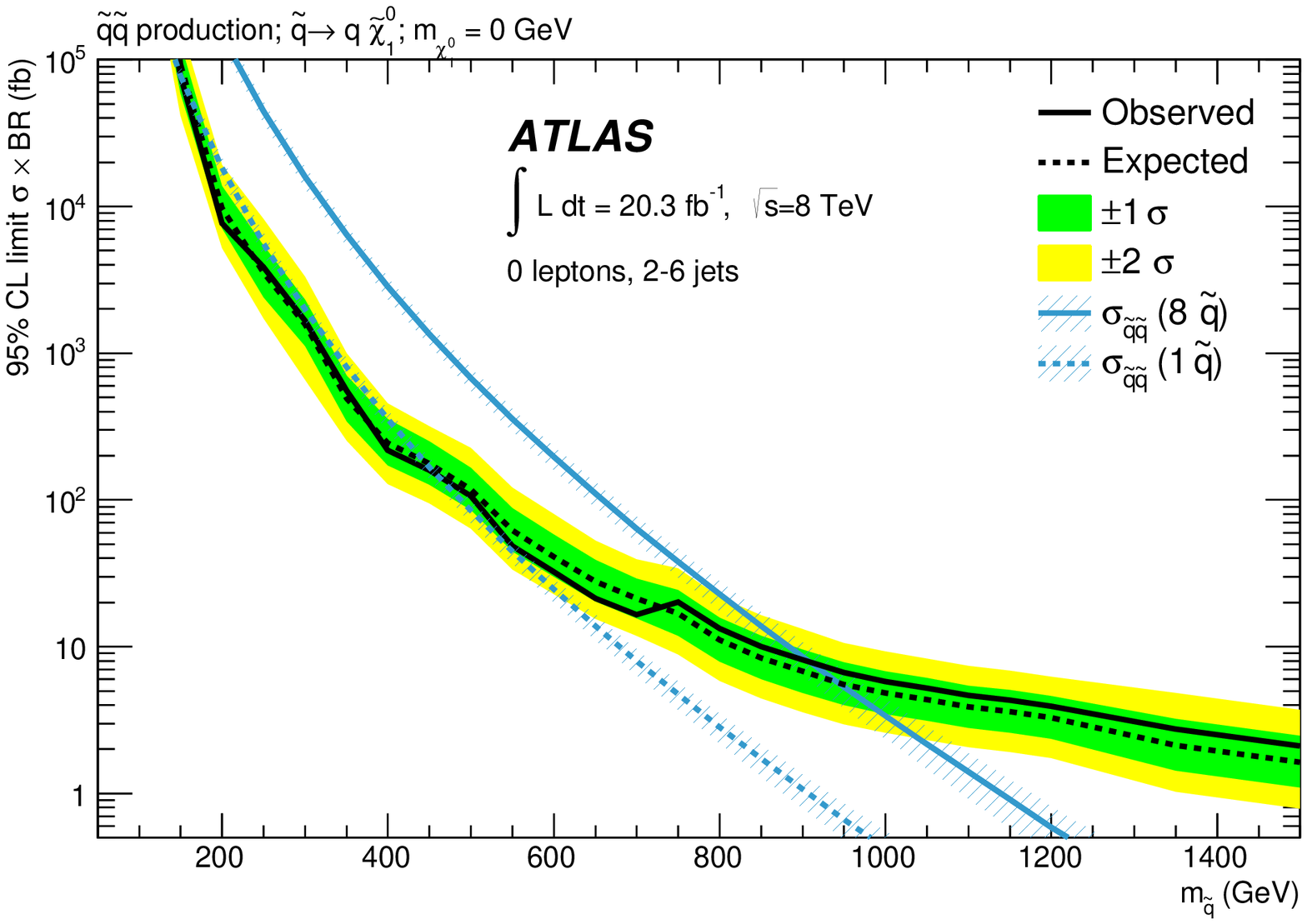}}
\subfigure{\includegraphics[width=0.49\textwidth]{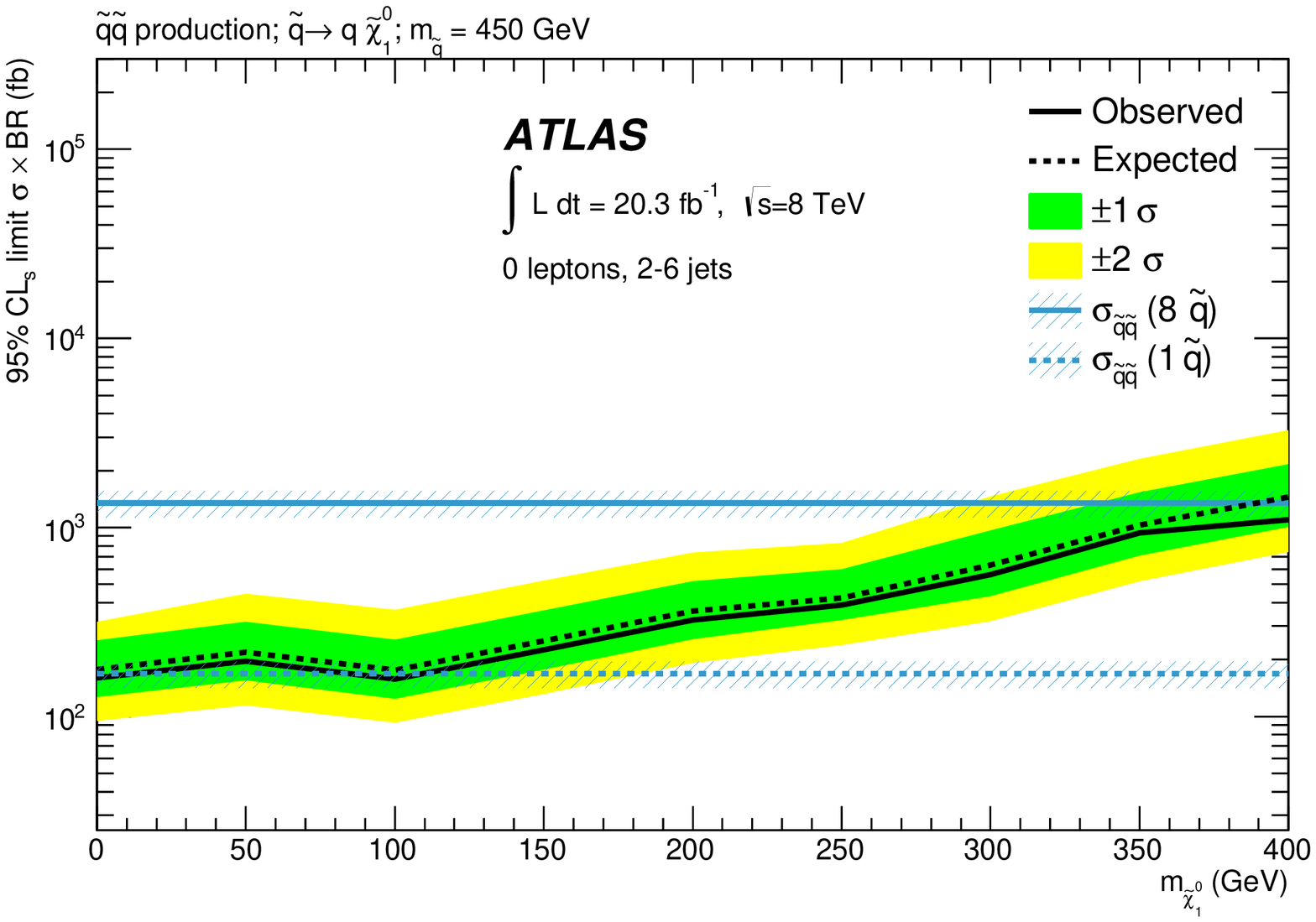}}
\caption{Limits on the light-flavour squark pair production cross-section times branching ratio for models with squark pairs decaying directly into quarks and $\ninoone$ (case (c) in figure~\ref{fig:directLimit}) as a function of $m_{\squark}$ for $m_{\ninoone}=0$ (left) and as a function of $m_{\ninoone}$ for $m_{\squark}=450$ GeV (right). Exclusion limits are obtained by using the signal region with the best expected sensitivity at each point. The medium dark (green) band indicates the 1$\sigma$ uncertainty on the expected upper limit, the light (yellow) band the 2$\sigma$ uncertainty. The solid medium dark (blue) line indicates the theoretical inclusive squark pair production cross-section times branching ratio for eight degenerate light-flavour squarks. The dashed medium dark (blue) line indicates the equivalent theoretical cross-section times branching ratio for models in which only one non-degenerate light-flavour squark is produced. The hatched (blue) bands around the theoretical $\sigma\cdot$BR curves denote the scale and PDF uncertainties.}
\label{fig:limitSMdirect_lines}
\end{center}
\end{figure}

In figure~\ref{fig:directLimit} limits are shown for three classes of simplified model in which only direct production of (a) gluino pairs, (b) light-flavour squarks and gluinos or (c) light-flavour squark pairs are considered. All other superpartners, except for the neutralino LSP $\ninoone$, are decoupled thereby forcing each light-flavour squark or gluino to decay directly to one or more quarks and a $\ninoone$. Cross-sections are evaluated assuming decoupled (masses set to 4.5 TeV) light-flavour squarks or gluinos in cases (a) and (c), respectively. In case (b) the masses of the light-flavour squarks are set to 0.96 times the mass of the gluino, matching the prescription used in refs.~\cite{LeCompte:2011cn,LeCompte:2011fh}. In case (c) limits are shown for scenarios with eight degenerate light-flavour squarks ($\squark_{\rm L}+\squark_{\rm R}$), or with only one non-degenerate light-flavour squark produced \cite{Mahbubani:2012qq}. For these models SUSY signal events are generated with \madgraph-5.0 interfaced to \pythia-6.426. Figure~\ref{fig:limitSMdirect_lines} presents upper limits for case (c) on the squark pair production cross-section times branching ratio, both as a function of $m_{\squark}$ for $m_{\ninoone}$= 0, and as a function of $m_{\ninoone}$ for $m_{\squark}=450$ GeV. In cases (a) and (c), when the $\ninoone$ is massless the lower limit on the gluino mass (case (a)) is 1330~GeV, and that on the light-flavour squark mass (case (c)) is 850~GeV (440~GeV) for mass degenerate (single light-flavour) squarks. 


\begin{figure}[htb]
\begin{center}
\subfigure{\includegraphics[width=0.48\textwidth]{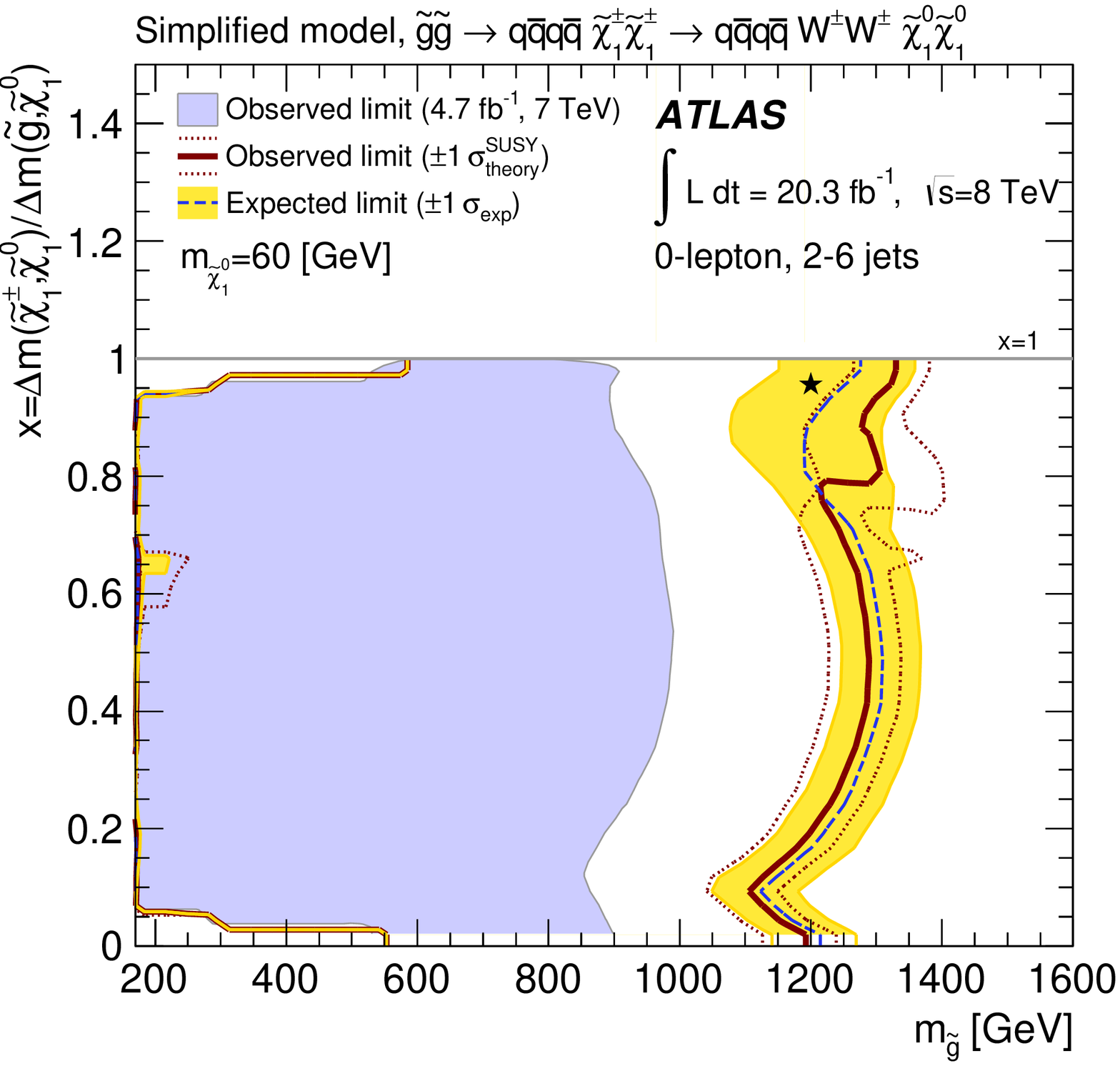}}
\subfigure{\includegraphics[width=0.48\textwidth]{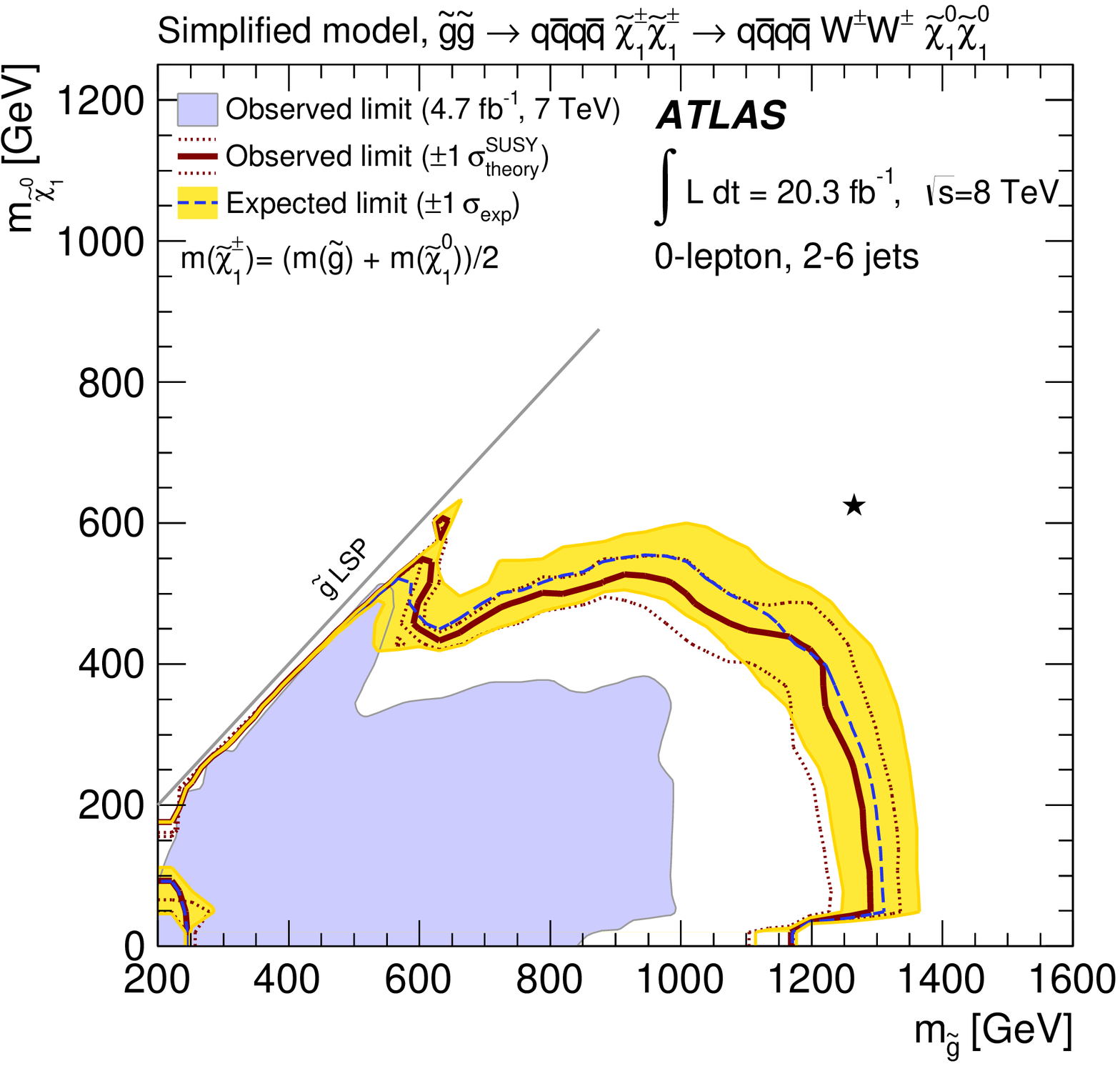}}
\subfigure{\includegraphics[width=0.48\textwidth]{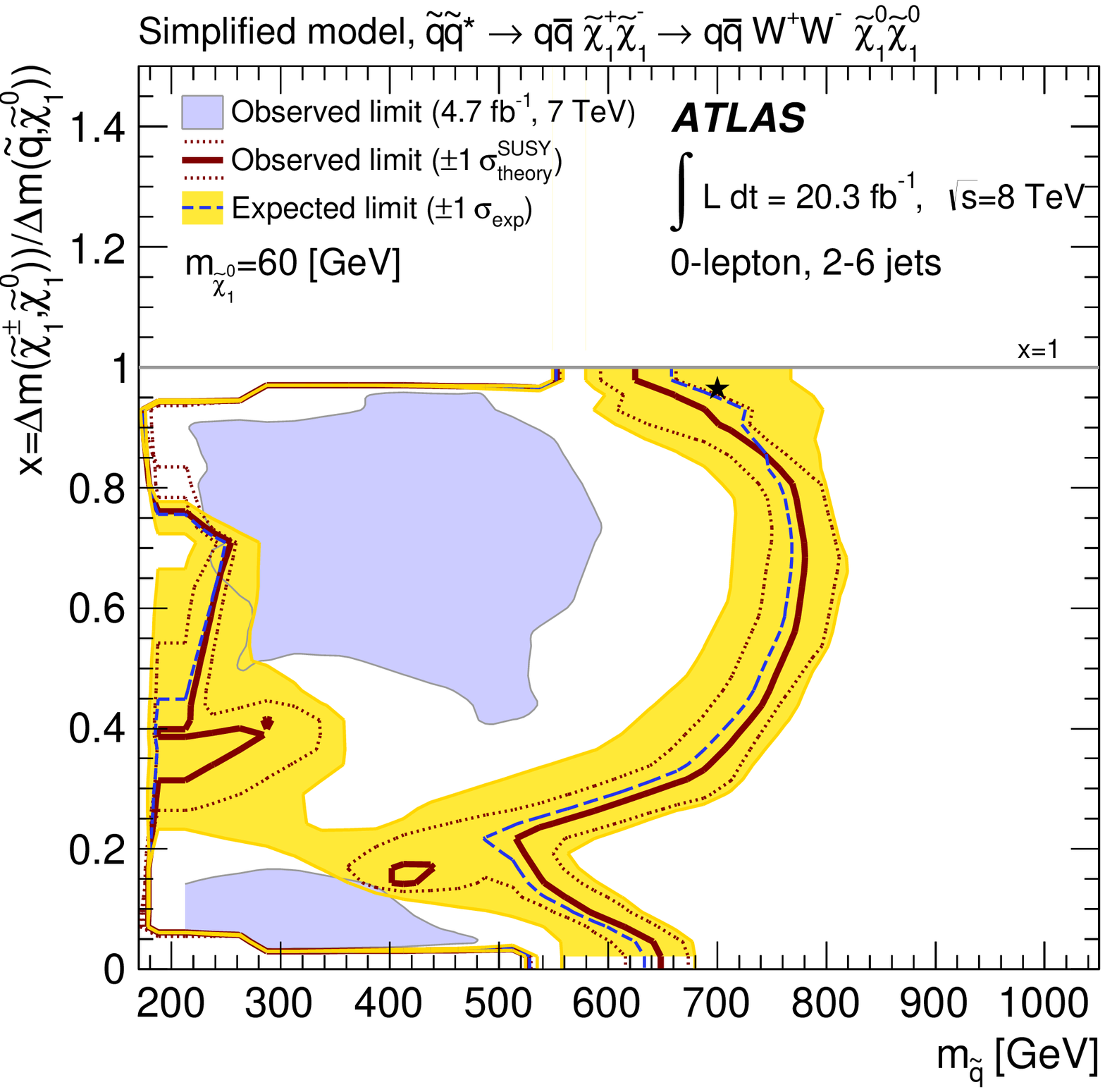}}
\subfigure{\includegraphics[width=0.48\textwidth]{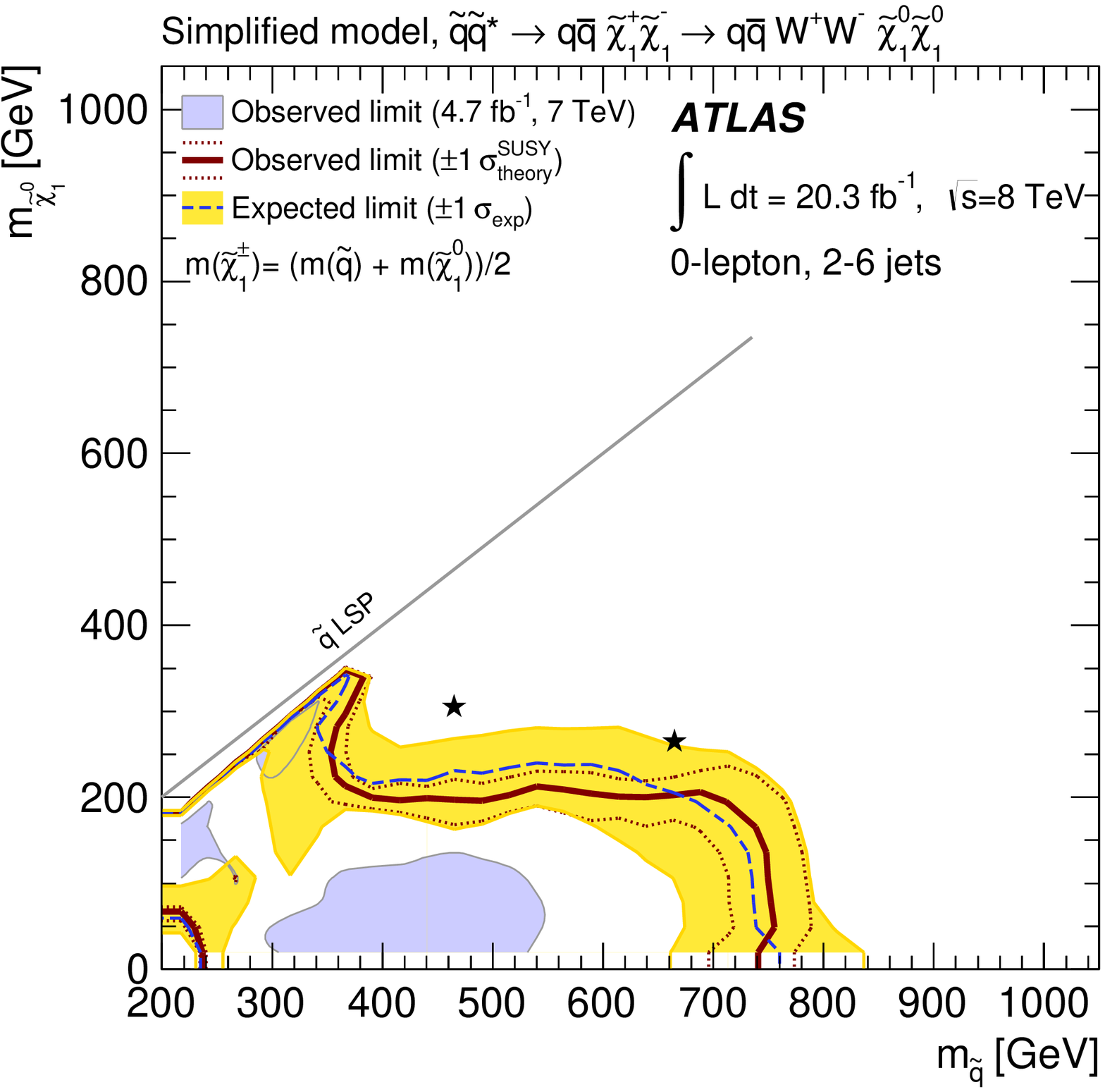}}
\caption{Exclusion limits for pair-produced gluinos each decaying via an intermediate $\chinoonepm$ to two quarks, a $W$ boson and a $\ninoone$ (top) or pair-produced light squarks each decaying via an intermediate $\chinoonepm$ to a quark, a $W$ boson and a $\ninoone$ (bottom). The left-hand figures show results for models with fixed $m(\ninoone) = 60$ GeV and varying values of 
$x = (m_{\chinoonepm} - m_{\ninoone})/(m_{y} - m_{\ninoone})$, where $y=\gluino$ ($y=\squark$) for the top (bottom) figure.
The right-hand plots show results for models with a fixed value of $x = 1/2$ and varying values of $m_{\ninoone}$. Exclusion limits are obtained by using the signal region with the best expected sensitivity at each point.  
The blue dashed lines show the expected limits at 95\% CL, with the light (yellow) bands indicating the $1\sigma$ excursions due to experimental and background-only theory uncertainties.
Observed limits are indicated by medium dark (maroon) curves, where the solid contour represents the nominal limit, and the dotted lines are obtained by varying the signal cross-section by the renormalisation and factorisation scale and PDF uncertainties. Previous results from ATLAS~\cite{Aad:2012fqa} are represented by the shaded (light blue) areas. The black stars indicate benchmark models used in figures~\ref{fig:sra}--\ref{fig:src}. }
\label{fig:limitSMonestep}
\end{center}
\end{figure}

In figure~\ref{fig:limitSMonestep} limits are shown for pair-produced gluinos each decaying via an intermediate $\chinoonepm$ to two quarks, a $W$ boson and a $\ninoone$, and pair-produced light squarks each decaying via an intermediate $\chinoonepm$ to a quark, a $W$ boson and a $\ninoone$. Results are presented for simplified models in which either the $\ninoone$ mass is fixed to 60~GeV, or the mass splitting ($x$) between the $\chinoonepm$ and the $\ninoone$, relative to that between the squark or gluino and the $\ninoone$, is fixed to $x=0.5$. These models illustrate the sensitivity of this analysis to events with multi-step decay chains involving intermediate $W$ bosons. SUSY signal events are generated with \madgraph-5.0 interfaced to \pythia-6.426. The lower limit on the gluino (squark) mass extends to 1100 GeV (700 GeV) for a massless $\ninoone$. The use of SRs 2jW and 4jW improves sensitivity to models with large $x$, for which the $\chinoonepm$ is nearly degenerate in mass with the squark or gluino. For $x\sim 1.0$ the use of these SRs improves the expected limit on the gluino (squark) mass by approximately 100 GeV (40 GeV).

\begin{figure}[h!]
\begin{center}
\subfigure{\includegraphics[width=0.49\textwidth]{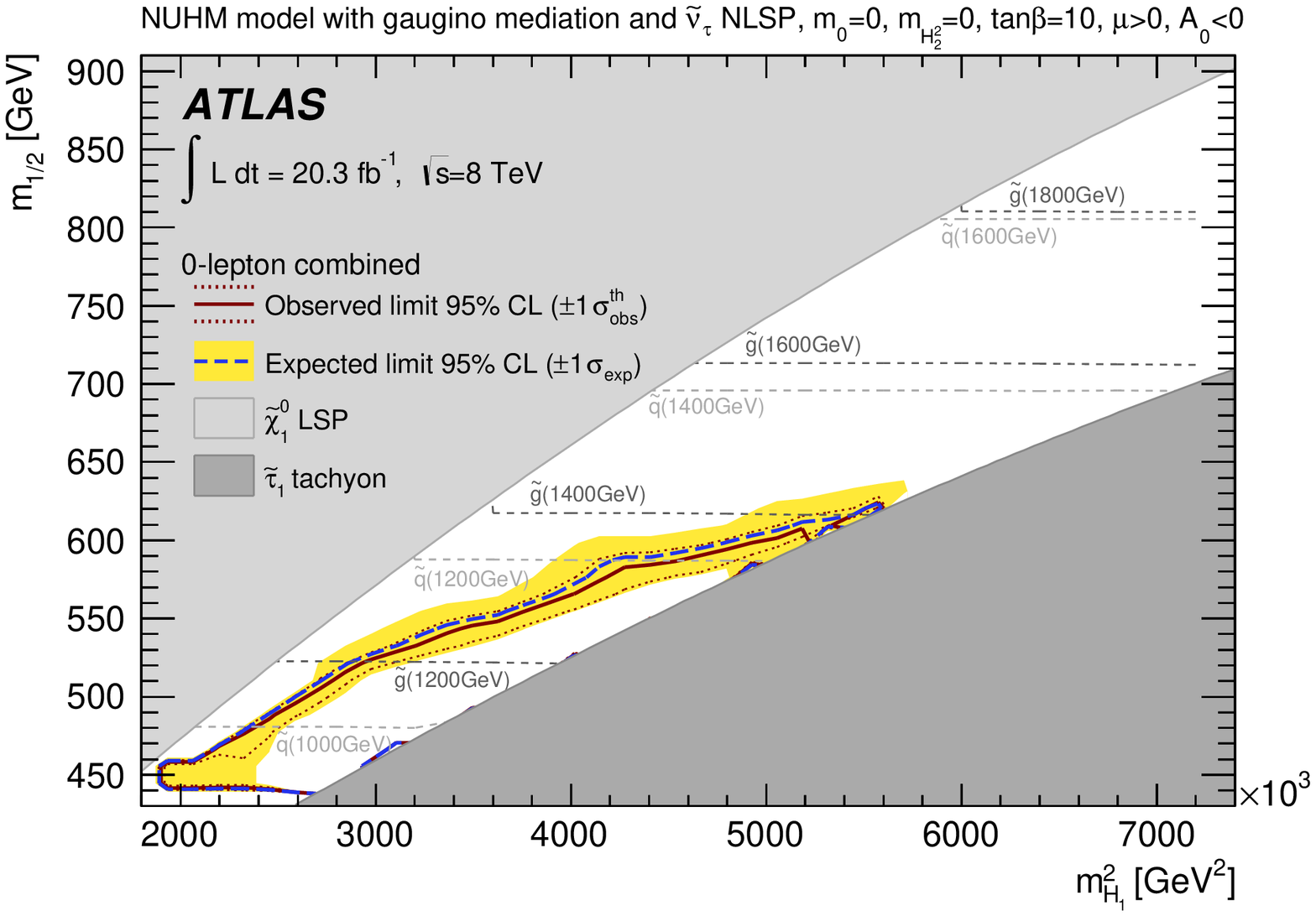}}
\subfigure{\includegraphics[width=0.49\textwidth]{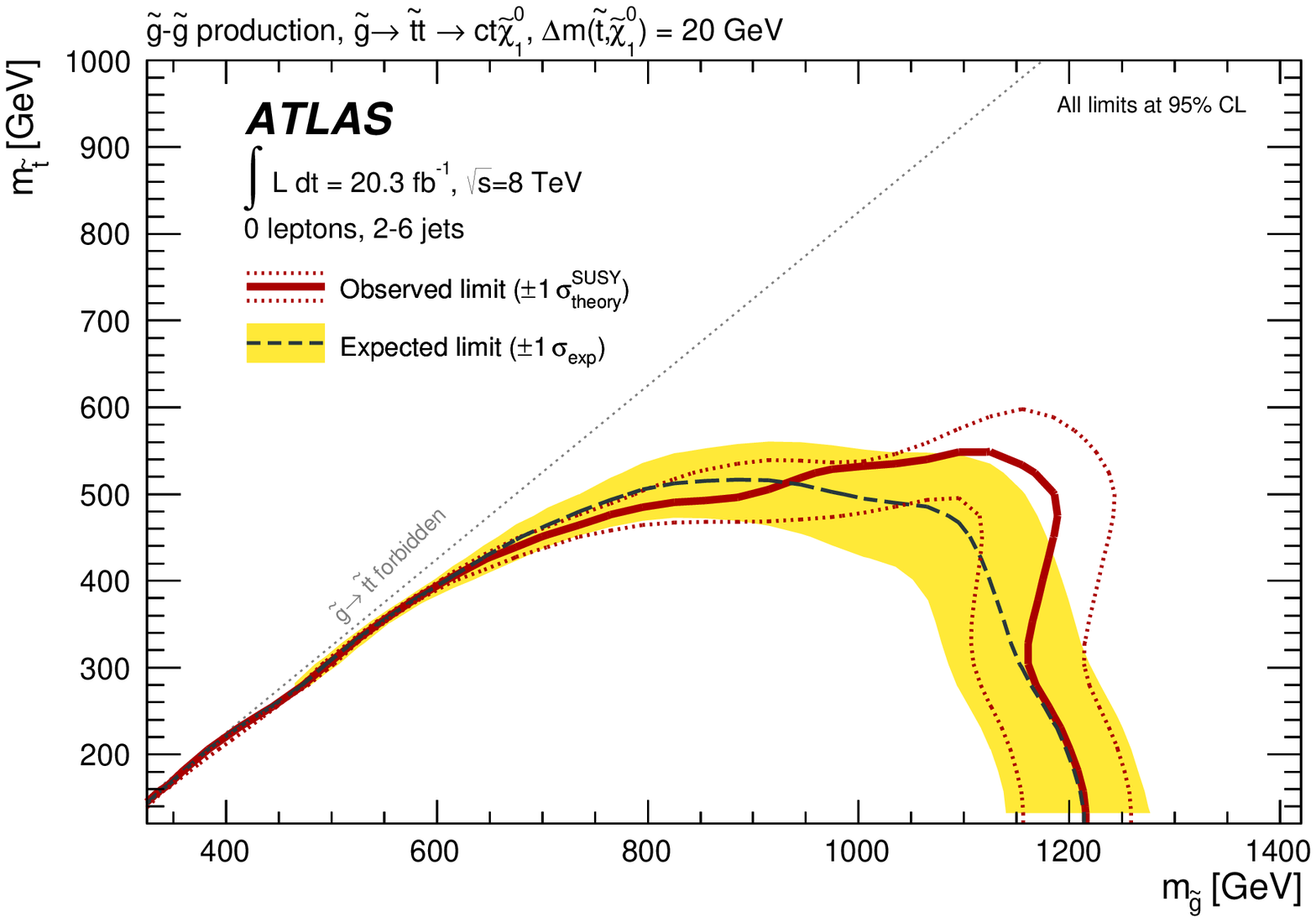}}
\caption{Exclusion limits in the 
$m_{1/2}$ versus $m_{H_{1}}^{2}$ plane for the NUHMG model described in the text (left), and
exclusion limits for pair-produced gluinos each decaying into a $\stop$ and a $\ninoone$, with the subsequent decay $\stop \to c~ \ninoone$ and $\Delta m(\stop,\ninoone)=20$ GeV (right). Exclusion limits are obtained by using the signal region with the best expected sensitivity at each point.  
The blue dashed lines show the expected limits at 95\% CL, with the light (yellow) bands indicating the $1\sigma$ excursions due to experimental and background-only theory uncertainties.
Observed limits are indicated by medium dark (maroon) curves, where the solid contour represents the nominal limit, and the dotted lines are obtained by varying the signal cross-section by the renormalisation and factorisation scale and PDF uncertainties. The $m_{\squark}$ contours in the left-hand figure are calculated using the mean of the masses of the light squarks, excluding those of the top and bottom squarks.}
\label{fig:limitNUHMG}
\end{center}
\end{figure}

In figure~\ref{fig:limitNUHMG} (left) the results are interpreted in the context of a non-universal Higgs mass model with gaugino mediation (NUHMG)~\cite{Covi:2007xj} with parameters $m_0=0$, $\tan \beta$ = 10, $\mu > 0$, $m_{H_{2}}^2$ = 0, and $A_{0}$ chosen to maximise the mass of the lightest Higgs boson. The ranges of the two remaining free parameters of the model, $m_{1/2}$ and $m_{H_{1}}^2$, are chosen such that the next-to-lightest SUSY particle (NLSP) is a tau sneutrino with properties satisfying Big Bang nucleosynthesis constraints. This model is characterised by significant cross-sections for $\squark$ and $\gluino$ production. SUSY signal events are generated with \pythia-6.426.

In figure~\ref{fig:limitNUHMG} (right) limits are presented for a simplified phenomenological SUSY model in which pairs of gluinos are produced, each of which then decays to a top squark and a top quark, with the top squark decaying to a charm quark and $\ninoone$. This model is motivated by `natural' SUSY scenarios with a light top squark and a small mass splitting between the top squark and the $\ninoone$ leading to co-annihilation between top squarks and $\ninoone$ dark matter particles in the early universe. SUSY signal events are generated with \herwig++-2.5.2. The lower limit on the gluino mass extends to 1110~GeV for a top squark of mass 400~GeV.

\clearpage

\section{Conclusions}

This paper reports a search for squarks and gluinos in final states containing
high-\pT{} jets, large missing transverse momentum and no electrons or muons, based on a 20.3 \ifb{} dataset of $\sqrt{s}=8$ TeV proton--proton collisions recorded by the ATLAS experiment at the LHC in 2012.
Good agreement is seen between the numbers of events observed in the data and the numbers of events expected from SM processes. 

Results are interpreted in terms of \mSUGRA{} models with $\tan\beta=30$, $A_0=-2m_0$ and $\mu>0$, and in terms of simplified models with only light-flavour squarks, or gluinos, or both, together with a neutralino LSP, with the other SUSY particles decoupled. The results are also interpreted in terms of several other SUSY models. In the \mSUGRA{}
models, the 95\% confidence level exclusion limit on $m_{1/2}$ is greater than 380 GeV for $m_0<6$ TeV and reaches 770 GeV for low $m_0$. Equal mass squarks and gluinos are excluded below 1700~GeV in this scenario. 
A lower limit of 1650 GeV for equal mass light-flavour squarks and gluinos is found for simplified MSSM models with a massless lightest neutralino.
For a massless lightest neutralino, gluino masses below 1330~GeV are excluded at the 95\% confidence level in a simplified model with only gluinos and the lightest neutralino.
For a simplified model involving the strong production of squarks of the first and second generations, with decays to a massless lightest neutralino, squark masses below 850~GeV (440~GeV) are excluded, assuming mass degenerate (single light-flavour) squarks. For simplified models involving the pair production of gluinos, each decaying to a top squark and a top quark, with the top squark decaying to a charm quark and a neutralino, the lower limit on the gluino mass extends to 1110~GeV for a top squark of mass 400~GeV. These results extend the region of supersymmetric parameter space excluded by previous searches with the ATLAS detector.

%

\section*{Acknowledgements} 
We thank CERN for the very successful operation of the LHC, as well as the
support staff from our institutions without whom ATLAS could not be
operated efficiently.

We acknowledge the support of ANPCyT, Argentina; YerPhI, Armenia; ARC,
Australia; BMWF and FWF, Austria; ANAS, Azerbaijan; SSTC, Belarus; CNPq and FAPESP,
Brazil; NSERC, NRC and CFI, Canada; CERN; CONICYT, Chile; CAS, MOST and NSFC,
China; COLCIENCIAS, Colombia; MSMT CR, MPO CR and VSC CR, Czech Republic;
DNRF, DNSRC and Lundbeck Foundation, Denmark; EPLANET, ERC and NSRF, European Union;
IN2P3-CNRS, CEA-DSM/IRFU, France; GNSF, Georgia; BMBF, DFG, HGF, MPG and AvH
Foundation, Germany; GSRT and NSRF, Greece; ISF, MINERVA, GIF, I-CORE and Benoziyo Center,
Israel; INFN, Italy; MEXT and JSPS, Japan; CNRST, Morocco; FOM and NWO,
Netherlands; BRF and RCN, Norway; MNiSW and NCN, Poland; GRICES and FCT, Portugal; MNE/IFA, Romania; MES of Russia and ROSATOM, Russian Federation; JINR; MSTD,
Serbia; MSSR, Slovakia; ARRS and MIZ\v{S}, Slovenia; DST/NRF, South Africa;
MINECO, Spain; SRC and Wallenberg Foundation, Sweden; SER, SNSF and Cantons of
Bern and Geneva, Switzerland; NSC, Taiwan; TAEK, Turkey; STFC, the Royal
Society and Leverhulme Trust, United Kingdom; DOE and NSF, United States of
America.

The crucial computing support from all WLCG partners is acknowledged
gratefully, in particular from CERN and the ATLAS Tier-1 facilities at
TRIUMF (Canada), NDGF (Denmark, Norway, Sweden), CC-IN2P3 (France),
KIT/GridKA (Germany), INFN-CNAF (Italy), NL-T1 (Netherlands), PIC (Spain),
ASGC (Taiwan), RAL (UK) and BNL (USA) and in the Tier-2 facilities
worldwide.

\clearpage


\bibliographystyle{atlasnote}
\bibliography{SUSY-2013-02}

\clearpage

\begin{flushleft}
{\Large The ATLAS Collaboration}

\bigskip

G.~Aad$^{\rm 84}$,
B.~Abbott$^{\rm 112}$,
J.~Abdallah$^{\rm 152}$,
S.~Abdel~Khalek$^{\rm 116}$,
O.~Abdinov$^{\rm 11}$,
R.~Aben$^{\rm 106}$,
B.~Abi$^{\rm 113}$,
M.~Abolins$^{\rm 89}$,
O.S.~AbouZeid$^{\rm 159}$,
H.~Abramowicz$^{\rm 154}$
% ...
\bigskip
\\
% ...
$^{11}$ Institute of Physics, Azerbaijan Academy of Sciences, Baku, Azerbaijan\\
% ...
$^{84}$ CPPM, Aix-Marseille Universit{\'e} and CNRS/IN2P3, Marseille, France\\
% ...
$^{89}$ Department of Physics and Astronomy, Michigan State University, East Lansing MI, United States of America\\
% ...
$^{106}$ Nikhef National Institute for Subatomic Physics and University of Amsterdam, Amsterdam, Netherlands\\
% ...
$^{112}$ Homer L. Dodge Department of Physics and Astronomy, University of Oklahoma, Norman OK, United States of America\\
$^{113}$ Department of Physics, Oklahoma State University, Stillwater OK, United States of America\\
% ...
$^{116}$ LAL, Universit{\'e} Paris-Sud and CNRS/IN2P3, Orsay, France\\
% ...
$^{152}$ Institute of Physics, Academia Sinica, Taipei, Taiwan\\
% ...
$^{154}$ Raymond and Beverly Sackler School of Physics and Astronomy, Tel Aviv University, Tel Aviv, Israel\\
% ...
$^{159}$ Department of Physics, University of Toronto, Toronto ON, Canada\\
% ...
\end{flushleft}


\end{document}